\documentclass[a4paper,11pt]{article}
\pdfoutput=1

\usepackage[T1]{fontenc}
\usepackage{csquotes}

\usepackage{xcolor}

\usepackage{jheppub}

\usepackage{tikz}
\usepackage{latexsym}
\usepackage{slashed}
\usepackage{braket}
\usepackage{float}
\usepackage[symbol]{footmisc}
\usepackage{multirow}
\usepackage{stmaryrd}

\def\myupsilon{\zeta}
\def\calF{{\mathcal F}}
\def\calG{{\mathcal G}}
\def\calS{{\mathcal S}}
\def\calW{{\mathcal W}}

\hypersetup{
    colorlinks=true,
    allcolors=blue
}

\title{Defect conformal field theories with free-in-the-bulk fermions}

\author{Christopher P.\ Herzog}
\author{and Daniel Lloyd-Jones}
\affiliation{Department of Mathematics, King's College London, \\  Strand, London, WC2R 2LS, UK}

\emailAdd{christopher.herzog@kcl.ac.uk}
\emailAdd{daniel.lloyd-jones@kcl.ac.uk}

\abstract{ 
We study codimension-two conformal defects for massless fermions that are free in a four-dimensional bulk, but interact on a two-dimensional defect. The bulk Dirac equation strongly constrains defect conformal field theory data: it fixes the dimensions of operators in the bulk-defect operator product expansion and, for bulk-defect-defect three-point functions, enforces a double-twist relation when only one member of a shadow pair is allowed by unitarity. The lowest modes of monodromy defects provide an important exception, since both shadow partners can be unitary and their three-point functions are instead related by a shadow transform.

We illustrate these structures with defect interactions, including an exactly tractable bilinear flow and Gross-Neveu-type quartic couplings.
At a perturbative mixed fixed point, we find an explicit interacting example in which the double-twist condition is violated and the two shadow channels combine to give a regular bulk-defect-defect correlation function.

}

\begin{document}

\maketitle

\section{Introduction}

Defect and boundary conformal field theories provide a conceptual foundation for many topics in theoretical physics,
from critical systems in condensed matter, to entanglement phenomena in many-body physics, to D-branes in string theory.
By definition, defect conformal field theories are $d$-dimensional quantum field theories where the presence of a $p$-dimensional defect reduces the (Lorentzian) conformal symmetry from $SO(d,2)$ to a $SO(p,2) \times SO(d-p)$ subgroup.
The emphasis in this work is on a computationally tractable yet still physically relevant subset, where the bulk is free and gapless, and the interactions
are confined to the defect or boundary.   
In the past, the majority of work on these free-in-the-bulk systems has focused on cases where the bulk field is a scalar \cite{Herzog:2017xha,Lauria:2020emq,Behan:2020nsf,Behan:2021tcn} or a Maxwell field \cite{Herzog:2017xha,DiPietro:2019hqe,Herzog_2022,Fraser-Taliente:2024lea}.  Here we investigate free-in-the-bulk fermions.

A few strands of previous work are particularly relevant.  In the codimension-two case, monodromy defects --- where a free bulk fermion picks up an Aharonov-Bohm-like phase when travelling around the defect --- were studied in detail \cite{Bianchi:2021snj}.  The codimension-two case with free bulk fermions was also studied in the context of crystalline defects in lattice systems \cite{Barkeshli:2025cjs}. We place these earlier results in a more general setting, allowing for interactions on the defect.  Also considered were long-range fermionic models, where a nonlocal kinetic term for the fermion in a conformal field theory can be interpreted as coming from integrating out a higher-dimensional, free bulk fermion.  
The choice $s=0$ of the nonlocality parameter in  \cite{Chai:2021wac} matches onto the ($d=4$, $p=2$) system we study in detail here, although this value is non-perturbative from the point of view of the epsilon expansion in that work. 
The codimension-one $s=1$ version has recently appeared naturally at Chern-insulator boundary critical points, where a localized chiral edge fermion delocalizes into a free critical bulk and acquires the scaling dimension of the bulk fermion \cite{Moy:2026iey}. 

While ideally it would be nice to consider the free-in-the-bulk fermions in a dimension-agnostic way, the sensitivity of the Clifford algebra to dimension makes this program difficult.  For simplicity, and also with a specific bulk-defect coupling in mind, we focus on the codimension-two case.  Classically, the bulk fermion $\psi$ will have scaling dimension $\Delta_\psi = \frac{p+1}{2}$, while a defect fermion $\chi$ would have $\Delta_\chi = \frac{p-1}{2}$.
Our principal case is $p=2$, although we keep $p$ general where useful.
Putting aside for the moment details of how to contract these different dimensional spinors, 
there is a classically marginal coupling of the schematic form,
\begin{equation}
\label{bilinearcoupling}
S_{\rm int} = g \int d^2 x (\bar \psi \chi + \bar \chi \psi) \ ,
\end{equation}
which allows for the application of conformal perturbation theory and the study of short renormalization group flows.  

Just as the Klein-Gordon equation and Maxwell's equations put severe restrictions on theories with a free-in-the-bulk scalar or photon \cite{Lauria:2020emq,DiPietro:2019hqe,Herzog_2022}, the Dirac equation places restrictions in our case. Consider the bulk-defect two-point function $\langle \psi(x) {\mathcal O}(x') \rangle$, which is nonzero only when ${\mathcal O}$ is a defect field in the defect operator expansion of the bulk fermion $\psi$.  As we demonstrate in section \ref{sec:bulkdefect}, that this two-point function is in the kernel of the Dirac operator $\langle\slashed{\partial} \psi(x) {\mathcal O}(x') \rangle = 0$ imposes a relation between the transverse $SO(d-p)$ 
angular momentum of ${\mathcal O}$ and its scaling dimension.  

Another way of understanding this restriction is through Kaluza-Klein reduction.   
As we are working with a conformal field theory, we are free to perform a Weyl transform to $AdS_{p+1} \times S^{d-p-1}$, where the boundary of $AdS_{p+1}$ can be thought of as the defect.  After this Weyl transformation, the bulk spinor $\psi$ decomposes into a tower of KK modes on the $S^{d-p-1}$ characterized by increasing values of transverse orbital angular momentum $l$, each satisfying its own massive Dirac equation, with two distinguished asymptotic modes.  The corresponding boundary data transform as the defect conformal primary operators found by imposing the Dirac equation on the bulk-defect two-point function in flat space.

From a geometric perspective, if we specify one boundary condition on $AdS$ and a regularity condition in the interior, the other boundary condition is fixed by the uniqueness of solutions to differential equations.  (In the scalar case, this map is sometimes called a Dirichlet-to-Neumann or Neumann-to-Dirichlet map.)  From the QFT point of view, this map gives us an effective equation of motion that specifies one defect primary (i.e.\ boundary condition) in terms of its shadow partner.  If we call one defect primary ${\mathcal W}$ and its shadow ${\mathcal S}$, this equation of motion has the schematic form
\begin{equation}
G_{\mathcal S}^{-1} {\mathcal S} = {\mathcal W}
\end{equation}
where $G_{\mathcal S}^{-1}$ is a non-local differential operator, whose inverse is the integral transform with kernel the $G_{\mathcal S}(x-x') = \langle {\mathcal S}(x) \bar{\mathcal S}(x') \rangle$ two-point function.  This equation of motion imposes a corresponding set of relations between correlation functions involving ${\mathcal W}$ and ${\mathcal S}$, up to contact terms.  The relation is the shadow transform, as was emphasized in the scalar case in ref.\ \cite{Behan:2020nsf}.

If we further impose unitarity, we find that for almost every member of the KK tower, one of the defect primaries is below the unitarity bound, forcing  ${\mathcal S}=0$.  
If the shadow transform had a trivial kernel, that would imply all the correlation functions involving ${\mathcal W}$ should also vanish.  In fact, however, the shadow transform has a nontrivial kernel.  For three-point functions involving ${\mathcal W}$  plus two more defect primaries, the shadow transform vanishes if a double twist condition is satisfied.  
Given the boundary condition ${\mathcal S} = 0$, for $\langle {\mathcal W}(x_1) {\mathcal O}_2(x_2) {\mathcal O}_3(x_3) \rangle$ to be nonzero, schematically ${\mathcal O}_3 = {\mathcal O}_2 (\slashed{\partial})^n {\mathcal W} + \ldots$ (or equivalently with ${\mathcal O}_2$ swapped with ${\mathcal O}_3$).  

For our 2d/4d fermions, ${\mathcal S}=0$ 
by unitarity, unless the defect is of monodromy type.  For a monodromy defect, the ${\mathcal W}$ and ${\mathcal S}$ pairs at the bottom of the KK tower can both sit above the unitarity bound. In this case, the shadow transform imposes a nontrivial relation between correlation functions involving them.  (Later in the text, we will call this pair of defect spinors $u$ and $d$.) These features of bulk-defect-defect correlation functions and the shadow transform are explained in section \ref{sec:bulkdefectdefect}.

Having used conformal symmetry and the Dirac operator shortening condition to place general constraints on these defect theories, we consider several examples that illustrate these constraints in section \ref{sec:examples}.  The first and simplest example is a free bulk fermion, $\psi$, coupled to a free defect fermion, $\chi$, through the bilinear (\ref{bilinearcoupling}).
By either introducing a monodromy defect or through dimensional regularization, the coupling can be made slightly relevant.  The IR fixed point is a theory where the bulk $\psi$ field absorbs the $\chi$ field in its defect operator product expansion, and changes its boundary conditions. 
The scale-dependent defect Green's function also exhibits marginal-Fermi-liquid scaling in the non-monodromy limit and its power-law generalization for nonzero monodromy. The second example further introduces a quartic coupling $h (\bar \chi \chi)^2$ on the defect.  
For a small range of couplings $h$, bosonization determines that at $g=0$ the bilinear coupling (\ref{bilinearcoupling}) remains marginally relevant, and we conjecture the IR fixed point is the same as before -- the $\chi$ field is absorbed by the bulk fermion $\psi$, and the boundary conditions for $\psi$ are altered.

Finally, we  introduce several flavours of defect fermion $\chi_i$, $i = 1, \ldots, N_f$, and study the RG behaviour perturbatively, now with the quartic coupling $\frac{h}{2} (\bar \chi_i \chi_i)^2$. 
At the mixed fixed point $g^*$, $h^* \neq 0$, the anomalous dimensions no longer obey the double-twist relation.
The set-up realizes the mechanism described above, where both members of a shadow pair contribute to a bulk-defect-defect three-point function, with coefficients related nontrivially via the shadow transform, so as to cancel a spurious bulk singularity.

An outline of our work is as follows.  In section \ref{sec:conformalconstraints}, we use embedding space to fix the conformal form of several two- and three-point correlation functions.  Next, sections \ref{sec:bulkdefect}, \ref{sec:bulkbulk} and \ref{sec:bulkdefectdefect} explore the consequences on these correlation functions when the bulk field satisfies the Dirac equation.  Concrete examples of these defect fermion theories are presented in section \ref{sec:examples}, coupling the bulk fermion to a defect fermion $\chi$ through the bilinear (\ref{bilinearcoupling}).
Section \ref{sec:discussion} concludes with a brief discussion.
Appendices \ref{sec:embeddingspace}, \ref{sec:OPEops}, \ref{sec:conformalblocks}, and \ref{sec:bddexpansions}
contain technical details about the embedding space formalism, the defect conformal block construction, and bulk-defect-defect correlation functions.

\section{Correlation functions} 
\label{sec:conformalconstraints}
To impose the constraints of conformal symmetry on the correlation functions, it is convenient to work in embedding space, where the action of the conformal group is linearized.  Our conventions for embedding space are spelled out in appendix \ref{sec:embeddingspace}.
 
Embedding space has two extra dimensions compared to the actual spacetime of the CFT.
Further, our spacetime has two more dimensions than our defect.
In our conventions, a spacetime point $x$, where an operator is inserted, is lifted to a null vector $P$.  We will also have a set of gamma matrices $\Gamma_M$, $M = 0, 1, \ldots, d+1$ in the embedding space, as well as  embedding space spinors $S$.
In real space, our gamma matrices $\gamma_\mu$ run from $\mu = 1, \ldots, d$, where $\mu=d-1$ and $d$ are normal to the defect, which we also label as $n_1$ and $n_2$.  We also have spinors $\psi$ in the bulk and $\rho$ (and eventually $\chi$, $u$ and $d$ as well) on the defect.  

In the following, we present the general forms for bulk-defect, defect-defect, bulk-bulk, bulk-defect-defect, and defect-defect-defect correlation functions consistent with codimension-two defect conformal symmetry.  Although we restrict to spinor and scalar operator insertions, the relevant behaviours we hope to capture are already present in this sub-class.

\subsection{Bulk-defect}
The most general bulk-defect correlation function compatible with conformal symmetry is
\begin{equation} \label{bulk-defect}   \left<\psi_\pm(x_1)\bar{\rho}_\alpha^{(l)}(x_2)\right>=\frac{{\mathbb P}_\pm^* \slashed{x}_{12}}{|z|^{\Delta_1-\Delta_2}|x_{12}|^{2\Delta_2+1}} \left( c_{\pm 1}+c_{\pm 2} \frac{\slashed{z}}{|z|}\right)\mathbb{P}^\perp_\alpha\left(\frac{z}{|z|}\right)^l \ , \\
\end{equation}
where the coordinates are labelled as $x_1=(\vec{x}_1,\vec{z})=(\vec{x}_1,z, \bar z)$ and $x_2=(\vec{x}_2,0)$. 
In even dimensions, we isolate the contributions of the two different chiral bulk spinors $\psi_\pm$ to the correlation function, using the projectors ${\mathbb P}_\pm^* = \frac{1}{2}(1 \pm \gamma_*)$, where $\gamma_*$ is ``gamma-five''.  In odd dimensions, ${\mathbb P}_\pm^*$ can be replaced with the identity.
Even with this chiral separation, there remain two allowed structures with c-number coefficients $c_{\pm 1}$ and $c_{\pm 2}$.  We need several further quantum numbers to specify the correlation function completely.  The real numbers $\Delta_1$ and $\Delta_2$ are the scaling dimensions of the bulk $\psi$ and defect $\rho_\alpha^{(l)}$ operators, respectively.  
The defect operator also carries angular momentum quantum numbers associated with the plane transverse to the defect, coordinatized by the complex number $z$.
The orbital angular momentum 
is $l$.  There is a further spin
angular momentum $-\alpha/2$, $\alpha = \pm 1$.
Here $-\alpha/2$ is the eigenvalue of $\rho_\alpha^{(l)}$ in the spinor representation, with respect to the Lorentz generator $-\frac{i}{2} \gamma_{n_1} \gamma_{n_2}$ in the $z$-plane.
The projector $\mathbb{P}^\perp_\alpha \equiv \frac{1}{2}(1 + i \alpha \gamma_{n_1} \gamma_{n_2})$ projects onto these defect spinors with spin $-\alpha/2$.

Our chirality notation here is idiosyncratic.  The subscript on the spinor indicates its eigenvalue under the action of the corresponding chirality generator, from the left for unbarred spinors, and from the right for barred spinors.  Because our bar notation includes the action of $\gamma_1$ from the right, one has $\overline {\psi_+} = \bar \psi_-$, while $\overline{\rho_+} = \bar \rho_+$.

This spacetime correlation function follows easily from an embedding space argument.
To keep things simple, let us specialize to the case $l=0$.  
The most general embedding space bulk-defect correlation function is then
\begin{equation} 
\label{embeddingbulkdefect}
\begin{split}    \left<\psi(P_1,\bar{S}_1)\bar{\rho}_\alpha(P_2,S_{2,\alpha})\right>=&\frac{c_1 \bar{S}_1 S_{2,\alpha}+c_2 \bar{S}_1\Gamma^*S_{2,\alpha}}{(P_1\circ P_1)^{\frac{1}{2}(\Delta_1-\Delta_2)}(P_1\bullet P_2)^{\Delta_2+\frac{1}{2}}}\\&+\frac{c_3 \bar{S}_1 \slashed{P}_{1\perp} S_{2,\alpha}+c_4 \bar{S}_1 \Gamma^*\slashed{P}_{1\perp} S_{2,\alpha}}{(P_1\circ P_1)^{\frac{1}{2}(\Delta_1-\Delta_2+1)}(P_1\bullet P_2)^{\Delta_2+\frac{1}{2}}} \ .\\
\end{split}
\end{equation}
The inner products $\bullet$ and $\circ$ are parallel and transverse to the defect respectively.

The exponents in the denominator are fixed by scaling symmetry
\begin{equation}\label{scaling}
\left<\psi(\lambda_1 P_1,\bar{S}_1)\bar{\rho}_\alpha(\lambda_2 P_2,S_{2,\alpha})\right>=\lambda_1^{-\Delta_1-\frac{1}{2}}\lambda_2^{-\Delta_2-\frac{1}{2}}\left<\psi(P_1,\bar{S}_1)\bar{\rho}_\alpha(P_2,S_{2,\alpha})\right>.
\end{equation}

We allow for transverse chirality $\alpha$ using the eigenvalue condition $i\Gamma_{n_1}\Gamma_{n_2}S_\pm(P)=\pm S_\pm(P)$.
We also allow for a bulk chirality through 
$\Gamma^* = \Gamma_1 \cdots 
\Gamma_{p+2} \Gamma_{n_1} \Gamma_{n_2}$.  
The chirality along the defect, established through the eigenvalue of $\Gamma_1 \cdots \Gamma_{p+2}$ is then fixed implicitly in terms of these other two chiralities.

The game reduces to writing spinor sandwiches of a bulk polarisation spinor $\bar S_1$, defect polarisation spinor $S_{2,\alpha}$ and all possible $P_1$, $P_2$, and $\Gamma^*$ objects in between.
Clearly the four numerators in (\ref{embeddingbulkdefect}) are allowed. 
The transversality condition \eqref{condition} means the objects,
\begin{equation}
     \bar{S}_1 \slashed{P}_1 S_{2,\alpha} =\bar{S}_1 \slashed{P}_2 S_{2,\alpha}=0 \ ,
\end{equation}
vanish.  
Thus the sandwiches involving $P_{1\parallel}$
\begin{equation}
      \bar{S}_1 \slashed{P}_{1\perp}S_{2,\alpha}=
    -\bar{S}_1 \slashed{P}_{1\parallel}S_{2,\alpha}    
\end{equation}
are not independent of those involving $P_{1 \perp}$.  Since $P_2$ has no transverse component, $P_{2} S_{2,\alpha} = 0$ on its own by transversality.  We claim the set of four numerators in (\ref{embeddingbulkdefect}) in fact exhausts the list of available conformally covariant structures. 

Reducing from embedding space to spacetime, we make the replacements
\begin{eqnarray}
\bar S_1  S_{2,\alpha} &\to &  \bar s_1 \slashed{x}_{12} s_{2,\alpha} \ , \\
\bar S_1 \slashed{P}_{1\perp} S_{2,\alpha} &\to & \bar s_1 \slashed{x}_{12} \slashed{z} s_{2,\alpha} \ ,
\end{eqnarray}
recovering (\ref{bulk-defect}) in the $l=0$ case.  
Adding the factor $(z / |z|)^l$ then generates generic orbital angular momentum dependence.

\subsection{Defect-defect}
Choosing a canonical normalization for the defect spinors $\rho$, the most general defect-defect correlation function is 
\begin{equation} \label{defect-defect}    \left<\rho^{(l)}_\alpha(x_1)\bar{\rho}^{(l)}_\alpha(x_2)\right>=\frac{ \mathbb{P}^*_{\pm}\slashed{x}_{12}\mathbb{P}_\alpha^\perp}{|x_{12}|^{2\Delta+1}}.
\end{equation}
The ${\mathcal P}^*_\pm$ is meant to indicate that we are writing this correlation function separately in the two bulk chirality sectors.
Since there is no coordinate $\vec{z}$ in this correlation function to carry the orbital angular momentum, both defect spinors must take the same value of $l$ so that under the rotation, the factors $e^{il\theta}$ and $e^{-il\theta}$ cancel out. Both spinors must also take the same value of $\alpha$, since the projectors satisfy $\mathbb{P}_\alpha^2=\mathbb{P}_\alpha$ but $\mathbb{P}_\alpha \mathbb{P}_{-\alpha}=0$.
\par Expression (\ref{defect-defect})  follows from the embedding space result
\begin{equation}    \left<\rho_\alpha(P_1,\bar{S}_{1,\alpha})\bar{\rho}_\alpha(P_2,S_{2,\alpha})\right>=\frac{c_1 \bar{S}_{1,\alpha}S_{2,\alpha}+c_2 \bar{S}_{1,\alpha}\Gamma^*S_{2,\alpha}}{(P_1\bullet P_2)^{\Delta+\frac{1}{2}}},
\end{equation}
where eq.\ \eqref{scaling} and the fact that only $P_1\bullet P_2$ is nonzero enforce that both spinors have the same scaling dimension $\Delta$.

\subsection{Bulk-bulk}
For the bulk-bulk correlation function, since both $x_1=(\vec{x}_1,\vec{z}_1)$ and $x_2=(\vec{x}_2,\vec{z}_2)$ are nonzero in the normal directions, we can introduce cross-ratios in real space that are invariant under the conformal group:
\begin{align} \label{cross-ratios}
    &\xi_1=\frac{(x_{12})^2}{4|z_1||z_2|},&\xi_2=\frac{z_1\cdot z_2}{|z_1||z_2|}.
\end{align}
The general correlation function is comprised of unknown functions $f_i(\xi_1, \xi_2)$, $i =1, \ldots, 4$, of these variables, and is given by

\begin{equation} \label{bulk-bulk}
\begin{split}
    &\left<\psi_a(x_1)\bar{\psi}_b(x_2)\right>
    = \\
    & \mathbb{P}^*_a\Bigg[\frac{1}{|z_1|^{\Delta_1+\frac{1}{2}}|z_2|^{\Delta_2+\frac{1}{2}}}\Big(f_1\slashed{x}_{12}+ f_2\gamma_{n_1}\gamma_{n_2} \slashed{\Tilde{x}}_{12}\Big)\\
    & \,\,+\frac{1}{|z_1|^{\Delta_1+\frac{3}{2}}|z_2|^{\Delta_2+\frac{1}{2}}}\Big(f_3
    (\slashed{x}_{12} \slashed{z}_1+2z_1\cdot z_2)+f_4\gamma_{n_1}\gamma_{n_2}(\slashed{\Tilde{x}}_{12} \slashed{z}_1+2z_1\cdot z_2)\Big)\Bigg]\mathbb{P}^*_b\\
\end{split}
\end{equation}
where 
\begin{equation*}
    \tilde{x}^\mu=x^\mu-2n_1^\mu \,x\cdot n_1-2n^\mu_2 \,x\cdot n_2=(\vec{x},-\vec{z}) \ , \; \; \; \slashed{\Tilde{x}}_{12} = \slashed{\Tilde{x}}_{1} - \slashed{x}_2 \ .
\end{equation*}
The projectors are only present in even dimensions (where $a$ and $b$ take the values of $\pm1$), in which this bulk-bulk correlation function decomposes into different chirality pieces. The terms proportional to $f_1$ and $f_2$ describe the chirality preserving correlators
$\langle \psi_\pm \bar \psi_{\mp} \rangle=\langle \psi_\pm \overline{\psi_{\pm}} \rangle$, while
$f_3$ and $f_4$ measure chirality conversion $\langle \psi_\pm \bar \psi_{\pm} \rangle = \langle \psi_\pm \overline{\psi_{\mp}} \rangle$ by the defect.

Eq.\ (\ref{bulk-bulk}) comes from the embedding space expression 
\begin{equation}
\begin{split}
   \langle \psi(P_1,\bar{S}_1) & \bar{\psi}(P_2, S_{2}) \rangle = \frac{1}{(P_1\circ P_1)^{\frac{1}{2}(\Delta_1+\frac{1}{2})}(P_2\circ P_2)^{\frac{1}{2}(\Delta_2+\frac{1}{2})}} \Biggl[ \\
    & \qquad \Big(f_1 \bar{S}_1 S_2+f_2 \bar{S}_1 \Gamma_{n_1 n_2} S_2 
    + g_1 \bar{S}_1 \Gamma^* S_2+g_2\bar{S}_1 \Gamma_{1\cdots p+2} S_2\Big)\\    
    &+\frac{1}{(P_1\circ P_1)^{\frac{1}{2} }}
    \Big(f_3 \bar{S}_1 \slashed{P}_{1\perp} S_2+f_4 \bar{S}_1 \Gamma_{n_1 n_2} \slashed{P}_{1\perp} S_2\\
    & \qquad + g_3\bar{S}_1 \Gamma^* \slashed{P}_{1\perp} S_2+g_4\bar{S}_1 \Gamma_{1 \cdots p+2}\slashed{P}_{1\perp} S_2\Big)\\
    &+\frac{1}{(P_2\circ P_2)^{\frac{1}{2} }}
    \Big(f_5 \bar{S}_1 \slashed{P}_{2\perp} S_2+f_6 \bar{S}_1 \Gamma_{n_1 n_2} \slashed{P}_{2\perp} S_2\\
    &\qquad + g_5 \bar{S}_1 \Gamma^* \slashed{P}_{2\perp} S_2+g_6\bar{S}_1 \Gamma_{1 \cdots p+2}  \slashed{P}_{2\perp} S_2\Big)\\
    &+\frac{1}{(P_1\circ P_1)^{\frac{1}{2}}(P_2\circ P_2)^{\frac{1}{2}}}
    \Big(f_7 \bar{S}_1 \slashed{P}_{1\perp}\slashed{P}_{2\perp} S_2+f_8\bar{S}_1 \Gamma_{n_1 n_2} \slashed{P}_{1\perp}\slashed{P}_{2\perp} S_2\\
    & \qquad + g_7\bar{S}_1 \Gamma^* \slashed{P}_{1\perp} \slashed{P}_{2\perp}S_2+g_8\bar{S}_1 \Gamma_{1 \cdots p+2}\slashed{P}_{1\perp} \slashed{P}_{2\perp}S_2\Big) \biggr] \ .
\end{split}
\end{equation}
Since there is coordinate dependence in the cross-ratios, we choose to just include $P_1\circ P_1$ and $P_2\circ P_2$, to ensure the expression has the correct scaling. The polarisation spinors are no longer eigenvectors of $\Gamma_{n_1}\Gamma_{n_2}$, so there is explicit dependence on these gamma matrices, as well as $\Gamma^*$ and $\Gamma_1...\Gamma_{p+2}$.
We can simplify the correlation function by only writing it in terms of $\Gamma_{n_1},\Gamma_{n_2}$ and $\Gamma^*$, since $\Gamma_1...\Gamma_{p+2}\propto\Gamma^*\Gamma_{n_1}\Gamma_{n_2}$.  This allows us to eliminate the functions $g_i$, and write the correlation function in terms of the bulk chirality projector.

\par It also turns out that the functions $f_1$ and $f_2$ are linearly dependent on $f_7$ and $f_8$:
\begin{equation}
\begin{split}
    &f_1=\frac{1}{|z_1||z_2|}(f_7 \,z_1\cdot z_2 - f_8\, z_1 \times z_2),\\
    &f_2=\frac{1}{|z_1||z_2|}(f_7\, z_1 \times z_2 +f_8\,z_1\cdot z_2 ).
\end{split}
\end{equation}
Similarly, $f_5$ and $f_6$ are linearly dependent on $f_3$ and $f_4$. Therefore, the most general real space correlation function only needs to include $f_1,f_2,f_3,f_4$. Depending on the bulk chirality, the remaining $g$-functions can be identified with the $f$'s via $f_i = \pm g_i$.
\par For the remaining terms, using eq.\ \eqref{s spinor relation}, we make the replacements from embedding space to spacetime: 
\begin{equation}
\begin{split}
    \bar{S}_1 \Gamma_{n_1}\Gamma_{n_2}S_2&\rightarrow\bar{s}_1(\slashed{x}_1\gamma_{n_1}\gamma_{n_2}-\gamma_{n_1}\gamma_{n_2}\slashed{x}_2)s_2=\bar{s}_1 \gamma_{n_1}\gamma_{n_2}(\slashed{\tilde{x}}_1-\slashed{x}_2)s_2 \ , \\
   \bar{S}_1 \slashed{P}_{1\perp} S_2&\rightarrow \bar{s}_1(\slashed{x}_{12} \slashed{z}_1+2z_1 \cdot z_2)s_2.\\   
\end{split}
\end{equation}

\subsection{Bulk-defect-defect}
We next consider the correlation function involving a bulk spinor $\psi$ with dimension $\Delta_1$, a defect spinor $\bar \rho$ with transverse orbital angular momentum $l_2$ and dimension $\Delta_2$, and a defect scalar $\bar \phi$ with transverse orbital angular momentum $l_3$ and dimension $\Delta_3$.

\begin{equation} \label{bulk-defect-defect}
    \begin{split} 
        &\left<\psi_\pm (x_1)\bar{\rho}^{(l_2)}_\alpha(x_2) \bar \phi^{(l_3)}(x_3)\right> =
       \frac{1}{|x_{12}|^{\Delta_1+\Delta_2-\Delta_3}|x_{13}|^{\Delta_1+\Delta_3-\Delta_2}|x_{23}|^{\Delta_2+\Delta_3-\Delta_1}}
        \times 
       \\
       &\quad \mathbb{P}_\pm^* \biggl[ \frac{f_1(\myupsilon) \slashed{x}_{12}}{|x_{12}|} + \frac{f_2(\myupsilon)\slashed{x}_{12}\slashed{z} |x_{23}|}{|x_{12}|^2 |x_{13}|}
        + \frac{f_3(\myupsilon)\slashed{x}_{13}\slashed{x}_{32}}{|x_{13}| |x_{23}|}
       + \frac{f_4(\myupsilon)\slashed{x}_{13}\slashed{x}_{32}\slashed{z}}{|x_{12}| |x_{13}|^2} \biggr]\mathbb{P}_\alpha^\perp \left(\frac{z}{|z|}\right)^{l_2+l_3}  \ . 
    \end{split}
\end{equation} 
There is now a single cross-ratio
\begin{equation}\label{u}
    \myupsilon =\frac{(x_{21})^2(x_{31})^2}{(x_{32})^2 |z|^2}
\end{equation}
with $x_1=(\vec{x}_1,\vec{z})$, $x_2=(\vec{x}_2,0)$ and $x_3=(\vec{x}_3,0)$. In embedding space, the most general correlation function is
\begin{equation}
    \begin{split} 
        &\langle \psi(P_1,\bar{S}_1)\bar{\rho}^{(l_2)}_\alpha(P_2,S_{2,\alpha}) \bar \phi^{(l_3)}(P_3)\rangle  \\
        &= \frac{1}{(P_1\bullet P_2)^{\frac{1}{2}(1+\Delta_1+\Delta_2-\Delta_3)}(P_1\bullet P_3)^{\frac{1}{2}(1+\Delta_1+\Delta_3-\Delta_2)}(P_2\bullet P_3)^{\frac{1}{2}(\Delta_2+\Delta_3 - \Delta_1)}} \times 
        \\
        & \quad \biggl[ (P_1 \bullet P_3)^\frac{1}{2} \left(f_1 \bar{S}_1 S_{2,\alpha}+g_1\bar{S}_1 \Gamma^* S_{2,\alpha} \right) \\
        &\quad +  (P_2 \bullet P_3)^\frac{1}{2} (P_1 \bullet P_2)^{-\frac{1}{2} } 
        \left(f_2 \bar{S}_1 \slashed{P}_{1\perp}S_{2,\alpha}+g_2 \bar{S}_1 \Gamma^* \slashed{P}_{1\perp}S_{2,\alpha} \right) \\
        & \quad + (P_1 \bullet P_2)^\frac{1}{2} (P_2 \bullet P_3)^{-\frac{1}{2} } 
        \left(
       f_3 \bar{S}_1 \slashed{P}_{3}S_{2,\alpha}+g_3 \bar{S}_1 \Gamma^* \slashed{P}_{3}S_{2,\alpha}
       \right) \\ 
       & \quad +  (P_1 \bullet P_3)^{-\frac{1}{2} } \left(
       f_4 \bar{S}_1  \slashed{P}_{1\perp}\slashed{P_3}S_{2,\alpha}+g_4 \bar{S}_1 \Gamma^* \slashed{P}_{1\perp}\slashed{P}_3S_{2,\alpha}
       \right) \biggl] \ . \\ 
    \end{split}
\end{equation}
The choice of denominator is not unique, given the functional dependence on the cross-ratio $\myupsilon$. In writing the real-space expression, we eliminated the functions $g_i$ in favour of a choice of bulk chirality.

\subsection{Defect-defect-defect}
The case of two defect fermions and a defect scalar follows similarly:
\begin{equation} \label{defect-defect-defect}
    \begin{split} 
         \langle \rho_{\alpha}^{(l)}(x_1)\bar{\rho}^{(l_2)}_\alpha(x_2) \bar \phi^{(l_3)}(x_3) \rangle = 
        \frac{ \mathbb{P}_\pm^* \biggl[c_{1 \pm}  \frac{\slashed{x}_{12}}{|x_{12}|} + c_{2 \pm} \frac{\slashed{x}_{13} \slashed{x}_{32}}{|x_{13}| |x_{23}|}
        \biggr] \mathbb{P}_\alpha^\perp }{|x_{12}|^{\Delta_1+\Delta_2-\Delta_3}|x_{13}|^{\Delta_1+\Delta_3-\Delta_2}|x_{23}|^{\Delta_2+\Delta_3-\Delta_1}} \ .
    \end{split}
\end{equation}
We include the ${\mathcal P}^*_\pm$ to allow for the two bulk chirality sectors.  
For the same reasons as the defect-defect correlator, we require that the two defect spinors have the same value of $\alpha$.
Since there are now no normal coordinates to carry the transverse orbital angular momentum in the correlation function, we require that $l=l_2+l_3$, so that the momenta cancel out under an $SO(2)$ transformation.
This real space expression follows from the embedding space correlator 
\begin{equation}
    \begin{split} 
        &\langle \rho^{(l)}_\alpha(P_1,\bar{S}_{1,\alpha})\bar{\rho}^{(l_2)}_\alpha(P_2,S_{2,\alpha}) \bar \phi^{(l_3)}(P_3)\rangle  \\
        &\quad=
       \frac{c_1\bar{S}_{1,\alpha}S_{2,\alpha}+c_2\bar{S}_{1,\alpha}\Gamma^* S_{2,\alpha}}{(P_1\bullet P_2)^{\frac{1}{2}(1+\Delta_1+\Delta_2-\Delta_3)}(P_1\bullet P_3)^{\frac{1}{2}(\Delta_1-\Delta_2+\Delta_3)}(P_2\bullet P_3)^{\frac{1}{2}(-\Delta_1+\Delta_2+\Delta_3)}}\\
        \\
        &\qquad +\frac{c_3\bar{S}_{1,\alpha}\slashed{P}_3 S_{2,\alpha}+c_4\bar{S}_{1,\alpha}\Gamma^* \slashed{P}_3 S_{2,\alpha}}{(P_1\bullet P_2)^{\frac{1}{2}(\Delta_1+\Delta_2-\Delta_3)}(P_1\bullet P_3)^{\frac{1}{2}(1+\Delta_1-\Delta_2+\Delta_3)}(P_2\bullet P_3)^{\frac{1}{2}(1-\Delta_1+\Delta_2+\Delta_3)}} \ . 
    \end{split}
\end{equation}
\section{Bulk-defect and the free Dirac equation}
\label{sec:bulkdefect}
Working in four dimensions, we solve the free Dirac equation for the bulk-defect correlation function (eq.\ \eqref{bulk-defect}), 
\begin{equation}
\slashed{\partial}_1\left<\psi_\pm(x_1)\bar{\rho}^{(l)}_\alpha(x_2)\right>=0  \ ,
\end{equation}
to obtain a relationship between the transverse orbital angular momentum $l$ of the defect spinor, its transverse spin $-\alpha/2$, and its scaling dimension $\Delta_\rho$. 

\begin{figure}
\begin{center}
\includegraphics[width=3in]{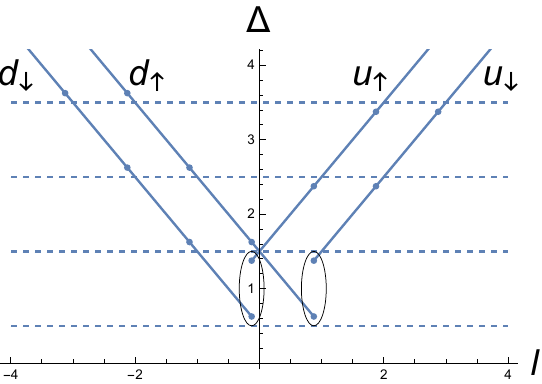}
\end{center}
\caption{
The first few defect operators in the defect OPE of a free fermion with a monodromy phase $\nu = 1/8$ associated with going around the defect. The shadow dual pairs have been circled.
\label{fig:spectrum}
}
\end{figure}

The relationship is more simply given in terms of the total helicity:
\begin{equation}
    j=l-\frac{\alpha}{2} \ .
\end{equation}
 That relationship is 
 \begin{eqnarray}
 \Delta_\rho= \begin{cases}
j+1 \ ,  & \;  j>-\frac{1}{2} \\
1-j \ , & \;  j<\frac{1}{2} 
 \end{cases}
 \end{eqnarray}
For monodromy defects, $l \in {\mathbb Z} - \nu$ where $0 \leq \nu < 1$. 
 In this case, the two ways of producing a given $j$ correspond to the two pairs $(l, \alpha) = (l ,-1)$ and $(l+1, 1)$, 
 giving each allowed value of $\Delta_\rho$ degeneracy two.  
Provided $\nu>0$, there will be one $j$ in the range $-\frac{1}{2}< j <\frac{1}{2}$ which produces two distinct values of $\Delta_\rho = j+1$ and $1-j$ satisfying the shadow dual relationship  $(1+j)+(1-j) = 2$.  
If we set $\nu=0$, then we have a standard defect where $l$ is an integer.  The relation for  the scaling dimension can be written more simply as $\Delta_\rho = 1+|j|$, now with degeneracy four.  
As we will see below, the defect spinors with $\Delta_\rho = \frac{1}{2}$ must decouple from the bulk.
A visual depiction of this spectrum in the case $\nu=\frac{1}{8}$ is offered in Fig.\ \ref{fig:spectrum}.

To see these relationships, we start with  eq.\ \eqref{bulk-defect} and the two spinor structures that are allowed by conformal invariance:
\begin{equation} \label{bulk-defect s family}
\left<\psi_a(x_1)\bar{\rho}_{\shortuparrow \alpha}^{(l)}(x_2)\right>=c^{(l)}_{\psi \shortuparrow}\frac{\mathbb{P}^*_a\, \slashed{x}_{12} \,\mathbb{P}^\perp_\alpha}{|z|^{\Delta_\psi-\Delta_\shortuparrow}|x_{12}|^{2\Delta_\shortuparrow+1}}\left(\frac{z}{|z|}\right)^{l}, \, 
\end{equation}
\begin{equation} \label{bulk-defect t family}   \left<\psi_a(x_1)\bar{\rho}^{(l)}_{\shortdownarrow \alpha}(x_2)\right>=
{c}^{(l)}_{\psi\shortdownarrow}\frac{\mathbb{P}^*_a\slashed{x}_{12}\slashed{z}\mathbb{P}^\perp_\alpha}{|z|^{\Delta_\psi-\Delta_\shortdownarrow+1}|x_{12}|^{2\Delta_\shortdownarrow+1}}\left(\frac{z}{|z|}\right)^{l}.\,
\end{equation}
For the moment, we have distinguished the two fermions that appear in these structures with $\uparrow$ and $\downarrow$.  We will be able to give a more physical meaning to these labels presently.   
While conformal symmetry would allow both structures to appear in a single bulk-defect correlation function, the Dirac constraint is consistent with the presence of either one or the other on the right-hand side, but not both at the same time.\footnote{%
As we will see, there is one exception to this claim, when $\Delta_\shortuparrow = \Delta_\shortdownarrow = 1$ and both structures are annihilated by the Dirac operator.
}

\par The scaling dimension $\Delta_\rho$ follows from the Dirac equation. We make use of a basis in the Clifford algebra to obtain four independent equations. In 4d, the basis is \cite{itzykson1980quantum}
\begin{equation} \label{clifford basis}
    M=a\mathbb{I}+b_\mu\gamma^\mu+\frac{i}{2}c_{\mu\nu}[\gamma^\mu,\gamma^\nu]+d_\mu\gamma^*\gamma^\mu+e\gamma^*=0.
\end{equation}
After applying the Dirac equation to eq.\ \eqref{bulk-defect s family}, we find two independent coefficients which we set to zero, and these equations are solved by taking 
\begin{equation}
    \begin{split}
        &- 3+\alpha l+\Delta_\psi+\Delta_\shortuparrow=0,\\
        &\alpha l-\Delta_\psi+\Delta_\shortuparrow=0,\\
    \end{split}
\end{equation}
We therefore have
\begin{equation}
\begin{split} \label{plus condition}
     \Delta_\psi &=\frac{3}{2} \ , \\
     \Delta_{\shortuparrow} &=\frac{3}{2}-\alpha l
     = \begin{cases}
         1 - j \ , & \alpha = 1 \\
         1 + j \ , & \alpha = -1
     \end{cases} \ ,
\end{split}
\end{equation}
giving the expected result for the scaling dimension of the bulk spinor. 
Repeating this with the other spinor structure eq.\ \eqref{bulk-defect t family} we find 
\begin{equation}
\begin{split}
     &-1+\alpha l+\Delta_\psi-\Delta_\shortdownarrow=0 \ , \\
        &2+\alpha l-\Delta_\psi-\Delta_\shortdownarrow=0 \ , \\
\end{split} \
\end{equation}
so we have 
\begin{equation}
\begin{split} 
     \Delta_\psi &=\frac{3}{2} \ ,\\
     \Delta_{\shortdownarrow} &=\frac{1}{2}+\alpha l
     = \begin{cases}
         1 + j \ , & \alpha =1 \\
         1-j \ , & \alpha =-1 
     \end{cases} \ .
\end{split}
\end{equation}
\par The unitarity bound for the defect spinor is 
\begin{equation}
    \Delta_\rho\geq\frac{p-1}{2}=\frac{1}{2}.
\end{equation}
At the values $j = \pm \frac{1}{2}$, the unitarity bound is saturated, so $\rho^{(\alpha)}_{\shortuparrow \alpha}(\vec{x})$ and $\rho^{(0)}_{\shortdownarrow  \alpha} (\vec{x})$ are free fields and must obey the Dirac equation. However, 
\begin{equation}
    \begin{split}
        &\left<\psi(x_1)\bar{\rho}_{\shortuparrow \alpha}^{(\alpha)}(x_2)\right>\overleftarrow{\slashed{\partial}}_2\neq0 \ , \; \; \;
        \left<\psi(x_1)\bar{\rho}_{\shortdownarrow \alpha}^{(0)}(x_2)\right>\overleftarrow{\slashed{\partial}}_2\neq0 \ ,\\
    \end{split}
\end{equation}
and so these two correlation functions must vanish.
The unitarity bound and decoupling introduce constraints on the possible values of $l$ for the two towers of defect spinors. For the spin-up channel, $\rho_{\shortuparrow \alpha}$, we must have
\begin{equation}  \label{sl1}  
     l <1, \,\alpha=+ \ ,
\end{equation}
\begin{equation} \label{sg-1}
    l >-1,\,\alpha=-.
\end{equation}
For the spin-down channel $\rho_{\shortdownarrow \alpha}$:
\begin{equation} \label{tg0}
     l>0, \,\alpha=+ \ ,
\end{equation}
\begin{equation} \label{tl0}
    l<0,\,\alpha=-.
\end{equation}
As we said above, each value of $j$ can be reached in two different ways. 
The $\uparrow$ and $\downarrow$ notation on $\rho$ was chosen to indicate whether the
spin is aligned or anti-aligned with the orbital angular momentum.

For what follows, it is helpful to introduce a different equivalent nomenclature.
\begin{eqnarray}
    u^{(l)}_{\shortuparrow} = \rho^{(l)}_{\shortuparrow -} \ , \; \; \;
     u^{(l)}_{\shortdownarrow} =  \rho^{(l)}_{\shortdownarrow +} \ , \; \; \;
      d^{(l)}_{\shortuparrow} = \rho^{(l)}_{\shortuparrow +} \ , \; \; \; 
     d^{(l)}_{ \shortdownarrow} =  \rho^{(l)}_{\shortdownarrow -} \ , \; \; \;
\end{eqnarray}
With this swap, the $u$ ``up-quarks'' are associated with the tower of positive helicities $j > -\frac{1}{2}$, while the $d$ ``down-quarks'' are associated with the tower of negative helicities $j < \frac{1}{2}$.
In the monodromy case, 
at the bottom of the up towers we have $u_\shortdownarrow^{(1-\nu)}$ 
and $u_\shortuparrow^{(-\nu)}$ with dimension $\frac{3}{2}-\nu$.
At the bottom of the down towers we have 
$d_\shortdownarrow^{(-\nu)}$ and $d_\shortuparrow^{(1-\nu)}$ with dimension $\frac{1}{2} + \nu$.  
The pair $u_\shortdownarrow^{(1-\nu)}$  and  $d_\shortuparrow^{(1-\nu)}$ 
are shadow dual, as is the pair $u_\shortuparrow^{(-\nu)}$ and $d_\shortdownarrow^{(-\nu)}$.

 \subsection{Defect operator product expansion}
 The defect operator product expansion takes a bulk field and expands it near the defect, giving a sum over defect primary operators and descendants:
 \begin{equation}
     \psi(\vec{x},\vec{z}) =\sum_{\hat{\Delta} }\mathcal{D}^{({\Delta})}(\vec{z},\partial_{\vec{x}})\rho^{({\Delta})}(\vec{x})
     \ .
 \end{equation}
 The sum here is schematic, suppressing the dependence on the Lorentz group and spinor quantum numbers.  The $\rho^{(\Delta)}$ are defect primary operators with scaling weight $ \Delta$, while $\mathcal{D}^{({\Delta})}$ applied to a primary reconstructs the sum over descendants.  
 
 Given the unitarity bounds and the relationship we have found between the defect scaling dimension, the transverse orbital angular momentum $l$, and the transverse spin $-\frac{\alpha}{2}$,
 we can be more specific about the particular defect operator expansion for the bulk $\psi$ field:
 \begin{equation}
 \begin{split}
\psi(\vec{x}, \vec{z}) &= \sum_{l>0}
a_\shortdownarrow^{(l)} \mathcal{U}_\shortdownarrow^{(l)} (\vec{z},\partial_{\vec{x}}) u_{\shortdownarrow}^{(l)}(\vec{x})
+
\sum_{l>-1} 
a_\shortuparrow^{(l)} 
\mathcal{U}_\shortuparrow^{(l)} (\vec{z},\partial_{\vec{x}}) u_{\shortuparrow}^{(l)}(\vec{x})
  \\
& +
\sum_{l<0}
b_\shortdownarrow^{(l)} 
\mathcal{D}_\shortdownarrow^{(l)} (\vec{z},\partial_{\vec{x}}) d_{\shortdownarrow}^{(l)}(\vec{x})
+
\sum_{l<1} 
b_\shortuparrow^{(l)} 
\mathcal{D}_\shortuparrow^{(l)}  (\vec{z},\partial_{\vec{x}}) d_{\shortuparrow}^{(l)}(\vec{x}) \ .
\end{split}
 \end{equation}
 We have taken the liberty of using the symbol ${\mathcal U}$ for the operator that reconstructs the sum over descendants when acting on an ``up-quark'' $u$.
 The constants $a^{(l)}_{\shortuparrow / \shortdownarrow}$ and $b^{(l)}_{\shortuparrow / \shortdownarrow}$ are defect-operator coefficients.  With our particular normalization choices for the defect-defect correlation functions and the ${\mathcal U}$ and ${\mathcal D}$ differential operators, they will also be  bulk-defect correlation function coefficients.

To make further progress, we need to be more specific about the spinor quantum numbers.  We shall decompose the bulk spinor into chiralities $\psi_\pm$, and discuss the corresponding consequences for the defect $u$ and $d$ fields in the defect operator expansion.  We will also spell out exactly what the ${\mathcal U}$ and ${\mathcal D}$ are.

\subsubsection*{Spin-up channel}
 Working in 4d with a 2d defect, we have three projectors
 \begin{equation}
     \begin{split}
         &\mathbb{P}^*_a=\frac{1}{2}(\mathbb{I}+a \gamma^*),\\
         &\mathbb{P}^\perp_\alpha=\frac{1}{2}(\mathbb{I}+i\alpha \gamma^3 \gamma^4),\\
          &\mathbb{P}^\parallel_\beta=\frac{1}{2}(\mathbb{I}+i\beta \gamma^1 \gamma^2),\\
     \end{split}
 \end{equation}
 with $a,\alpha,\beta=\pm1$. 
 The projectors ${\mathbb P}^*_a$ and ${\mathbb P}^\perp_\alpha$ were introduced before, while
 $\mathbb{P}^\parallel_\beta$ determines  the chirality of the defect spinors within the defect.

 Introducing the defect chirality projector (which adds another label to the defect spinor) to the spin-up bulk-defect two-point function (eq.\ \eqref{bulk-defect s family}) gives 
 \begin{equation} \label{projected bulk-defect} 
    \left<\psi_a(x_1)\bar{\rho}_{\shortuparrow\alpha\beta}^{(l)}(x_2)\right>=c^{(l)}_{\psi \shortuparrow}\frac{\mathbb{P}^*_a\, \slashed{x}_{12} \,\mathbb{P}^\perp_\alpha\,\mathbb{P}^\parallel_\beta}{|z|^{\Delta_\psi-\Delta_\rho}|x_{12}|^{2\Delta_\rho+1}}\left(\frac{z}{|z|}\right)^{l}.
\end{equation}
We will shortly replace the general bulk-defect coefficients $c_{\psi \shortuparrow}^{(l)}$ with the $a_{\shortuparrow/\shortdownarrow}^{(l)}$, $b_{\shortuparrow/\shortdownarrow}^{(l)}$ notation from above.
Moving $\mathbb{P}^*_a$ past $\gamma_\mu$, we have the product $\mathbb{P}^*_{-a}\mathbb{P}^\perp_\alpha\mathbb{P}^\parallel_\beta$. Expanding $\gamma^*$ in terms of the other $\gamma$-matrices, it is clear that this two-point function is only nonzero for the following cases:
\begin{center}
\begin{tabular}{ |c|c|c| } 
\hline
$a$ & $\alpha$ & $\beta$ \\
\hline
\multirow{2}{1em}{$+$} & $+$ & $+$ \\ 
& $-$ & $-$ \\  
\hline
\multirow{2}{1em}{$-$} & $+$ & $-$ \\ 
& $-$ & $+$ \\  
\hline
\end{tabular}
\end{center}
For the defect-defect two-point function (eq.\ \eqref{defect-defect}) after expanding $\gamma^*$, the correlation function can be written as 
\begin{equation}                        \left<\rho_{\shortuparrow \alpha,-\beta}^{(l)}(x_1)\bar{\rho}_{\shortuparrow \alpha \beta}^{(l)}(x_2)\right>=
\frac{\mathbb{P}^\perp_{\alpha}\mathbb{P}^\parallel_{-\beta}\,\slashed{x}_{12}\,\mathbb{P}^\perp_\alpha\mathbb{P}^\parallel_{\beta}}{|x_{12}|^{2\Delta_\rho+1}}\\    
\end{equation}
where $\beta=a\alpha$. In our notation, $\overline{\rho_{++}} = \bar \rho_{+-}$ and so the nonzero defect-defect correlation functions require the spinors to have the same value of $\alpha$ and the same defect chiralities. 
\par We can choose a basis for the gamma matrices where all three projectors are diagonal:
\begin{equation} \label{gamma basis}
    \gamma^1=\begin{pmatrix}
        0&-i\mathbb{I}_{2\times2}\\i\mathbb{I}_{2\times2}&0 
    \end{pmatrix},
    \ \gamma^2=\begin{pmatrix}
        0&\sigma_3\\\sigma_3 &0
    \end{pmatrix},
     \ \gamma^3=\begin{pmatrix}
        0&\sigma_1\\\sigma_1 &0
    \end{pmatrix},
     \ \gamma^4=\begin{pmatrix}
        0&\sigma_2\\\sigma_2 &0
    \end{pmatrix}
\end{equation}
where $\sigma_i$ are the Pauli matrices. Since the projectors are diagonal, the components of a general bulk spinor and defect spinor can be written in this basis as \begin{equation}
    \psi=\begin{pmatrix}
        \psi_+\\\psi_- 
    \end{pmatrix},
    \,\rho=\begin{pmatrix}
        \rho_{-+}\\\rho_{+-}\\\rho_{--}\\\rho_{++}\\ 
    \end{pmatrix}, 
    \bar \rho = 
    \begin{pmatrix}
    \bar \rho_{-+} \\
    \bar \rho_{+-} \\
    \bar \rho_{--} \\
    \bar \rho_{++}
    \end{pmatrix}^T
    =
     \begin{pmatrix}
    i \rho_{--}^\ddagger \\
    i  \rho_{++}^\ddagger \\
    -i  \rho_{-+}^\ddagger \\
    -i \rho_{+-}^\ddagger
    \end{pmatrix}^T
\end{equation}
where $\psi_+$ and $\psi_- $ have two components. 
Here $\ddagger$ implies a semi-Lorentzian version of conjugation which combines complex conjugation with time reversal, sending $z \to \bar z$ but $w \to -w$.
Working in complex coordinates with $(\vec{x},\vec{z})=(w,\bar{w},z,\bar{z})$, the only nonzero bulk-defect correlation functions in this basis are 
\begin{equation}
    \begin{aligned}
        &\braket{\psi_+(x_1)\bar{\rho}^{(l)}_{\shortuparrow ++}(x_2)}
        =
        \braket{\psi_+(x_1)\bar{d}^{(l)}_{\shortuparrow -}(x_2)}=\frac{b^{(l)}_{\shortuparrow+}}{\bar{z}^{l}|x_{12}|^{4-2l}}\begin{pmatrix}
            \bar{z}\\i\bar{w}_{21}
        \end{pmatrix} \ , \\
        &\braket{\psi_+(x_1)\bar{\rho}^{(l)}_{\shortuparrow --}(x_2)}
        =
        \braket{\psi_+(x_1)\bar{u}^{(l)}_{\shortuparrow -}(x_2)}=\frac{a^{(l)}_{\shortuparrow +}}{z^{-l}|x_{12}|^{4+2l}}\begin{pmatrix}
            i w_{21}\\z
        \end{pmatrix} \ , \\
        &\braket{\psi_-(x_1)\bar{\rho}^{(l)}_{\shortuparrow +-}(x_2)}=
        \braket{\psi_-(x_1)\bar{d}^{(l)}_{\shortuparrow +}(x_2)}=\frac{b^{(l)}_{\shortuparrow -}}{\bar{z}^{l}|x_{12}|^{4-2l}}\begin{pmatrix}
            \bar{z}\\i w_{12}
        \end{pmatrix} \ , \\
         &\braket{\psi_-(x_1)\bar{\rho}^{(l)}_{\shortuparrow -+}(x_2)}=
         \braket{\psi_-(x_1)\bar{u}^{(l)}_{\shortuparrow +}(x_2)}=\frac{a^{(l)}_{\shortuparrow -}}{z^{-l}|x_{12}|^{4+2l}}\begin{pmatrix}
            i\bar{w}_{12}\\z
        \end{pmatrix}  \ , \\
    \end{aligned}
\end{equation}
where we have evaluated eq.\ \eqref{projected bulk-defect} with the different projections and set $\Delta_\psi=\frac{3}{2}$ and $\Delta_\rho=\frac{3}{2}-\alpha l$.
The defect OPE coefficients from before, $a_{\shortuparrow }$ and $b_{\shortuparrow}$, have gained a bulk chirality index $a_{\shortuparrow \pm}$ and $b_{\shortuparrow \pm}$.
The nonzero defect-defect correlation functions are
\begin{equation}
\label{defectdefectspinup}
    \begin{aligned}
        &\braket{\rho^{(l)}_{\shortuparrow+-}(x_1)\bar{\rho}^{(l)}_{\shortuparrow++}(x_2)}=
        \braket{d^{(l)}_{\shortuparrow+}(x_1)\bar{d}^{(l)}_{\shortuparrow -}(x_2)}=
        \frac{i\bar{w}_{21}}{|x_{12}|^{4-2l}} \ , \\
        &\braket{\rho^{(l)}_{\shortuparrow-+}(x_1)\bar{\rho}^{(l)}_{\shortuparrow--}(x_2)}=
        \braket{u^{(l)}_{\shortuparrow +}(x_1)\bar{u}^{(l)}_{\shortuparrow-}(x_2)}=
        \frac{i w_{21}}{|x_{12}|^{4+2l}} \ , \\
        &\braket{\rho^{(l)}_{\shortuparrow++}(x_1)\bar{\rho}^{(l)}_{\shortuparrow+-}(x_2)}=
        \braket{d^{(l)}_{\shortuparrow -}(x_1)\bar{d}^{(l)}_{\shortuparrow+}(x_2)}=
        \frac{i w_{12}}{|x_{12}|^{4-2l}} \ , \\
        &\braket{\rho^{(l)}_{\shortuparrow--}(x_1)\bar{\rho}^{(l)}_{\shortuparrow-+}(x_2)}=
        \braket{u^{(l)}_{\shortuparrow -}(x_1)\bar{u}^{(l)}_{\shortuparrow +}(x_2)}=
        \frac{i\bar{w}_{12}}{|x_{12}|^{4+2l}}.\\
    \end{aligned}
\end{equation}
By comparing the nonzero defect-defect and bulk-defect correlation functions, we see that the OPE for the two bulk chiralities must take the form (including for now only the spin-up channel)
\begin{equation}
\label{OPEspinup}
\begin{split}
    \psi_{\pm}(x) &= 
    \sum_{l>-1} a_{\shortuparrow \pm}^{(l)} {\mathcal U}_{\shortuparrow \pm}^{(l)}(z,\bar{z},\partial_{\vec{x}}) u_{\shortuparrow \pm}^{(l)}
    (\vec{x})
    +
    \sum_{l<1} b_{\shortuparrow \pm}^{(l)} {\mathcal D}_{\shortuparrow \pm}^{(l)}(z,\bar{z},\partial_{\vec{x}}) d_{\shortuparrow \pm}^{(l)}
    (\vec{x}) \ .
\end{split}
\end{equation}
The added $\pm$ subscript is the 4d chirality.  
The differential operators ${\mathcal U}_{\shortuparrow \pm}$ and ${\mathcal D}_{\shortuparrow \pm}$, which we will now make precise, have two remaining components.  The details of their construction can be found in Appendix \ref{sec:OPEops}.
The results are
\begin{equation}
\begin{split}
    {\mathcal U}^{(l)}_{\shortuparrow+}(z, \bar z, \partial_w, \partial_{\bar w})
    &= 
     \sum_{n=0}^\infty
\mathbb{{D}}_n^{(l)}  \left( \begin{array}{c}
1 \\
\frac{-i  z}{1+l+n} \partial_{ w} 
\end{array}
\right)
\ , 
    \\
    {\mathcal U}^{(l)}_{\shortuparrow  -}(z, \bar z, \partial_w, \partial_{\bar w})
    &=
     \sum_{n=0}^\infty
\mathbb{{D}}_n^{(l)}  \left( \begin{array}{c}
1 \\
\frac{i  z}{1+l+n} \partial_{\bar w} 
\end{array}
\right)
\ , 
    \\
    {\mathcal D}^{(l)}_{\shortuparrow  +}(z, \bar z, \partial_w, \partial_{\bar w})
    &= \sum_{n=0}^\infty
\left( {\mathbb D}_n^{(-l)} \right)^* \left( \begin{array}{c}
\frac{-i \bar z}{1-l+n} \partial_{\bar w} \\
1
\end{array}
\right)  = -i \sigma_2 \, \left( {\mathcal U}_{\shortuparrow +}^{(-l)}\right)^* \ ,
    \\
    {\mathcal D}^{(l)}_{\shortuparrow  -}(z, \bar z, \partial_w, \partial_{\bar w})
    &=
    \sum_{n=0}^\infty
\left( {\mathbb D}_n^{(-l)}\right)^* \left( \begin{array}{c}
\frac{i \bar z}{1-l+n} \partial_w \\
1
\end{array}
\right) = -i \sigma_2 \, \left( {\mathcal U}_{\shortuparrow -}^{(-l)} \right)^*
\ ,
\end{split}
\end{equation}
where 
\begin{equation} \label{DDopdef}
{\mathbb D}_n^{(l)} \equiv  z^{l} \frac{(-|z|^2)^n}{(1+l)_n n!} \partial_w^n \partial_{\bar w}^n \ .
\end{equation}
Notice that $\left( {\mathcal U}_{\shortuparrow+}^{(l)}\right)^\ddagger = \left( {\mathcal U}_{\shortuparrow-}^{(l)}\right)^*$ and
$\left( {\mathcal D}_{\shortuparrow+}^{(l)}\right)^\ddagger = \left( {\mathcal D}_{\shortuparrow-}^{(l)}\right)^*$.

\subsubsection*{Spin-down channel}
Working in the same way as before, we introduce the defect chirality projector in eq.\ \eqref{bulk-defect t family}, giving
\begin{equation}
    \left<\psi_a(x_1)\bar{{\rho}}^{(l)}_{\shortdownarrow \alpha\beta}(x_2)\right>=
{c}^{(l)}_{\psi \shortdownarrow}\frac{\mathbb{P}^*_a\slashed{x}_{12}\slashed{z}\mathbb{P}^
\perp_\alpha\mathbb{P}^\parallel_\beta}{|z|^{\Delta_\psi-\Delta_\rho+1}|x_{12}|^{2\Delta_
\rho+1}}\left(\frac{z}{|z|}\right)^{l}.
\end{equation}
Due to the presence of $\gamma_i$ in this correlation function, it is nonzero for the opposite choice of 4d chirality, as compared with the $\shortuparrow$ case:
\begin{center}
\begin{tabular}{ |c|c|c| } 
\hline
$a$ & $\alpha$ & $\beta$ \\
\hline
\multirow{2}{1em}{$-$} & $+$ & $+$ \\ 
& $-$ & $-$ \\  
\hline
\multirow{2}{1em}{$+$} & $+$ & $-$ \\ 
& $-$ & $+$ \\  
\hline
\end{tabular} \ ,
\end{center}
and so the only nonzero bulk-defect correlation functions in the basis are 
\begin{equation}
    \begin{split}
        &\braket{\psi_-(x_1)\bar{\rho}^{(l)}_{\shortdownarrow++}(x_2)}=
        \braket{\psi_-(x_1)\bar{u}^{(l)}_{\shortdownarrow +}(x_2)}={a}^{(l)}_{ \shortdownarrow -}\frac{z^{l}}{|x_{12}|^{2+2l}}\begin{pmatrix}
            iz^{-1}\bar{w}_{12}\\1
        \end{pmatrix} \ , \\
        &\braket{\psi_-(x_1)\bar{{\rho}}^{(l)}_{\shortdownarrow--}(x_2)}=
        \braket{\psi_-(x_1)\bar{{d}}^{(l)}_{\shortdownarrow +}(x_2)}=
        {b}^{(l)}_{ \shortdownarrow -}\frac{\bar{z}^{-l}}{|x_{12}|^{2-2l}}\begin{pmatrix}
            1\\i\bar{z}^{-1}w_{12}
        \end{pmatrix} \ , \\
        &\braket{\psi_+(x_1)\bar{\rho}^{(l)}_{\shortdownarrow+-}(x_2)}=
        \braket{\psi_+(x_1)\bar{u}^{(l)}_{\shortdownarrow -}(x_2)}=
        {a}^{(l)}_{ \shortdownarrow +}\frac{z^{l}}{|x_{12}|^{2+2l}}\begin{pmatrix}
            iz^{-1}w_{21}\\1
        \end{pmatrix} \ , \\
         &\braket{\psi_+(x_1)\bar{\rho}^{(l)}_{\shortdownarrow-+}(x_2)}=
         \braket{\psi_+(x_1)\bar{d}^{(l)}_{\shortdownarrow-}(x_2)}=
         {b}^{(l)}_{ \shortdownarrow +}\frac{\bar{z}^{-l}}{|x_{12}|^{2-2l}}\begin{pmatrix}
            1\\i\bar{z}^{-1}\bar{w}_{21}
        \end{pmatrix}\ .
    \end{split}
\end{equation}
Here we have set $\Delta_\psi=\frac{3}{2}$ and $\Delta_
\rho=\frac{1}{2}+ \alpha l$. 
These bulk-defect correlation functions must be compared against the spin-down defect-defect correlation functions:  
\begin{equation}
    \begin{aligned}
        &\braket{\rho^{(l)}_{\shortdownarrow+-}(x_1)\bar{\rho}^{(l)}_{\shortdownarrow++}(x_2)}=
        \braket{u^{(l)}_{\shortdownarrow-}(x_1)\bar{u}^{(l)}_{\shortdownarrow +}(x_2)}=
        \frac{i\bar{w}_{21}}{|x_{12}|^{2+2l}} \ , \\
        &\braket{\rho^{(l)}_{\shortdownarrow-+}(x_1)\bar{\rho}^{(l)}_{\shortdownarrow--}(x_2)}=
        \braket{d^{(l)}_{\shortdownarrow -}(x_1)\bar{d}^{(l)}_{\shortdownarrow+}(x_2)}=
        \frac{i w_{21}}{|x_{12}|^{2-2l}} \ , \\
        &\braket{\rho^{(l)}_{\shortdownarrow++}(x_1)\bar{\rho}^{(l)}_{\shortdownarrow+-}(x_2)}=
        \braket{u^{(l)}_{\shortdownarrow +}(x_1)\bar{u}^{(l)}_{\shortdownarrow-}(x_2)}=
        \frac{i w_{12}}{|x_{12}|^{2+2l}} \ , \\
        &\braket{\rho^{(l)}_{\shortdownarrow--}(x_1)\bar{\rho}^{(l)}_{\shortdownarrow-+}(x_2)}=
        \braket{d^{(l)}_{\shortdownarrow +}(x_1)\bar{d}^{(l)}_{\shortdownarrow -}(x_2)}=
        \frac{i\bar{w}_{12}}{|x_{12}|^{2-2l}}.\\
    \end{aligned}
\end{equation}
Including only the spin-down channel of defect fermions, the defect operator expansion, analogous to eq.\ (\ref{OPEspinup}), is  
\begin{equation}
\label{OPEspindown}
\begin{split}
    \psi_{\pm}(x) &= 
    \sum_{l>0} a_{\shortdownarrow \pm}^{(l)} {\mathcal U}_{\shortdownarrow \pm}^{(l)}(z,\bar{z},\partial_{\vec{x}}) u_{\shortdownarrow \pm}^{(l)}
    (\vec{x})
    +
    \sum_{l<0} b_{\shortdownarrow \pm}^{(l)} {\mathcal D}_{\shortdownarrow \pm}^{(l)}(z,\bar{z},\partial_{\vec{x}}) d_{\shortdownarrow \pm}^{(l)}
    (\vec{x}) \ ,
\end{split}
\end{equation}
with defect OPE coefficients $a_{\downarrow \pm}^{(l)}$ and $b_{\downarrow \pm}^{(l)}$. 
We can find these operators the same way as before, and the results are

\begin{equation}
\begin{split}
{\mathcal U}^{(l)}_{\shortdownarrow +}
(z, \bar z, \partial_w, \partial_{\bar w}) &=
 \sum_{n=0}^\infty {\mathbb D}_n^{(l-1)}  
\left(
\begin{array}{c}
-1 \\
\frac{i z}{l+n} \partial_w
\end{array}
\right) \ , \\
 {\mathcal U}^{(l)}_{\shortdownarrow -} (z, \bar z, \partial_w, \partial_{\bar w}) &=
 \sum_{n=0}^\infty {\mathbb D}_n^{(l-1)}  
\left(
\begin{array}{c}
-1 \\
-\frac{i  z}{l+n} \partial_{\bar w}\\
\end{array}
\right) \ , \\
 {\mathcal D}^{(l)}_{\shortdownarrow +}(z, \bar z, \partial_w, \partial_{\bar w}) &=
 \sum_{n=0}^\infty \left({{\mathbb D}}_n^{(-l-1)}  \right)^*
\left(
\begin{array}{c}
\frac{i \bar z}{-l+n} \partial_{\bar w} \\
-1
\end{array}
\right) 
=
-i \sigma_2 \, \left({{\mathcal U}}^{(-l)}_{\downarrow +}\right)^*
\ , \\
{\mathcal D}^{(l)}_{\shortdownarrow -} (z, \bar z, \partial_w, \partial_{\bar w}) &=
 \sum_{n=0}^\infty \left({{\mathbb{D}}}_n^{(-l-1)}  \right)^*
\left(
\begin{array}{c}
-\frac{i \bar z}{-l+n} \partial_{ w} \\
-1
\end{array}
\right) 
=
-i \sigma_2 \,  \left( {\mathcal U}^{(-l)}_{\downarrow -} \right)^*
\ ,
\end{split}
\end{equation}
where ${\mathbb D}^{(l)}_n$ is defined as before (\ref{DDopdef}).
Note that ${\mathcal U}^{(l)}_{\shortdownarrow \pm} = 
-{\mathcal U}^{(l-1)}_{\shortuparrow \pm}$
and 
${\mathcal D}^{(l)}_{\shortdownarrow \pm} =-{\mathcal D}^{(l+1)}_{\shortuparrow \pm}$.

Putting together the spin down (\ref{OPEspindown}) and spin up (\ref{OPEspinup}) channels, the final result for the defect operator expansion of the bulk fermion is
\begin{equation}
\label{OPEtotal}
\begin{split}
    \psi_{\pm}(x) &=  \sum_{l>-1} a_{\shortuparrow \pm}^{(l)} {\mathcal U}_{\shortuparrow \pm}^{(l)}(z,\bar{z},\partial_{\vec{x}}) u_{\shortuparrow \pm}^{(l)}
    (\vec{x})
    +
    \sum_{l<1} b_{\shortuparrow \pm}^{(l)} {\mathcal D}_{\shortuparrow \pm}^{(l)}(z,\bar{z},\partial_{\vec{x}}) d_{\shortuparrow \pm}^{(l)}
    (\vec{x}) \\
    &+
    \sum_{l>0} a_{\shortdownarrow \pm}^{(l)} {\mathcal U}_{\shortdownarrow \pm}^{(l)}(z,\bar{z},\partial_{\vec{x}}) u_{\shortdownarrow \pm}^{(l)}
    (\vec{x})
    +
    \sum_{l<0} b_{\shortdownarrow \pm}^{(l)} {\mathcal D}_{\shortdownarrow \pm}^{(l)}(z,\bar{z},\partial_{\vec{x}}) d_{\shortdownarrow \pm}^{(l)}
    (\vec{x}) \ .
\end{split}
\end{equation}

\begin{table}
\[
\renewcommand{\arraystretch}{1.8}
\begin{array}{|c|c|c|c|}
\hline
 u^{(l)}_{\shortuparrow +} = \rho^{(l)}_{\shortuparrow - +}&
      u^{(l)}_{\shortdownarrow +} =  \rho^{(l)}_{\shortdownarrow + +} &
      d^{(l)}_{\shortuparrow +} = \rho^{(l)}_{\shortuparrow +-} &
     d^{(l)}_{ \shortdownarrow+ } =  \rho^{(l)}_{\shortdownarrow --} \\
     \hline
     u^{(l)}_{\shortuparrow -} = \rho^{(l)}_{\shortuparrow --}&
      u^{(l)}_{\shortdownarrow -} =  \rho^{(l)}_{\shortdownarrow +-} &
      d^{(l)}_{\shortuparrow -} = \rho^{(l)}_{\shortuparrow ++} &
     d^{(l)}_{ \shortdownarrow -} =  \rho^{(l)}_{\shortdownarrow -+}  \\
     \hline
\end{array}
\]
\caption{The dictionary between the $\rho$ spinors and the $u$ and $d$ spinors.
\label{table:rhotoudtranslation}
}
\end{table}

\section{Bulk-bulk and the free Dirac equation}
\label{sec:bulkbulk}

The purpose of this section is to compute the bulk-bulk correlation function $\langle \psi(x_1)\bar{\psi}(x_2) \rangle$ in two complementary ways, first by solving the Dirac equation and next by employing the defect OPE.
The first approach is technically simpler, while the second offers a physical understanding of the integration constants in the first approach, as bulk-defect operator product coefficients.  
The final result is a sum over conformal blocks associated with the defect primaries in the defect OPE of $\psi(x)$. 
One nice feature of the result is that the conformal block contribution 
of a defect primary to $\langle \psi(x_1)\bar{\psi}(x_2) \rangle$ depends only on the total transverse angular momentum $j$, and not on the orbital $l$ and spin angular momentum separately.

\subsection{Solving Dirac}

We now  solve the free Dirac equation of the bulk-bulk correlation function (eq.\ \eqref{bulk-bulk})
\begin{equation}
    \slashed{\partial}_1\left<\psi(x_1)\bar{\psi}(x_2)\right>=0
\end{equation}
for the four functions $f_i$. To simplify the problem, conformal transformations can be made after differentiating to set $x_1=(\vec{x}_1,\vec{z}_1)=(\vec{0},1,0)$ and $x_2=(\vec{x}_2,\vec{z}_2)=(\vec{0},r\cos{\theta},r\sin\theta)$, where $\theta$ is the angle between $\vec{z}_1$ and $\vec{z}_2$. The cross-ratio $\xi_1$ in \eqref{cross-ratios} has both $r$ and $\theta$ dependence, so we instead consider the $\theta$ independent combination, given by  
\begin{equation} \label{chi}
   \chi \equiv 2\xi_1+\xi_2= \frac{(\vec{x}_1-\vec{x}_2)^2+\vec{z}_1^2+\vec{z}_2^2}{2|\vec{z}_1||\vec{z}_2|}=\frac{1+r^2}{2r}.
\end{equation}
The second cross-ratio is $\xi_2=\cos{\theta}$, and can be kept. 

\par We can make use of the same basis in the Clifford algebra (eq.\ \eqref{clifford basis}) to obtain four independent equations and solve these for the four functions $f_i$. It turns out that $f_1$ and $f_2$ satisfy two coupled partial differential equations, and $f_3$ and $f_4$ satisfy two others. 
\par The PDEs for $f_i$ can be solved by combining each pair of functions into 
\begin{equation}
    \begin{split}
        &G_\pm(r,\theta)\equiv f_1(r,\theta)\pm i f_2(r,\theta)
        \ , \; \; \;
        H_\pm(r,\theta)\equiv f_3(r,\theta)\pm if_4(r,\theta)
    \end{split}
\end{equation}
to give two pairs of coupled PDEs
\begin{equation} \label{F pde}
\begin{split}
    0 &= r e^{\mp i \theta} (2 + D_\pm) G_\pm + (2  - D_\pm )G_\mp \ , \\
    0 &= r e^{\pm i \theta} (1 + D_\pm)H_\pm - (1 - D_\pm) H_\mp \ , 
\end{split}
\end{equation}
where $D_\pm \equiv r \partial_r \pm i \partial_\theta$. 

We next make the separation of variables ansatz 
\begin{align}
    G_+(r,\theta)&=R_1(r)e^{i l\theta}  \ , \; \; \; &H_+(r,\theta)&=R_3(r)e^{i (l-1)\theta} \ ,
    \\ 
    G_-(r,\theta)&=R_2(r)e^{i(l-1)\theta} \ ,  \; \; \; 
      &H_-(r,\theta)&=R_4(r)e^{i l \theta} \ ,  \nonumber
\end{align}
as this allows us to factor out the exponentials and solve the resulting ordinary differential equations in $R_1(r), R_2(r)$, and $R_3(r), R_4(r)$ to find $f_1,f_2$, and $f_3,f_4$. 
We are anticipating a relation between the exponential factor here and the orbital angular momentum $l$ introduced earlier.
The results are
\begin{equation} \label{bulk-bulk solutions}
\begin{split}
f_{1} &= \sum_l e^{il\theta}
\left[
(F_{l-1} e^{-i \theta} + F_{l}) c_{1,l} + (F_{-l+1}e^{-i\theta} +  F_{-l}) c_{2,l}
\right] \ , 
\\
f_{2} &= i \sum_l e^{il\theta} 
\left[
(F_{l-1} e^{-i \theta} - F_{l}) c_{1,l} + (F_{-l+1}e^{-i\theta} -  F_{-l}) c_{2,l}
\right] \ , 
\\
f_{3} &= -i\sum_l  e^{il\theta} 
\left[
(F_{l-1} - e^{-i \theta} F_l ) c_{3,l} + (F_{-l+1} - e^{-i \theta} F_{-l}) c_{4,l}
\right]\ , 
\\
f_{4} &= \sum_l e^{il\theta} 
\left[
(F_{l-1} + e^{-i \theta} F_l) c_{3,l} + (F_{-l+1} + e^{-i \theta} F_{-l})c_{4,l}
\right] \ , 
\end{split}
\end{equation}
with integration constants $c_{i,l}$, and 
\begin{equation}
   F_l(r) \equiv r^{2-l}(1+r^2 + l(r^2-1))(r^2-1)^{-3},
\end{equation}
which has the property $F_{-l}(r) = -F_{l}(r^{-1})$.  

We next want to consider the allowed values of $l$ in the sum. 
The function $F_l \sim r^{2-l}$ at small $r$ and $F_l \sim r^{-2-l}$ at large $r$. 
These scalings are related to the dimension of a defect operator contributing to the bulk two-point function. The $r$ scaling behaviour suggests that in the defect OPE expansion we keep negative $l$ in the series for $c_{1,l}$ if we expand at small $r$, and positive $l$ if we expand at large $r$. However, we can avoid this issue by replacing $F_l(r)$ in our functions with
\begin{equation}
I_l(\chi) \equiv (l+1) (2\chi)^{-l-2} {}_2 F_1 \left( \frac{2+l}{2}, \frac{3+l}{2}, 1+l, \chi^{-2} \right) = \begin{cases}
    F_{l}(r) \ , \; \; \; r>1 \\
     -F_{-l}(r) \ , \; \; \; 0<r<1 
\end{cases}
\end{equation}
For the terms in the $f_1$ sum to correspond to exchanges of operators above the unitarity bound, we require
\begin{equation}
\label{f1result}
    f_{1} = \sum_{l>0} e^{il\theta}
(I_{l-1} e^{-i \theta} + I_{l}) c_{1,l} + 
\sum_{l<1} e^{il\theta}(I_{-l+1}e^{-i\theta} +  I_{-l}) c_{2,l}
 \ ,
\end{equation}
where we can identify the dimension of the exchanged operator as 
$\Delta_\rho = \frac{1}{2} + l$ in the first sum and 
$\Delta_\rho = \frac{3}{2}-l$ in the second.
There are similar expressions for the remaining $f_i$.

\subsection{Conformal blocks}
We now re-analyze this bulk-bulk correlation function using the defect OPE.
The goal is to reinterpret each term in the $l$-sums, for instance in (\ref{f1result}), as a conformal block, i.e.\ the contribution of a defect primary and its descendants to the correlation function.  As a result, we will be able to make a direct identification between the integration constants $c_{1,l}$ and $c_{2,l}$
 and the defect OPE coefficients $a_{\shortuparrow / \shortdownarrow, \pm}$ and  $b_{\shortuparrow / \shortdownarrow, \pm}$.

Decomposed into chiral sectors, eq.\ \eqref{bulk-bulk} becomes the pair 
\begin{equation}
    \begin{split}
        \left<\psi_a(x_1)\bar{\psi}_{-a}(x_2)\right>&=\frac{1}{|z_1|^{\Delta_1+\frac{1}{2}}|z_2|^{\Delta_2+\frac{1}{2}}}\Big(f_1(\xi_1,\xi_2)\slashed{x}_{12}\\
    &\qquad+ f_2(\xi_1,\xi_2)\gamma_{n_1}\gamma_{n_2}(\slashed{\tilde{x}}_1-\slashed{x}_2)\Big)\\
    \left<\psi_a(x_1)\bar{\psi}_{a}(x_2)\right>&= \frac{1}{|z_1|^{\Delta_1+\frac{3}{2}}|z_2|^{\Delta_2+\frac{1}{2}}}\Big(f_3(\xi_1,\xi_2)(\slashed{x}_{12} \slashed{z}_1+2z_1\cdot z_2)\\
    &\qquad+f_4(\xi_1,\xi_2)\gamma_{n_1}\gamma_{n_2}((\slashed{\Tilde{x}}_1-\slashed{x}_2) \slashed{z}_1+2z_1\cdot z_2)\Big) \ .
    \end{split}
\end{equation}
The $f_3$ and $f_4$ functions measure chirality conversion by the defect, and for a defect preserving chiral symmetry, they should vanish.

We henceforth restrict to the chirality preserving pair $\left<\psi_+(x_1)\bar{\psi}_{-}(x_2)\right>$. After substituting in the OPEs, we find
\begin{equation}
    \label{conformal block matrix}
    \begin{split}
\left<\psi_+(x_1)\bar{\psi}_{-}(x_2)\right>=&
\sum_{l>-1} 
a^{(l)}_{\shortuparrow +} \bar a^{(l)}_{\shortuparrow -}
\left( {\mathcal U}^{(l)}_{\shortuparrow +}(x_1) * \overline{\mathcal U}^{(l)}_{\shortuparrow -}(x_2)  \right)
\langle u^{(l)}_{\shortuparrow + }(\vec x_1) \bar u^{(l)}_{\shortuparrow -} (\vec x_2) \rangle \\
& +
\sum_{l>0} 
a^{(l)}_{\shortdownarrow +} \bar a^{(l)}_{\shortdownarrow -}
\left( {\mathcal U}^{(l)}_{\shortdownarrow +}(x_1) * \overline{\mathcal U}^{(l)}_{\shortdownarrow -}(x_2)  \right)
\langle u^{(l)}_{\shortdownarrow +}(\vec x_1) \bar u^{(l)}_{\shortdownarrow -} (\vec x_2) \rangle
\\
& +
\sum_{l<1}
b^{(l)}_{\shortuparrow +} \bar b^{(l)}_{\shortuparrow -}
\left( {\mathcal D}^{(l)}_{\shortuparrow +}(x_1) * \overline{\mathcal D}^{(l)}_{\shortuparrow -}(x_2)  \right)
\langle d^{(l)}_{\shortuparrow +}(\vec x_1) \bar d^{(l)}_{\shortuparrow -} (\vec x_2) \rangle\\
& +
\sum_{l<0}  
b^{(l)}_{\shortdownarrow +} \bar b^{(l)}_{\shortdownarrow -}
\left( {\mathcal D}^{(l)}_{\shortdownarrow +}(x_1) * \overline{\mathcal D}^{(l)}_{\shortdownarrow -}(x_2)  \right)
\langle d^{(l)}_{\shortdownarrow +}(\vec x_1) \bar d^{(l)}_{\shortdownarrow -} (\vec x_2) \rangle \ .
    \end{split}
\end{equation}
where the bar on ${\mathcal U}$ and ${\mathcal D}$ denotes complex conjugation.
The two $u$-channel sums  are related under the shift $l \to l-1$, as are the two $d$-channel sums:
\begin{equation}
    \label{conformalblocksimp}
    \begin{split}
\langle \psi_+(x_1) & \bar{\psi}_{-}(x_2)\rangle = \\
&
\sum_{l>0} 
\left( a^{(l)}_{\shortdownarrow +} \bar a^{(l)}_{\shortdownarrow -}
-a^{(l-1)}_{\shortuparrow +} \bar a^{(l-1)}_{\shortuparrow -}
\right)
\left( {\mathcal U}^{(l)}_{\shortdownarrow +}(x_1) * \overline{\mathcal U}^{(l)}_{\shortdownarrow -}(x_2)  \right)
\langle u^{(l)}_{\shortdownarrow +}(\vec x_1) \bar u^{(l)}_{\shortdownarrow -} (\vec x_2) \rangle \\
& + \sum_{l<1} 
\left( b^{(l)}_{\shortuparrow +} \bar b^{(l)}_{\shortuparrow -} -  b^{(l-1)}_{\shortdownarrow +} \bar b^{(l-1)}_{\shortdownarrow -}
\right) 
\left( {\mathcal D}^{(l)}_{\shortuparrow +}(x_1) * \overline{\mathcal D}^{(l)}_{\shortuparrow -}(x_2)  \right)
\langle d^{(l)}_{\shortuparrow +}(\vec x_1) \bar d^{(l)}_{\shortuparrow -} (\vec x_2) \rangle \ .
\end{split}
\end{equation}
This relation under the shift $l \to l-1$ means that the two spinor modes with the same total transverse angular momentum contribute with the same functional dependence to the correlation function.  Both the up-spin and down-spin up-quarks contribute in the same way.  Similarly, the up-spin and down-spin down-quarks contribute in the same way.  
If we carefully follow through the implications of the definition of the bar, we find
\begin{equation}
    a_{\shortuparrow \pm}^* = \bar a_{\shortuparrow \mp} \ , \;  \; \;
    a_{\shortdownarrow \pm}^* = -\bar a_{\shortdownarrow \mp} \ , \; \; \;
    b_{\shortuparrow \pm}^* = \bar b_{\shortuparrow \mp} \ , \;  \; \;
    b_{\shortdownarrow \pm}^* = -\bar b_{\shortdownarrow \mp} \ .
\end{equation}

\par In Appendix \ref{sec:conformalblocks}, we show how the $_{12}$ component of these $2\times 2$ matrices, ${\mathcal U} * {\mathcal U}$ and ${\mathcal D} * {\mathcal D}$, can be evaluated and written in terms of hypergeometric functions.  We can repeat the exercise with all the matrix elements:
\begin{equation}
    \begin{split}
     \left<\psi_+(x_1)\bar{\psi}_{-}(x_2)\right>_{11} 
    = &
         \sum_{l<1}
       \left( \left| b^{(l)}_{\shortuparrow +} \right|^2
       + \left|  b^{(l-1)}_{\shortdownarrow +} \right|^2
       \right)
        \frac{i w_{21}}{\bar z_1^{l-1} z_2^{l-1}}\frac{2-l}{1-l} \calF_{3-l}(\chi)
        \\
        &
         - \sum_{l>0}
         \left(
        \left| a^{(l)}_{\shortdownarrow +} \right|^2  +
         \left| a^{(l-1)}_{\shortuparrow +} \right|^2
         \right)
        i w_{12} z_1^{l-1} \bar z_2^{l-1} \calF_{1+l}(\chi) . \\
    \end{split}
\end{equation}
where
\begin{equation}
\calF_l(\chi) \equiv \frac{1}{(2 |z_1| |z_2| \chi)^l} {}_2 F_1
\left( \frac{l}{2}, \frac{l+1}{2}, l-1, \frac{1}{\chi^2} \right) \ .
\end{equation}
Similarly, for the other components, we find
\begin{eqnarray}
\left<\psi_+(x_1)\bar{\psi}_{-}(x_2)\right>_{12}
&=&
\sum_{l<1} \left( \left| b^{(l)}_{\shortuparrow +} \right|^2 + \left| b^{(l-1)}_{\shortdownarrow +} \right|^2  \right) \frac{1}{\bar z_1^{l-1} z_2^l} \frac{1-l+\frac{1}{2}(z_2 \partial_{z_2} + \bar z_2 \partial_{\bar z_2})}{1-l} \calG_{2-l} (\chi) \nonumber \\
&&
-\sum_{l>0} \left( \left| a^{(l)}_{\shortdownarrow +} \right|^2+ \left| a^{(l-1)}_{\shortuparrow +}\right|^2\right) z_1^{l-1} \bar z_2^l  \frac{l+\frac{1}{2}(z_1 \partial_{z_1} + \bar z_1 \partial_{\bar z_1})}{l} \calG_{1+l}(\chi) \nonumber
\\
\left<\psi_+(x_1)\bar{\psi}_{-}(x_2)\right>_{21} &=&
-\sum_{l<1} \left( \left| b^{(l)}_{\shortuparrow +}\right|^2 + \left| b^{(l-1)}_{\shortdownarrow +} \right|^2 \right) \frac{1}{\bar z_1^{l} z_2^{l-1}} \frac{1-l+\frac{1}{2}(z_1 \partial_{z_1} + \bar z_1 \partial_{\bar z_1})}{1-l} \calG_{2-l} (\chi) \nonumber \\
&&
+\sum_{l>0} \left( \left| a^{(l)}_{\shortdownarrow +} \right|^2 + \left| a^{(l-1)}_{\shortuparrow +}\right|^2 \right) z_1^{l} \bar z_2^{l-1}  \frac{l+\frac{1}{2}(z_2 \partial_{z_2} + \bar z_2 \partial_{\bar z_2})}{l} \calG_{1+l}(\chi) \nonumber
\\
\left<\psi_+(x_1)\bar{\psi}_{-}(x_2)\right>_{22} 
&=& \sum_{l<1} \left( \left| b^{(l)}_{\shortuparrow +}\right|^2 
+ \left| b^{(l-1)}_{\shortdownarrow +}\right|^2 \right)\frac{i \bar w_{21}}{\bar z_1^l z_2^l} \calF_{2-l}(\chi)  
\nonumber\\
&& - \sum_{l>0} \left( \left|a^{(l)}_{\shortdownarrow +}\right|^2 + \left| a^{(l-1)}_{\shortuparrow +} \right|^2 \right)  i \bar w_{12} z_1^{l} \bar z_2^{l} \frac{1+l}{l} \calF_{2+l}(\chi) , 
\end{eqnarray}
where
\begin{equation}
\calG_l(\chi) \equiv \frac{1}{(2 |z_1| |z_2| \chi)^{l}} {}_2 F_1 \left( \frac{l}{2}, \frac{l+1}{2}, l, \frac{1}{\chi^2} \right)
\end{equation}

As argued previously, 
we have strict inequalities in the sums ($l>0$ and $l<1$).
The terms at the limits $l=0$ and $l=1$
are excluded, because the defect fields are free at these values and should decouple. 
Here this decoupling is enforced by the simple poles in $l$ at these values.  Scaling the bulk-defect coefficients to zero at these values to keep a finite contribution in the sum decouples the defect fields from the bulk.

\par We next compare summation of the defect OPE to the solution to the Dirac equation. For simplicity, we make the coordinate transformations used when finding the solution, $x_1=(\vec{x}_1,\vec{z}_1)=(\vec{0},1,0)$ and $x_2=(\vec{x}_2,\vec{z}_2)=(\vec{0},r\cos{\theta},r\sin{\theta})$. The results become:
\begin{equation}
    \left<\psi_+(x_1)\bar{\psi}_{-}(x_2)\right>_{11}=\left<\psi_+(x_1)\bar{\psi}_{-}(x_2)\right>_{22}=0,
\end{equation}
\begin{equation}
    \begin{split}
       \left<\psi_+(x_1)\bar{\psi}_{-}(x_2)\right>_{12}= 
       & \sum_{l<1} \left( \left| b^{(l)}_{\shortuparrow+} \right|^2 + \left| b^{(l-1)}_{\shortdownarrow +} \right|^2 \right) \frac{e^{-i l \theta} r^l}{(l-1) (r^2-1)^2} \\
       & -\sum_{l>0} \left( \left| a^{(l)}_{\shortdownarrow+}\right|^2 + \left| a^{(l-1)}_{\shortuparrow +}\right|^2 \right) \frac{e^{-i l \theta} r^{-l} (1 + l(r^2-1))}{l(r^2-1)^2} \ , 
    \end{split}
\end{equation}
and a similar expression for the $_{21}$ component.
By comparing the $_{12}$ component to the bulk-bulk two-point function written using $f_1$ and $f_2$, and the solution to the Dirac equation, we see that
\begin{eqnarray}
\frac{1}{1-l}\left( \left|b^{(l)}_{\shortuparrow+} \right|^2 + \left|b^{(l-1)}_{\shortdownarrow +}\right|^2 \right) &=& 2 c_{1,1-l}  \ , \\
 \frac{1}{l}\left( \left| a^{(l)}_{\shortdownarrow+} \right|^2 + \left| a^{(l-1)}_{\shortuparrow +} \right|^2 \right)
 &=& 2 c_{2,1-l}  \ .
\end{eqnarray}
This identification establishes a relation between the solution to the Dirac equation and the OPE expansion under the mapping $l \leftrightarrow 1-l$.

Note that this bulk-bulk two-point function is a useful starting point for computing spinor one-point functions.  Using point-splitting regularization, it is straightforward to compute both a current operator $\langle \bar \psi \gamma^\mu \psi (x) \rangle$ and the stress-tensor, which through this formalism can be expressed as a weighted sum over the bulk-defect coefficients $a_{\shortuparrow/ \shortdownarrow, \pm}^{(l)}$ and $b_{\shortuparrow/ \shortdownarrow , \pm}^{(l)}$.

Before moving on, we give the result for the  two-point function in the absence of a defect
\begin{equation}
\langle \psi(x_1) \bar \psi(x_2) \rangle = 
\kappa \frac{\slashed{x}_{12}}{|x_{12}|^{4}} \ ,
\end{equation}
decomposed into partial waves.
Comparing this with the conformal result (\ref{bulk-bulk}) demonstrates that $f_1 = \frac{\kappa}{16 \xi_1^2}$, with all other $f_i$ set to zero.
Expanding at large or small $r$, we find
\begin{eqnarray}
\frac{\kappa}{16 \xi_1^2} &=&  
\frac{\kappa}{2} \sum_{l=1}^\infty e^{i l \theta}(e^{-i \theta} I_{l-1}(\chi) + I_{l}(\chi)) + c.c., \\
0 &=& 
\frac{i \kappa}{2} \sum_{l=1}^\infty e^{i l \theta}(e^{-i \theta} I_{l-1}(\chi) - I_{l}(\chi)) + c.c. \ , 
\end{eqnarray}
and so we set $c_{1,l} = \frac{\kappa}{2} = c_{2,1-l}$ for positive $l$, with all other $c_{a,l} = 0$. 
We anticipate that locality means that at large spin $l$, the $c_{a,l}$ should approach these values.  In other words, high angular momentum fermions should miss the defect and see the trivial theory.

\section{Three-point functions, shadows, and double twist}
\label{sec:bulkdefectdefect}

We now move to a consideration of the bulk-defect-defect three-point function 
\begin{equation}
\langle \psi(x_1) O_2(x_2) O_3(x_3) \rangle \ .
\end{equation} 
 While nominally an arbitrary function of a cross-ratio,  its structure is in fact greatly constrained by Lorentz invariance.  
 If we replace the bulk field $\psi$ with its defect OPE (\ref{OPEtotal}), Lorentz invariance picks out just two terms in this sum.
 In this 2d/4d case, where angular momentum is represented by helicity parallel and transverse to the defect, the helicities of $O_2$ and $O_3$ determine a unique set of Lorentz quantum numbers.  Given that $\psi$ is a free field, there are a maximum of two possible defect primaries that have a nonzero three-point function with $O_2$ and $O_3$.  Let us call one of these defect primaries ${\mathcal W}$ with dimension $\Delta_{\mathcal W}= 1+|j|$, and its shadow dual ${\mathcal S}$ with dimension $\Delta_{\mathcal S} = 1-|j|$.

The shadow-dual nature of ${\mathcal W}$ and ${\mathcal S}$ 
provides a simple way to constrain their three-point functions with $O_2$ and $O_3$. 
For most transverse angular momenta, unitarity allows only the ${\mathcal W}$ member of the pair. The corresponding three-point function must then lie in the kernel of the shadow transform, which occurs at a discrete set of double-twist dimensions. In the monodromy window, both shadow partners are unitary.  In this case, the three-point functions need not obey a double-twist condition, but their OPE coefficients are related by the shadow transform.

We first derive these statements directly from the defect shadow transform. We then verify them by solving the bulk Dirac equation for the representative correlator $\langle\psi\,\bar\rho\,\bar \phi\rangle$, where $\phi$ is a defect scalar and $\rho$ a defect spinor. In this example, the same constraint appears as cancellation of an otherwise spurious bulk singularity at $\zeta=1$.

\subsection{Shadow transform}

As we will see later in a specific example, 
because of the Lorentz invariance constraints, solving
\begin{equation}
\slashed{\partial}_{x_1} \langle \psi(x_1) O_2(x_2) O_3(x_3) \rangle = 0  
\end{equation} 
reduces to solving a coupled set of ODEs in a radial variable -- the separation of $\psi(x_1)$ from the defect.  In this context, the two correlation functions
\begin{equation}
 \langle {\mathcal W}(x) O_2(x_2) O_3(x_3) \rangle  \ , \; \; \;
 \langle {\mathcal S}(x) O_2(x_2) O_3(x_3) \rangle
\end{equation}
are boundary data.  Given a regularity condition in the interior, the solution of the bulk Dirac equation allows us to determine one of these three-point functions from the other.  (The map is the spinor analog of the usual Dirichlet-to-Neumann map in solving the Klein-Gordon equation for a scalar field in the presence of a boundary.)
This map can be written in the form of an equation of motion relating ${\mathcal W}$ and ${\mathcal S}$:
\begin{equation}
G_{\mathcal S}^{-1} {\mathcal S} \sim {\mathcal W} \ .
\end{equation}
Here $G_{\mathcal S}^{-1}$ is a nonlocal differential operator whose inverse is the integral transform with kernel
$G_{\mathcal S}(x, x') = \langle {\mathcal S}(x) \bar {\mathcal S}(x') \rangle$.

Given this equation of motion, the relation between the boundary data is
\begin{equation}
\label{shadowtransform}
\langle {\mathcal S}(x_1) O_2(x_2) O_3(x_3) \rangle = 
  \kappa_{\calS \calW} \int d^2x \langle {\mathcal S}(x_1) \bar {\mathcal S}(x) \rangle \langle {\mathcal W}(x) O_2(x_2) O_3(x_3) \rangle \ 
\end{equation}
(and similar expressions for higher point functions).  
For now, we assume that ${\mathcal W}$ and ${\mathcal S}$ have canonical CFT normalized two-point functions, and allow for an, at this point, arbitrary normalization constant $\kappa_{\calS \calW}$ in the definition of the shadow transform.
In the three-point function case, as the correlation functions are fixed up to numbers,
we can hope to find a relation between these three-point function coefficients.
To carry out the integral,
by conformal invariance let us choose special points $x_1 = 0$, $x_2 = 1$ and $x_3 \to \infty$.  The integral (\ref{shadowtransform}) reduces in complex coordinates to
\begin{eqnarray}
   c_{{\mathcal S} 23}&=&  \kappa_{\calS \calW} c_{{\mathcal W}23} \int d^2 z \, z^{a-1}  \bar z^{a'-1}
    (1-z)^{b-1} (1-\bar z)^{b'-1 } \\
    &=&\pi 
    \frac{\Gamma(a) \Gamma(b) \Gamma(1-a'-b')}{\Gamma(1-a')\Gamma(1-b')\Gamma(a+b)} \kappa_{\calS \calW} c_{{\mathcal W}23}
    \end{eqnarray}
where $a = 2h-1$, $a' = 2h'-1$, $b=1-h-h_2+h_3$ and $b' = 1-h'-h_2'+h_3'$.  
(For this integral, see for example (1.7) of \cite{Neretin:2024}.)  The differences $a-a'$ and $b-b'$ need to be integer for the result to hold, which here is guaranteed by the integrality of the spin.  We are using standard 2d CFT notation where $h+h' = \Delta$ is the scaling weight and $h'-h$ is the spin of ${\mathcal W}$, and similarly for $O_2$ and $O_3$. The shadow $S$ has weights $1-h$ and $1-h'$. 

If we perform the shadow transform twice with these canonical normalized  ${\mathcal S}$ and ${\mathcal W}$, we get back the original three-point function up to a multiplicative factor
\begin{equation}
\label{invertibilityfailure}
\frac{\pi^2}{(1-2h)(1-2h')} \kappa_{\calS \calW} \kappa_{\calW \calS} \ . 
\end{equation}
The poles at $h=1/2$ and $h'=1/2$ signal points where the shadow transform ceases to be invertible. For a spin-$\frac12$ operator, these correspond to $\Delta=\frac12$ and $\Delta=\frac32$: the former is the spinor unitarity bound, where the operator becomes free, while the latter is its shadow.
Away from these special values, we should choose $\kappa_{\calS \calW}$ and $ \kappa_{\calW \calS}$ such that the product (\ref{invertibilityfailure}) is equal to one.

Now we ask, what happens if unitarity forbids ${\mathcal S}$?  Naively, the shadow transform would seem to imply that $c_{{\mathcal W} 23}$ must vanish as well.  However, this conclusion is too fast because the $\Gamma$-function prefactor may vanish, but only when a double-twist condition is satisfied.
For $O_2$ to be double twist, it should have the schematic form $O_3  \partial_w^n \partial_{\bar w}^{n'} {\mathcal W}$ (and correspondingly for $O_3$).   From this construction, the weights of $O_2$ must be
\begin{equation}
h_2 = h_3 + h + n \ , \; \; \; h_2' = h_3' + h' + n' \ .
\end{equation}
Now let us compare this data with the ratio of $\Gamma$-functions in the shadow transform.  
The poles in $\Gamma(a+b) = \Gamma(h-h_2+h_3)$ occur for the double twist values $h_2 = h_3 + h +n$, $n=0,1,2, \ldots$.  
Suppose we are given the spins of these three operators, $h'-h= s$, $h_2'- h_2 = s_2$ and $h_3'-h_3 = s_3$.
Then we can write an equivalent relation for $h_2'$:
\begin{equation}
h_2' = h_3' + h' + n + s_2 - s_3 - s \ .
\end{equation}
The integer $n'=n+s_2 - s_3 - s$ is no longer obviously non-negative however.  This problem is cured by a $\Gamma$-function in the numerator.
The numerator $\Gamma(1-a'-b') = \Gamma(1-h'+h_2'-h_3')$ has poles at $h_2' = h_3' +h'-1 -n$ for $n=0,1,2, \ldots$.  Thus the values of $n$
for which $n + s_2 - s_3 - s$ may be negative have poles in the denominator, which are canceled by corresponding poles in the numerator, leaving
the double twist condition for both $h_2$ and $h_2'$.  
A similar argument goes through for $h_3$ and $h_3'$, through an analysis of the ratio $\Gamma(b) / \Gamma(1-b') = \Gamma(1-h-h_2+h_3)/\Gamma(h'+h_2'-h_3')$.

Sometimes, however, ${\mathcal S}$ is allowed.
In the case of the monodromy defect,
we have a shadow dual pair of operators in the defect OPE of $\psi$, one called ${\mathcal W} = u$ with dimension $\Delta_u = \frac{3}{2} - \nu$ and one called ${\mathcal S} = d$ with dimension $\Delta_d = \frac{1}{2} + \nu$, restricting the range $0 \leq \nu < \frac{1}{2}$.  
If we have a nonzero three-point function $\langle d(x_1) O_2(x_2) O_3(x_3) \rangle$ on the defect, we can determine a three-point function involving the shadow operator from
\begin{equation}
\langle u(x_1) O_2(x_2) O_3(x_3) \rangle = 
  \kappa_{ud} \int d^2x \langle u(x_1) \bar u(x) \rangle \langle d(x) O_2(x_2) O_3(x_3) \rangle \ .
\end{equation}

For future reference, we give the result for the example we will consider next, 
where $h-h'= \pm 1/2$ and $h_2 - h_2'= 1/2$. Using Table \ref{table:spins}, the ratio of three-point function coefficients becomes
    \begin{eqnarray}
    \label{cu23overcd23rel}
 \frac{c_{u \shortdownarrow \pm}}{c_{d \shortuparrow \pm}} =
    \frac{c_{u23}}{c_{d23}} &=&
    \mp
    \pi  \kappa_{ud}
    \frac{\Gamma(\nu-1)}{\Gamma(1-\nu)}
    \frac{\Gamma (\frac{1}{4}(2\mp 1 - 2 \Delta_{23}-2\nu))
    \Gamma (\frac{1}{4}(2\pm 1 + 2 \Delta_{23}-2\nu))}
    {\Gamma (\frac{1}{4}(\pm 1 - 2 \Delta_{23}+2\nu))
    \Gamma (\frac{1}{4}(\mp 1 + 2 \Delta_{23}+2\nu))} \ .
\end{eqnarray}

There are further poles and zeroes associated with the ratio $\Gamma(a)/ \Gamma(1-a') = \Gamma(2h-1)/\Gamma(2-2h')$.  The zeroes occur at $h' = 1, \frac{3}{2}, 2, \ldots$.  The corresponding shadows have weights $\tilde h' = 1 - h' = 0, -\frac{1}{2}, -1, \ldots$.  These operators are exceptional; they have descendants from acting with $\bar \partial$ which are also primary at the values  $h' = 1, \frac{3}{2}, 2, \ldots$.  There is a corresponding story for the poles, involving the inverse transform.  For these special shadow pairs, the shadow transform is not well-behaved, but these details are not important for the main thrust of our discussion.

\subsection{A simple example}

We now verify these general statements directly by studying the simplest nontrivial bulk-defect-defect correlator, with a defect spinor $O_2(x_2) = \bar \rho^{(l_2)}_\alpha(x_2)$ and scalar  $O_3(x_3) = \bar \phi^{(l_3)}(x_3)$:
\begin{equation}
\left<\psi_\pm(x_1)\bar{\rho}^{(l_2)}_\alpha(x_2)\bar \phi^{(l_3)}(x_3)\right> \ .
\end{equation}
The strategy is similar to the bulk-bulk analysis above, where we first constrain the three-point function using the Dirac equation and then pin down the meaning of the integration constants using the OPE.

\subsubsection*{Solving Dirac}

By applying the Dirac equation to the bulk-defect-defect correlation function (eq.\ \eqref{bulk-defect-defect}), we can solve for the four functions $f_i(\myupsilon)$: 
\begin{equation}
    \slashed\partial_1\left<\psi_\pm(x_1)\bar{\rho}^{(l_2)}_\alpha(x_2)\bar \phi^{(l_3)}(x_3)\right>=0
\end{equation}
Again we work in 4d and use the basis in the Clifford algebra. After applying the Dirac equation, we can choose a special point: $x_1=(0,0,r\cos{\theta},r\sin\theta),x_2=(0,0,0,0),x_3=(0,1,0,0)$. By rewriting $r$ in terms of $\myupsilon$ (eq.\ \eqref{u}), we can express the Clifford basis coefficients in terms of $\myupsilon$ and $\theta$. In fact, we choose to replace $\myupsilon$ with $v=\frac{1}{1-\myupsilon}$, because it is the variable that naturally appears in the OPE expansions. 

\par We obtain four independent basis coefficients that we set to zero. The $\theta$ dependence can be cancelled in all of these equations and we find that $f_1(v)$ and $f_4(v)$ satisfy two coupled ordinary differential equations and $f_2(v)$, and $f_3(v)$ satisfy two others. In the following, we take $\alpha=+$. The results are the same for the $\alpha=-$ case, except with $l\rightarrow-l$.
\par The differential equations for $f_1(v)$ and $f_4(v)$ are
\begin{eqnarray}
\label{f1f4diffeqone}
4(1-v) f_1'(v) + \left( 3 -2\Delta_{23} -2l \left(1-\frac{1}{v}\right)\right)f_1(v) +(1+2 \Delta_{23} + 2l) f_4(v) &=& 0 \\
8v(v-1) (f_1'(v) - f_4'(v)) +4(2-l) f_4(v) +(4 \Delta_{23} - 6) f_1(v) &=& 0 \ .
\end{eqnarray}
Here we have set the bulk fermion dimension $\Delta_1=\frac{3}{2}$, but left the defect fermion and scalar dimensions $\Delta_2$ and $\Delta_3$ arbitrary. The general solution is 
\begin{eqnarray} 
f_1(v) &=& C_{1+} H(-l, v) + C_{2+} H(l,v) \ , \label{bulk-defect-defect f1}\\
f_4(v) &=& \frac{4(v-1) f_1'(v)
+ (2l(1-\frac{1}{v}) + (2 \Delta_{23}-3  )) f_1(v) }{ 2\Delta_{23}+2l+1} \label{bulk-defect-defect f4}\ ,
\end{eqnarray}
where
\begin{equation}
H(l,  v) \equiv (-v)^{\frac{l}{2}} (1-v)^{\frac{5}{4} + \frac{\Delta_{23}}{2}} {}_2 F_1 \left( \frac{1+2l}{4} + \frac{\Delta_{23}}{2} ,  \frac{5+2l}{4} + \frac{\Delta_{23}}{2}, 1+l, v \right) \ .
\end{equation}
Given the denominator in the expression for $f_4(v)$, this general solution is invalid when $1+2 \Delta_{23} + 2l = 0$.  
In this case, we can think of $\bar \phi^{(l_3)}(x_3)$ as $\bar \rho_\alpha^{(l_2)} (x_2)$ multiplied by one of the defect operators in the spin-down channel of the defect expansion of $\psi$. 
The $f_4$ dependence drops from the first differential equation (\ref{f1f4diffeqone}), which can then be solved directly to give
\begin{equation} \label{n=0 f1}
f_1(v) = c_{1+} (v-1)^{1+\frac{l}{2}} v^{-\frac{l}{2}} \ .
\end{equation}
As the three-point function should be regular in the limit $v \to \infty$, we can set $c_{1+} = 0$.  The remaining solution for $f_4$ is then
\begin{equation}
\label{f4specialcase}
f_4(v) = c_{2+} (v-1)^{1-\frac{l}{2}} v^{-1 + \frac{l}{2}} = c_{2+} \myupsilon^{1 - \frac{l}{2}}\ ,
\end{equation}
which is regular in this limit.

The cases where $\frac{1}{4}(1 + 2l + 2\Delta_{23}) = -n$, $n = 1, 2, 3, \ldots$ are also special.  Here the regular solution in the limit $v \to \infty$ is a polynomial (up to the $(-v)^{l/2} (1-v)^{1-l/2-n}$ prefactor)
\begin{equation}
f_1(v) = c_{2+} (-v)^{\frac{l}{2}} (1-v)^{1 - \frac{l}{2} -n} {}_2 F_1(-n, -n+1, 1+l, v) \ .
\end{equation}
Similarly for $\frac{1}{4}(1 - 2l + 2\Delta_{23}) = -n$, $n = 1, 2, 3, \ldots$, the solution is also a polynomial
\begin{equation}
f_1(v) = c_{1+} (-v)^{-\frac{l}{2}} (1-v)^{1 + \frac{l}{2} -n} {}_2 F_1(-n, -n+1, 1-l, v) \ .
\end{equation}

We go through the same exercise with $f_2(v)$ and $f_3(v)$.  The differential equations are
\begin{eqnarray}
\left(\frac{5  - 2 \Delta_{23} v}{v-1}+ 2 l +1 \right) f_2(v) + (3 + 2 \Delta_{23}- 2 l)  f_3(v) - 
 4  v f_2'(v) &=& 0 \ , \\
 4v(v-1)((v-1) f_3'(v) - v f_2'(v)) -2l(v-1) f_3(v) +(5 - 2 \Delta_{23}) v f_2(v) &=& 0 \ ,
\end{eqnarray}
with general solution
\begin{eqnarray}
f_2(v) &=& (1-v)^{\frac{1}{2}}  (-v)^{-\frac{1}{2}}\left[  C_{3+} H(1-l, v)+  C_{4+} H(l-1,v) \right] \ , \\
f_3(v) &=&\frac{4v(v-1) f_2'(v)
+ ((2 \Delta_{23}-2l-1) v+2(l-2)) f_2(v) }{(2\Delta_{23}-2l+3)(v-1)} \ .
\end{eqnarray}
Again, something special happens when $3 + 2 \Delta_{23} -2l = 0$.  
In this case, we can think of operator 3 as operator 2 multiplied by one of the defect operators in the spin-up channel of the defect expansion of $\psi$.  
In this case, the single solution for $f_2(v)$ is
\begin{equation} \label{n=0 f2}
f_2(v) = c_{3+} (v-1)^{2 -\frac{l}{2}} v^{-1+\frac{l}{2}} \ .
\end{equation}
which can be safely thrown out because of the singularity in the limit $v \to \infty$. 
The remaining solution for $f_3(v)$ is
\begin{equation} \label{n=0 f3}
f_3(v) = c_{4+} (v-1)^{\frac{l}{2}} v^{-\frac{l}{2}}  = c_{4+} \zeta^{\frac{l}{2}} \ ,
\end{equation}
which is regular in this same limit. 

There are analogous solutions when $\alpha = -1$ with integration constants $C_{1-}$, $C_{2-}$, $C_{3-}$, and $C_{4-}$ given by sending $l \to -l$.  The expressions for the $f_i$ are at the end of appendix \ref{sec:bddexpansions}.

\subsubsection*{Comparison to the OPE}
Applying the OPE to the defect-defect-defect correlation function (eq.\ \eqref{defect-defect-defect}) should give the same functions as obtained via the Dirac equation, and allow us to find the meaning of the integration constants. We can rewrite the defect-defect-defect correlation function in terms of the defect chirality projector, for which there are two different cases:
\begin{equation}
    \begin{split}
        &\left<\rho_{\alpha\beta}^{(l_2+l_3)}(x_1)\bar{\rho}^{(l_2)}_{\alpha,-\beta}(x_2)\bar \phi^{(l_3)}(x_3)\right>=c_{\rho \bar{\rho}\bar\phi}\frac{\mathbb{P}^\perp_\alpha\mathbb{P}^\parallel_\beta\slashed{x}_{12}\mathbb{P}^\perp_\alpha\mathbb{P}^\parallel_{-\beta}}{|x_{12}|^{1+\Delta_1+\Delta_2-\Delta_3}|x_{13}|^{\Delta_1-\Delta_2+\Delta_3}|x_{23}|^{-\Delta_1+\Delta_2+\Delta_3}},\\
        &\left<\rho_{\alpha\beta}^{(l_2+l_3)}(x_1)\bar{\rho}^{(l_2)}_{\alpha\beta}(x_2)\bar \phi^{(l_3)}(x_3)\right>=c'_{\rho \bar{\rho}\bar\phi}\frac{\mathbb{P}^\perp_\alpha\mathbb{P}^\parallel_\beta\slashed{x}_{13}\slashed{x}_{32}\mathbb{P}^\perp_\alpha\mathbb{P}^\parallel_\beta}{|x_{12}|^{\Delta_1+\Delta_2-\Delta_3}|x_{13}|^{1+\Delta_1-\Delta_2+\Delta_3}|x_{23}|^{1-\Delta_1+\Delta_2+\Delta_3}}.\\
    \end{split}
\end{equation}
Therefore, we see that the correlation function is only nonzero if both defect fermions have the same value of $\alpha$, but they do not have to have the same defect chirality.

\par 
To fix  the relation between the integration constants we found in solving the Dirac equation and the OPE coefficients, 
it is enough to perform the analysis in the
$x_3\rightarrow\infty$ limit. In this limit, the defect-defect-defect correlation function takes the form
\begin{equation}
    \begin{split}
        &\left<\rho_{\alpha\beta}^{(l_2+l_3)}(x_1)\bar{\rho}^{(l_2)}_{\alpha,-\beta}(x_2)\bar \phi^{(l_3)}(x_3\rightarrow\infty)\right>=|x_3|^{-2\Delta_3} c_{\rho \bar{\rho} \bar\phi} \frac{\mathbb{P}^\perp_\alpha\mathbb{P}^\parallel_\beta\slashed{x}_{12}\mathbb{P}^\perp_\alpha\mathbb{P}^\parallel_{-\beta}}{|x_{12}|^{1+\Delta_1+\Delta_2-\Delta_3}},\\
        &\left<\rho_{\alpha\beta}^{(l_2+l_3)}(x_1)\bar{\rho}^{(l_2)}_{\alpha\beta}(x_2)\bar \phi^{(l_3)}(x_3\rightarrow\infty)\right>=-|x_3|^{-2\Delta_3}c'_{\rho \bar{\rho} \bar \phi}\frac{\mathbb{P}^\perp_\alpha\mathbb{P}^\parallel_\beta\mathbb{P}^\perp_\alpha\mathbb{P}^\parallel_\beta}{|x_{12}|^{\Delta_1+\Delta_2-\Delta_3}}.\\
    \end{split}
\end{equation}
The bulk-defect-defect correlation function becomes
\begin{equation} \label{bulk-defect-defect infinite limit} 
    \begin{split} 
        &\left<\psi_a(x_1)\bar{\rho}^{(l_2)}_{\alpha\beta}(x_2)\bar \phi^{(l_3)}(x_3\rightarrow\infty)\right>\\
        &\quad=|x_3|^{-2\Delta_3}\mathbb{P}_a^*\Bigg(\frac{f_1(\myupsilon)\slashed{x}_{12}}{|x_{12}|^{1+\Delta_1+\Delta_2-\Delta_3}}+\frac{f_2(\myupsilon)\slashed{x}_{12}\slashed{z}}{|x_{12}|^{2+\Delta_1+\Delta_2-\Delta_3}}\\    
        &\quad -\frac{f_3(\myupsilon)\mathbb{I}}{|x_{12}|^{\Delta_1+\Delta_2-\Delta_3}} -\frac{f_4(\myupsilon)\slashed{z}}{|x_{12}|^{1+\Delta_1+\Delta_2-\Delta_3}}\Bigg)\mathbb{P}_\alpha^\perp \mathbb{P}_\beta^\parallel\left(\frac{z}{|z|}\right)^{l_2+l_3}\\
    \end{split}
\end{equation}
with 
\begin{equation}
    \myupsilon=\frac{(x_2-x_1)^2}{|z|^2}.
\end{equation}
From now on we ignore the common prefactor of 
$|x_3|^{-2\Delta_3}$. Moving into complex coordinates and choosing specific projections, we find that 
\begin{equation}
\begin{split}
     & \left<u_{\shortdownarrow +}^{(l_2+l_3)}(x_1)\bar{\rho}^{(l_2)}_{++}(x_2)\bar \phi^{(l_3)}(\infty)\right>=-c_{u \shortdownarrow +}\,(w_{12})^{\frac{1}{2}(-\Delta_1 - \Delta_2 + \Delta_3)}
(\bar{w}_{12})^{\frac{1}{2}(-\Delta_1 - \Delta_2 + \Delta_3)}\\
& \left<d_{\shortuparrow +}^{(l_2+l_3)}(x_1)\bar{\rho}^{(l_2)}_{++}(x_2)\bar \phi^{(l_3)}(\infty)\right>=-i c_{d \shortuparrow +}\,(w_{12})^{\frac{1}{2}(-1 - \Delta_1 - \Delta_2 + \Delta_3)}
(\bar{w}_{12})^{\frac{1}{2}(1 - \Delta_1 - \Delta_2 + \Delta_3)}.
\end{split}
\end{equation}
where, as per Table \ref{table:rhotoudtranslation}, $d_{\shortuparrow+}^{(l)} = \rho^{(l)}_{\shortuparrow + -}$ and $u_{\shortdownarrow +}^{(l)} = \rho_{\shortdownarrow ++}^{(l)}$.
We are using a short-hand for the three-point function coefficients, suppressing the angular momentum label $l$ and the identity of the other two defect operators $\rho$ and $\phi$.
When we apply the OPE (eq.\ \eqref{OPEtotal}) to these, only one value of $l$ is picked out from the sum, such that $l=l_2+l_3$. 

In order to fix the integration constants $C_{1\pm}$, $C_{2\pm}$, $C_{3\pm}$ and $C_{4\pm}$ from the previous subsection, 
we look at four different bulk-defect-defect correlation functions:  $\left<\psi_+\bar{\rho}^{(l_2)}_{++}\bar \phi^{(l_3)}\right>$,
$\left<\psi_-\bar{\rho}^{(l_2)}_{++}\bar \phi^{(l_3)}\right>$, $\left<\psi_+\bar{\rho}^{(l_2)}_{-+}\bar \phi^{(l_3)}\right>$, and $\left<\psi_-\bar{\rho}^{(l_2)}_{-+}\bar \phi^{(l_3)}\right>$.  The detailed expressions are in appendix \ref{sec:bddexpansions}.  
We then compare these expressions against the relevant projections of the bulk-defect-defect correlation function 
(eq.\ \eqref{bulk-defect-defect infinite limit})
in terms of the $f_i$ functions we have solved for.  Dropping  the factor of $|x_3|^{-2 \Delta_3}$, we are specifically trying to match
\begin{equation}   
\begin{split}
    &\left<\psi_+(x_1)\bar{\rho}^{(l_2)}_{++}(x_2)\bar \phi^{(l_3)}(\vec{x}_3\rightarrow\infty)\right>_2= -i \bar w_{12} \left|x_{12}\right|^{-\frac {5} {2}-  \Delta_{23}}  f_1 (v)\left(\frac{z}{|z|} \right)^{l},\\
    &\left<\psi_-(x_1)\bar{\rho}^{(l_2)}_{++}(x_2)\bar \phi^{(l_3)}(\vec{x}_3\rightarrow\infty)\right>_1=i \bar z \bar w_{12}\left|x_{12}\right|^{-\frac {7} {2}-  \Delta_{23}} f_2(v)\left(\frac{z}{|z|} \right)^{l},\\
    &\left<\psi_+(x_1)\bar{\rho}^{(l_2)}_{-+}(x_2)\bar \phi^{(l_3)}(\vec{x}_3\rightarrow\infty)\right>_2=-i  z \bar w_{12}\left|x_{12}\right|^{-\frac {7} {2}-  \Delta_{23}} f_2(v)\left(\frac{z}{|z|} \right)^{l},\\
    &\left<\psi_-(x_1)\bar{\rho}^{(l_2)}_{-+}(x_2)\bar \phi^{(l_3)}(\vec{x}_3\rightarrow\infty)\right>_1=i \bar w_{12} \left|x_{12}\right|^{-\frac {5} {2}-  \Delta_{23}}  f_1 (v)\left(\frac{z}{|z|} \right)^{l}.
\end{split}
\end{equation}

Comparing the expressions in appendix \ref{sec:bddexpansions} with the solution to the Dirac equation, we are able to fix the relations
\begin{equation}
\label{C1to4pm}
\begin{alignedat}{3}
    &C_{1 +} =b^{(l)}_{\shortuparrow +} c_{d \shortuparrow +}, \qquad
    &C_{1-} &=a^{(l)}_{\shortuparrow -}  c_{u \shortuparrow -},
    \\
    &C_{2 +} = - {a}^{(l)}_{\shortdownarrow +} c_{u \shortdownarrow +} \frac{\left(1+ 2l+2\Delta_{23}\right)}{4l}, \qquad
    &C_{2-} &=b^{(l)}_{\shortdownarrow -} c_{d \shortdownarrow -}\,\frac{\left(1-2l+2\Delta_{23}\right)}{4l},
     \\
     &C_{3 +} =b^{(l)}_{\shortuparrow -} c_{d \shortuparrow -}\frac{\left(3 + 2 \Delta_{23} -2l\right)}{4(1-l)}, \qquad
     &C_{3-} &=a^{(l)}_{\shortuparrow +} c_{u \shortuparrow +}\frac{\left(3 + 2 \Delta_{23} +2l\right)}{4(1+l)}, 
     \\
     &C_{4 +} ={a}^{(l)}_{\shortdownarrow -} c_{u \shortdownarrow -}, \qquad
     &C_{4-} &=b^{(l)}_{\shortdownarrow +} c_{d \shortdownarrow +}.
\end{alignedat}
\end{equation}
As these associations make clear, each $C_{a\pm}$ is associated with an allowed range of $l$, and, in general, only one member of each pair $(C_{1+}, C_{2+})$, $(C_{1-}, C_{2-})$, $(C_{3+}, C_{4+})$, 
and $(C_{3-}, C_{4-})$  can be nonzero because of unitarity constraints.  In the monodromy window, however, both members of the pair are allowed by unitarity.
For example, in this window, $C_{1 +}$ and $C_{3+}$ are associated with operators of dimension $\frac{1}{2}+\nu$ in the defect OPE of $\psi$, while $C_{2 +}$ and $C_{4+}$ are associated with operators of dimension $\frac{3}{2}-\nu$.

\subsubsection*{Divergence at $\myupsilon=1$}

The bulk-defect-defect three-point functions are expressed in terms of hypergeometric functions of
\[
v=\frac{1}{1-\myupsilon},
\]
which have singular points at $(v=0,1,\infty)$, corresponding respectively to $(\myupsilon\to\infty,0,1)$. In Euclidean signature, the physical domain is $\myupsilon \geq 1$.  The limit $\myupsilon\to\infty$ includes both the coincident limit $x_2\to x_3$ and the defect limit $z\to0$, where singular behaviour is physically expected. The locus 
$\myupsilon=0$ is excluded. By contrast, $\myupsilon = 1$ corresponds to the bulk point $x_1$  lying on the semicircle whose diameter joins $x_2$ and $x_3$, and there is no physical reason for the correlator to become singular there. Regularity at $\myupsilon=1$ therefore imposes a nontrivial constraint on solutions of the bulk Dirac equation.

There are two ways to satisfy this constraint. When both members $u$ and $d$ of a shadow-related pair are allowed in the defect OPE, the singular terms can cancel between the corresponding solutions. Requiring this cancellation fixes their relative coefficients and hence relates the defect three-point functions involving $u$ and $d$; this is the shadow-transform relation. When unitarity permits only one member of the pair to appear, such a cancellation is unavailable. Regularity then requires the coefficient of the singular term itself to vanish, which occurs only at special values of the defect scaling dimensions. These are precisely the double-twist conditions.

\par 
We will look at this bulk singularity for the four three-point functions we considered above: 
\[
\left<\psi_+\bar{\rho}^{(l_2)}_{++}\bar \phi^{(l_3)}\right> \ , \; \; \; 
\left<\psi_-\bar{\rho}^{(l_2)}_{++}\bar \phi^{(l_3)}\right> \ , \; \; \; \left<\psi_+\bar{\rho}^{(l_2)}_{-+}\bar \phi^{(l_3)}\right> \ , \; \; \; \left<\psi_-\bar{\rho}^{(l_2)}_{-+}\bar \phi^{(l_3)}\right> \ .
\]
In the main text, we give the details just for $\left<\psi_+\bar{\rho}^{(l_2)}_{++}\bar \phi^{(l_3)}\right>$.  Formulae relevant for the other three can be found in appendix \ref{sec:singularity}

We  investigate the singular limit by choosing the point (in real coordinates) $x_1=(0,0,1+\epsilon,0)$, $x_2=(1,0,0,0)$ and $x_3=(-1,0,0,0)$ since $\myupsilon=1$ for $\epsilon=0$. Using the solution to the Dirac equation and expanding one projection of the bulk-defect-defect correlation function about $\epsilon=0$ gives to leading order
\begin{equation} \label{hypergeom series}
\begin{split}
&\left<\psi_+\bar{\rho}^{(l_2)}_{++}\bar \phi^{(l_3)}\right>_1
\sim\\
\\
&\quad\begin{cases}
\begin{aligned}
C_{2+}\ \frac{2^{-\frac{1}{2}-\Delta_{2}-\Delta_{3}}\, \Gamma\!\left(1+l\right)}
{\Gamma\!\left(\frac{1}{4}(5+2l+2\Delta_{23})\right)\,
 \Gamma\!\left(\frac{1}{4}(3+2l-2\Delta_{23})\right)\,\epsilon}
\end{aligned}
& \text{if }l=l_2+l_3>0,
\\[1.2em]
\begin{aligned}
C_{1+}\ \frac{2^{-\frac{1}{2}-\Delta_{2}-\Delta_{3}}\, \Gamma\!\left(1-l\right)}
{\Gamma\!\left(\frac{1}{4}(5-2l+2\Delta_{23})\right)\,
 \Gamma\!\left(\frac{1}{4}(3-2l-2\Delta_{23})\right)\,\epsilon}
\end{aligned}
& \text{if }l=l_2+l_3<1,
\end{cases}
\end{split}
\end{equation}
\begin{equation}
\left<\psi_+\bar{\rho}^{(l_2)}_{++}\bar \phi^{(l_3)}\right>_2\sim -i\left<\psi_+\bar{\rho}^{(l_2)}_{++}\bar \phi^{(l_3)}\right>_1.
\end{equation}
In the monodromy case, where there can exist a pair of shadow-dual operators with transverse angular momentum $0<l<1$, we see that the singularity can be removed by choosing the ratio $C_{1+}/C_{2+}$ appropriately.
Similar ratios for  $C_{1-}/C_{2-}$ and $C_{3\pm}/C_{4\pm}$ can be deduced from the formulae in appendix \ref{sec:singularity}.
Comparing these ratios against the relations (\ref{C1to4pm}) and the shadow transform relation (\ref{cu23overcd23rel}), we find
the following ratios of  bulk-defect OPE coefficients:
\begin{equation}
\label{bovera}
    \frac{b^{(1-\nu)}_{\shortuparrow \pm}}{a^{(1-\nu)}_{\shortdownarrow \pm}} = \frac{\pi \kappa_{ud}}{\nu-1}= -     \frac{b^{(-\nu)}_{\shortdownarrow \pm}}{a^{(-\nu)}_{\shortuparrow \pm}}  \ ,
\end{equation}
where all dependence on  $\Delta_2$ and $\Delta_3$ has dropped out, as it should.
This relation (\ref{bovera}) suggests a more natural normalization for the operators appearing in the shadow transform.  Instead of using the canonically normalized $u^{(l)}$ and $d^{(l)}$ as we have done so far, we could use the normalization suggested by the defect OPE of $\psi$, namely the rescaled fields
$a^{(l)} u^{(l)}$ and $b^{(l)} d^{(l)}$.  Including these $a^{(l)}$ and $b^{(l)}$ coefficients in the definition of the shadow pair would correspondingly remove the coefficients from the relation (\ref{bovera}).  In this rescaling, it is convenient to distinguish between the normalization of the boundary data $u^{(l)}$ and $d^{(l)}$ and the normalization of the conformal kernel   $\langle u(x) \bar u(0) \rangle = \slashed{x} |x|^{-2 \Delta_{u}-1}$. (Equivalently, we can absorb any rescaling of $\langle u(x) \bar u(0) \rangle$ into a redefinition of $\kappa_{ud}$.)  
With the OPE normalized $u^{(l)}$ and $d^{(l)}$, 
eq.\ (\ref{bovera}) determines
\begin{equation}
\kappa_{ud} =  \pm \frac{\nu-1}{\pi} \ .
\end{equation}
The condition that the product (\ref{invertibilityfailure}) equals one then sets
\begin{equation}
\kappa_{du} = \pm\frac{\nu}{\pi} \ . 
\end{equation}
The $\pm$ phase can be absorbed with one last rescaling without affecting the two-point function normalizations, but we will stop here.

There is another way to remove the singularity at $\epsilon=0$:  we choose the $\Gamma$-functions in the denominators also to  be singular by setting their arguments to be zero or a negative integer.  In other words, one of the following holds for $n=1,2,3...$
\begin{equation} \label{l hypergeom conditions}
\begin{alignedat}{2}
     \frac{1}{4}(5+2l+2\Delta_{23})&=-n+1\ , \qquad 
     &\frac{1}{4}(3+2l-2\Delta_{23})&=-n+1  \ ,  \qquad  l>0 \ , \\
      \frac{1}{4}(5-2l+2\Delta_{23})&=-n+1\ , \qquad 
     &\frac{1}{4}(3-2l-2\Delta_{23})&=-n+1\ , \qquad  l<1 \ .
\end{alignedat}
\end{equation}
Replacing $l$ with the corresponding scaling dimension of the operator in the defect OPE of $\psi$,  $l=\Delta_{{u\shortdownarrow}\pm}-\frac{1}{2}$ or  $l=\frac{3}{2}-\Delta_{{d\shortuparrow}\pm}$,  these conditions become
\begin{equation}
\begin{alignedat}{2}
        \Delta_3&=-2+\Delta_{u \shortdownarrow +}+\Delta_2+2n\ , \qquad
        &\Delta_2&=-1+\Delta_{u \shortdownarrow +}+\Delta_3+2n \ , \\
        \Delta_3&=-1+\Delta_{d \shortuparrow +}+\Delta_2+2n\ , \qquad 
        &\Delta_2&=-2+\Delta_{d \shortuparrow +}+\Delta_3+2n\ .
\end{alignedat}
\end{equation}
The $n=1$ member of the first condition, $\Delta_3=\Delta_{u\shortdownarrow+}+\Delta_2$, is not captured by \eqref{l hypergeom conditions}. It is the special case $1+2\Delta_{23}+2l=0$, for which the general solution for $f_4$ \eqref{bulk-defect-defect f4} is invalid, but the solution \eqref{f4specialcase} is regular at $\zeta=1$. As the series expansion \eqref{hypergeom series} is derived from the general solution, it does not see this case.

\par Since we are working with a 2d defect, we can make use of the conformal weights $h'+h=\Delta$ and $h'-h=s$, where $s$ is the spin parallel to the defect. This allows us to make a comparison with \cite{Herzog_2022}, where it is proved that if operators of any spin have a generalized free field spectrum, then all the $n$-point functions are of generalized free fields as well. 
\begin{table}
\begin{center}
\renewcommand{\arraystretch}{1.8}
\begin{tabular}{|c|c|}
\hline
   Field & $s$ \\
    \hline
       $\bar{\rho}^{(l_2)}_{++}$  & $-\frac{1}{2}$\\
        $\bar \phi^{(l_3)}$& 0\\
        $u_{\shortdownarrow+}^{(l)}=\rho_{\shortdownarrow++}^{(l)}$&$\frac{1}{2}$\\
        $d_{\shortuparrow+}^{(l)}=\rho_{\shortuparrow+-}^{(l)}$&$-\frac{1}{2}$\\
        $u_{\shortdownarrow-}^{(l)}=\rho_{\shortdownarrow+-}^{(l)}$&$-\frac{1}{2}$\\
        $d_{\shortuparrow-}^{(l)}=\rho_{\shortuparrow++}^{(l)}$&$\frac{1}{2}$\\
     \hline
\end{tabular}
\caption{The parallel spins of some of the defect fields.
\label{table:spins}
}
\end{center}
\end{table}
Using Table \ref{table:spins}, the constraints can be written as
\begin{equation}
    \begin{alignedat}{2}
        h_3&=-1+h_{u \shortdownarrow +}+h_2+n\ , \qquad
        &h_2&=h_{u \shortdownarrow +}+h_3+n\\
        h_3&=-1+h_{d \shortuparrow +}+h_2+n\ , \qquad
        &h_2&=-1+h_{d \shortuparrow +}+h_3+n \ .\\
    \end{alignedat}
\end{equation}
We can make sense of these conditions in terms of a double twist construct.  
Consider $h_3 = -1 + h_{u \shortdownarrow +} + h_2 + n$.  
From the helicities of $\bar\phi$, $\bar \rho$ and $u_{\shortdownarrow+}$, we have
$h_3' = h_3$, $h_2 = h_2' +\frac{1}{2}$ and $h_{u \shortdownarrow+} = h_{u \shortdownarrow+}' - \frac{1}{2}$ and can infer
$h_3' = -1 + h_{u \shortdownarrow +}' + h_2' + n$.  
These conditions on $h_3$ and $h_3'$ are 
compatible with an operator of the schematic form
$\bar \phi^{(l_3)} = \rho_{+-}^{(l_2)} \, \partial_w^{n-1} \partial_{\bar w}^{n-1} \bar u_{\shortdownarrow-}^{(l)}$.  
A less trivial example is the relation $h_2  = h_{u \shortdownarrow +}+h_3+n$.  
Here, the helicities instead impose on the primed weights $h_2' =  h_{u \shortdownarrow +}'+h'_3+n -1$.
The corresponding double twist operator then needs to be
$\bar \rho_{++}^{(l_2)} = \phi^{(l_3)} \, \partial_w^{n} \partial_{\bar w}^{n-1} \bar u_{\shortdownarrow -}^{(l)}$, instead,
with a relative shift in the order of the $\partial_w$ and $\partial_{\bar w}$ derivatives necessary to preserve parallel spin. 

Related formulae for the remaining three examples 
\[
\left<\psi_-\bar{\rho}^{(l_2)}_{++}\bar \phi^{(l_3)}\right> \ , \; \; \; \left<\psi_+\bar{\rho}^{(l_2)}_{-+}\bar \phi^{(l_3)}\right> \ , \; \; \; \left<\psi_-\bar{\rho}^{(l_2)}_{-+}\bar \phi^{(l_3)}\right> 
\]
are in appendix \ref{sec:singularity}.
\par In the non-monodromy case, we can draw the same conclusions as in \cite{Herzog_2022}: given the double twist condition we have just derived,  any operators in the defect OPE of $\psi$ must have a generalized free field spectrum. As a consequence, all of their $n$-point functions are of generalized free fields as well, i.e.\ for any operators in the defect OPE of $\psi$, their correlation functions will follow from Wick's theorem. Note that this is only true if all $n$ operators are in the OPE. For example, the 3-point function $\left<u_{\shortdownarrow +}^{(l_1)}(x_1)\bar{d}_{\shortdownarrow +}^{(l_2)}(x_2)\, {:}\bar u_{\shortdownarrow -}^{(l_1)}{d}_{\shortdownarrow -}^{(l_2)}(x_3){:}\right>$ does not vanish, since the composite scalar operator ${:}\bar u_{\shortdownarrow -}^{(l_1)}{d}_{\shortdownarrow -}^{(l_2)}{:}$ is not in the OPE of $\psi$.

\section{Examples}
\label{sec:examples}
To produce examples of the correlation functions considered earlier in this work,
we couple bulk fermions $\psi$ to defect fermions $\chi$. 
(We assume $\chi$ is an eigenvector under the action of the two orthogonal gamma matrices $i\gamma_{n_1}\gamma_{n_2}$ with eigenvalue $\alpha = \pm 1$.)

While there is essentially just one example under discussion, we will approach it with increasing levels of sophistication and generality.
Consider the action
\begin{equation}
    \int d^dx \,\bar{\psi}(x) \slashed{\partial}\psi(x) \,+\,\int d^{d}x\, \delta^{(2)}(x_\perp) \left(\bar{\chi}(x_\parallel)\slashed{\partial}\chi(x_\parallel) \,+\,g\left(\bar{\psi}(x)\chi(x_\parallel)+\bar{\chi}(x_\parallel)\psi(x)\right)\right) \ .
\end{equation}
\begin{itemize}

\item
We begin with a perturbative analysis in $g$.  At leading order, the bulk-defect and bulk-defect-defect correlation functions are of precisely the form determined earlier from conformal symmetry and the Dirac shortening condition.

\item
We generalize the model by imposing monodromy-type boundary conditions on the bulk $\psi$ field, producing a family of these absorbing flows.
As the action is Gaussian, the model can be solved exactly, and we discuss the IR fixed point, where $\psi$ changes its boundary condition by absorbing the $\chi$ defect field into its defect operator expansion.

\item
We include a quartic coupling $\frac{1}{2}h( \bar \chi \chi)^2$, which can be solved in the absence of $g $ exactly
through bosonization. Including $g$, we find a range of allowed $h$ where $g$ is relevant
in the UV; we conjecture in this range that the absorbing flow persists.

\item
We introduce several flavours of defect fermions $\chi \to \chi_i$ with $i = 1, \ldots, N_f$.  This theory does not have an exact solution, but we discuss its RG flow in perturbation theory. At a perturbative fixed point where both $h$ and $g$ are nonzero, it provides
a family of three-point functions that do not satisfy the double-twist condition and
that are related by the shadow transform.

\end{itemize}

\subsection{Perturbation theory}

Using the Feynman rules in Euclidean space, we can find the bulk-defect correlation function to first order in $g$:
\begin{equation}
    \left<\psi(x)\bar{\chi}_\alpha(x'_\parallel)\right>= g \int d^p y_\parallel \, S_F^\psi(x-y_\parallel)\, S_F^\chi (y_\parallel-x'_\parallel)\,\mathbb{P}_\alpha
\end{equation}
where $S_F^\psi(w-z)$ is the free bulk fermion propagator and $S_F^\chi(w_\parallel-z_\parallel)$ is the free defect fermion propagator. We are considering a $p=d-2$ -dimensional defect. Substituting in explicit expressions for the propagators in terms of the momenta $k$ and $q_\parallel$, we find
\begin{equation}
\label{bulkdefectsetup}
    \begin{split}    \left<\psi(x)\bar{\chi}_\alpha(0)\right>
        &=g \int \frac{d^2 k_\perp}{(2\pi)^2} \int \frac{d^p q_\parallel}{(2\pi)^p}\,\frac{\slashed{k}_\perp+\slashed{q}_\parallel}{k_\perp^2+q_\parallel^2}\,\frac{\slashed{q}_\parallel}{q_\parallel^2} \,e^{ik_\perp x_\perp} \,e^{iq_\parallel x_\parallel}\,\mathbb{P}_\alpha \ .
    \end{split}
\end{equation}
This expression can be split into two momentum integrals, which can be performed separately:
\begin{equation} \label{example bulk-defect}
    \begin{split}
        \left<\psi(x)\bar{\chi}_\alpha(0)\right>
        &=g \int \frac{d^2 k_\perp}{(2\pi)^2} \int \frac{d^p q_\parallel}{(2\pi)^p}\,\frac{1}{k_\perp^2+q_\parallel^2} \,e^{ik_\perp x_\perp} \,e^{iq_\parallel x_\parallel}\,\mathbb{P}_\alpha\\
        &\quad-g \,\slashed{\partial}_{x_\perp}\slashed{\partial}_{x_\parallel}\int \frac{d^2 k_\perp}{(2\pi)^2} \int \frac{d^p q_\parallel}{(2\pi)^p}\,\frac{1}{(k_\perp^2+q_\parallel^2)q_\parallel^2} \,e^{ik_\perp x_\perp} \,e^{iq_\parallel x_\parallel}\,\mathbb{P}_\alpha.\\
    \end{split}
\end{equation}

The first of these terms 
gives the standard relation between the momentum space and real space Green's function for a massless scalar field:

\begin{equation}
\label{result1}
    g \int \frac{d^2 k_\perp}{(2\pi)^2} \int \frac{d^p q_\parallel}{(2\pi)^p}\,\frac{1}{k_\perp^2+q_\parallel^2} \,e^{ik_\perp x_\perp} \,e^{iq_\parallel x_\parallel}
    =\frac{g}{(d-2) {\rm Vol(}S^{d-1})}\frac{1}{|x|^{d-2}} \ ,
\end{equation}
where $d=p+2$. 
For the second term, the $k_\perp$ integral is equivalent to the massive scalar propagator with mass-squared $q_\parallel^2$ in 2d, and is known to be: 
\begin{equation}
    \int \frac{d^2 k_\perp}{(2\pi)^2}\frac{e^{ik_\perp x_\perp}}{k_\perp^2+q_\parallel^2} =\frac{1}{2\pi}K_0(|q_\parallel||x_\perp|) \ ,
\end{equation}
where $K_0$ is a Bessel function. The remaining integral can be expressed as a hypergeometric function 
\begin{equation}
\label{result2}
\begin{split}
     &\frac{1}{2\pi} \int \frac{d^p q_\parallel}{(2\pi)^p}\frac{e^{iq_\parallel x_\parallel}}{q_\parallel^2}K_0(|q_\parallel||x_\perp|)
=\frac{|x_\perp|^{2-p}}{2^4\pi^{\frac{p+2}{2}}} \Gamma\left(\frac{p}{2}-1\right)^2\,{}_2F_1^\text{reg}\left(\frac{p-2}{2},\frac{p-2}{2},\frac{p}{2};-\frac{|x_\parallel|^2}{|x_\perp|^2}\right).
     \end{split}
\end{equation}
Substituting results (\ref{result1}) and (\ref{result2}) into eq.\ \eqref{example bulk-defect} and applying the derivatives, the hypergeometric disappears leaving the simple expression
\begin{equation} \label{bulk-defect interacting theory}
     \left<\psi(x)\bar{\chi}_\alpha(0)\right>=\frac{g \Gamma(\frac{p}{2})}{4\pi^\frac{p+2}{2}}\left(\frac{1}{|x|^p}-\frac{\slashed{x}_\perp \slashed{x}_\parallel}{|x_\perp|^2|x|^p}\right)\,\mathbb{P}_\alpha=\frac{g}{p {\rm Vol(}S^{p+1})}\frac{\slashed{x} \slashed{x}_\perp}{|x_\perp|^2|x|^p}\,\mathbb{P}_\alpha \ .
\end{equation}
Using the free field values $\Delta_1=\frac{p+1}{2}$ and $\Delta_2=\frac{p-1}{2}$, 
this correlation function has the same form as the spin-down channel term in the bulk-defect correlation function (eq.\ \eqref{bulk-defect t family}) without the transverse orbital angular momentum.

Next we consider a bulk-defect-defect correlation function in perturbation theory.
We can make a defect scalar by contracting two defect fermions at the same point, and normal ordering them so that they do not couple with each other, preventing divergences. Therefore, the bulk fermion, defect fermion and defect scalar correlation function is
\begin{equation}
\left<\psi(x_1)\bar{\chi}_\alpha(x_{2})\bar \phi(x_{3})\right>= \left<\psi(x_1)\bar{\chi}_\alpha(x_{2})\, {:}\bar{\chi}_\alpha(x_3)\chi_\alpha(x_{3}){:}\right>.
\end{equation}
The $\phi$ operator is precisely of the double-twist form discussed earlier, so we expect the correlation function to be well-behaved when $x_1$ lies along the special geodesic locus of points connecting $x_2$ and $x_3$, without special tuning of the defect OPE coefficients.

Using the Feynman rules, the three-point function becomes to leading order
\begin{equation}
    \left<\psi(x_1)\bar{\chi}_\alpha(x_2)\,{:}\bar{\chi}_\alpha(x_{3})\chi_\alpha(x_{3}){:}\right>=g \int d^p y_\parallel \, S_F^\psi(x_1-y_\parallel)\,  S_F^\chi (y_\parallel-x_{3})\,\mathbb{P}_\alpha S_F^\chi (x_{32})\,\mathbb{P}_\alpha.
\end{equation}
The $y_\parallel$ integral is the same as in the bulk-defect case just examined, so we immediately obtain the result
\begin{equation} \label{example bulk-defect-defect}
\left<\psi(x_1)\bar{\chi}_\alpha(x_{2})\,{:}\bar{\chi}_\alpha(x_{3})\chi_\alpha(x_{3}){:}\right>=c \frac{\slashed{x}_{13} \slashed{z}}{|z|^2|x_{13}|^p}\frac{\slashed{x}_{32}}{|x_{32}|^p}\,\mathbb{P}_\alpha.
\end{equation}
In eq.\ \eqref{bulk-defect-defect} we had the structure 
\begin{equation}
f_4(\myupsilon)\frac{\slashed{x}_{13}\slashed{x}_{32}\slashed{z}}{|x_{12}|^{1+\Delta_1+\Delta_2-\Delta_3}|x_{13}|^{2+\Delta_1-\Delta_2+\Delta_3}|x_{23}|^{-\Delta_1+\Delta_2+\Delta_3}}\mathbb{P}_\alpha. 
\end{equation}
Setting 
\begin{equation}
    f_4(\myupsilon)=-c \,\myupsilon^{\frac{1}{2}(1+\Delta_1+\Delta_2-\Delta_3)}
\end{equation}
we obtain
\begin{equation} 
    -c\,\frac{\slashed{x}_{13}\slashed{x}_{32}\slashed{z}}{|z|^{1+\Delta_1+\Delta_2-\Delta_3}|x_{13}|^{1-2\Delta_2+2\Delta_3}|x_{23}|^{2\Delta_2+1}}\mathbb{P}_\alpha.
\end{equation}
If we set $\Delta_1=\frac{p+1}{2}$, $\Delta_2=\frac{p-1}{2}$ (the free field dimensions) and $\Delta_3=2\Delta_2=p-1$, we recover eq.\ \eqref{example bulk-defect-defect}.
Indeed, this solution is the special $n=0$ double twist example of (\ref{f4specialcase}).

\subsection{Monodromy defects and $b$-theorem}

We now narrow the scope of the previous example in one manner, but broaden it in another.  We set $p=2$, but allow the bulk fermion to pick up a phase upon circling the codimension-two defect:
\begin{equation}
\psi(\theta+2 \pi) = e^{-2\pi i \nu} \psi(\theta) \ , \; \; \; 0 \leq \nu < 1 \ .
\end{equation}
The field $\psi$ is still assumed to satisfy the Dirac equation.  Such a situation can be realized physically by giving the fermion a charge and threading the defect with magnetic flux, and is often called a monodromy defect \cite{Bianchi:2021snj}.  
Our previous example corresponds to $\nu=0$.
While in the usual case the free bulk fermion couples to a tower of defect operators of total helicity $|j| = \frac{1}{2}, \frac{3}{2}, \ldots$, the monodromy shifts the spin by $\nu$: 
\begin{equation}
j \in {\mathbb Z} + \frac{1}{2} - \nu \ .
\end{equation}
  The relation between the spin and defect dimension continues to be fixed by the Dirac shortening condition, $\Delta = 1 \pm  |j|$.
  The choice $1 - |j|$ lies below the unitarity bound for all $j$ except the bottom rung of the ladder $j = \frac{1}{2} - \nu$.    
  In the non-monodromy case, we discarded the bottom rung based on a decoupling argument.  For $\nu>0$, however, this choice is allowed.
  Physically, the spectrum of the defect OPE of the bulk fermion is not completely fixed.  The lowest lying elements can include both  $\Delta = \frac{3}{2}-\nu$ and $\Delta = \frac{1}{2} +\nu$ operators, with a relative admixture.\footnote{%
  In  ref.\  \cite{Bianchi:2021snj}, this admixture was characterized by the parameter $\xi$ where $\xi = 0$ is $\Delta = \frac{3}{2}-\nu$ only and $\xi = 1$ is $\Delta = \frac{1}{2}+\nu$ only.}
    Spectral flow exchanges the role of these two operators once $\nu > \frac{1}{2}$, and so 
  in what follows, we restrict to the range $0 \leq \nu \leq \frac{1}{2}$.\footnote{%
  An interesting variant of this theory would be an axial monodromy defect, where the shift in helicity depends on whether the initial helicity is positive or negative.  With an axial monodromy defect, both the $j = \frac{1}{2} - \nu$ and $j = -\frac{1}{2} + \nu$  operators can have a $\Delta = 1- |j|$ above the unitarity bound.
  }
  
As the defect value $\psi(0)$ in the previous example is precisely the $\Delta = 3/2$ dimension operator in the defect OPE of the bulk $\psi$, we can think of the $\bar \psi \chi$ coupling purely in 2d terms, as a bilinear between this $\Delta =3/2$ dimension operator and the introduced defect operator $\chi$.  
In the expanded monodromy setting, let us assume the $\Delta = \frac{1}{2} +\nu$ operator is absent in the defect OPE of $\psi$.  We replace $\psi(0)$ in the previous example with the dimension $\Delta = \frac{3}{2} -\nu$ operator, $u$, at the bottom of the defect OPE tower of $\psi$:
\begin{equation}
    g \int d^2 x \left( \bar{u} \chi + \bar \chi u \right) \ .
\end{equation}
In contrast to the previous example where $g$ was marginal, this coupling is relevant when $\nu > 0$.  
Note that the equation of motion for $\chi$ anticipates the multiplet recombination that happens in the IR where 
$u$ becomes a descendant of $\chi$: $\slashed{\partial} \chi = - g u$.  

Without the coupling, by conformal symmetry the Fourier transform of the two-point function of $u$ should have the form
\begin{equation}
G_{u}(q) = -ic \frac{ \slashed{q}}{|q|^{2\nu}} \ , 
\end{equation}
where $c$ is a real number and the exponent $2\nu$ is fixed by scaling symmetry.\footnote{%
Recall if in real space, the two-point function is
$C\slashed{x}/|x|^{4-2 \nu}$, then $c =   2^{2\nu-1}\pi C \Gamma(\nu) / \Gamma(2-\nu)$.
}
The overall two-point function for $\chi$ can then be constructed by summing a geometric series:
\begin{eqnarray}
\label{coupledchichi}
G_{\chi}(q) = -i\frac{ \slashed{q}}{q^2(1 + c g^2 |q|^{-2 \nu})} \ .
\end{eqnarray}
This two-point function demonstrates the RG flow between a free fermion $\chi$ of dimension 
$\frac{1}{2}$ in the UV, and a generalized free fermion of dimension $\frac{1}{2} + \nu$ in the IR.
As this IR defect fermion is one of the allowed boundary conditions for the bulk fermion, the RG flow takes on a physical significance -- the defect fermion $\chi$ has been absorbed by the bulk $\psi$ and changed the boundary condition from the $\frac{3}{2} - \nu$ operator to the shadow $\frac{1}{2} + \nu$.  The old 
$\frac{3}{2} - \nu$ operator gets an anomalous dimension $2 \nu$ and becomes the descendant of the captured and dimensionally shifted $\chi$, which, to align with the previous notation, we could rename $d$. 
Indeed, we can repeat the same calculation for $u$, recovering the resummed
\begin{equation}
    G_{u}(q) = - i c\frac{\slashed{q}}{|q|^{2\nu}(1 + c g^2 |q|^{-2\nu})} \ .
\end{equation}
Expanding this out at small momentum, we find
\begin{equation}
G_{u}(q) = \frac{-i \slashed{q}}{g^2} + \frac{i \slashed{q}}{c g^4 |q|^{-2 \nu}} + \ldots 
\end{equation}
The first term is an analytic contact term, and the second is the descendant, with scaling dimension $\frac{3}{2} + \nu$.  

The limit $\nu \to 0$ is the non-monodromy model that we started with.  The denominators of $G_{u}$ and $G_{\chi}$ expand out as
\begin{equation}
    1 + c g^2 |q|^{-2 \nu} = 1+ \frac{C\pi g^2}{2 \nu} -C\pi g^2 \log |q| + \ldots \ . 
\end{equation}
The $1/\nu$ divergence and finite local pieces can be absorbed into a defect counterterm, leaving a logarithmic dependence on momentum.
The remaining logarithmic dependence in a fermionic Green's function has been associated with a marginal Fermi liquid 
\cite{PhysRevLett.63.1996}, originally introduced to describe aspects of the normal state of the cuprate superconductors.

The mechanism here is closely related to that of the semi-holographic Fermi liquid of ref.~\cite{Faulkner:2010tq}. In that construction, a dynamical fermion near a Fermi surface is coupled to a strongly interacting fermionic operator, whose low-energy Green's function is determined holographically by an $AdS_2\times{\mathbb R}^2$ region. The spatial momentum is retained as a label of the $AdS_2$ operator, but does not scale in the infrared. Hence,  its critical Green's function behaves as
$\mathcal G(\omega)\sim\omega^{2\Delta-1}$, becoming $\omega\log\omega$ at $\Delta=1$.

In our case, the Weyl transformation instead produces $AdS_3\times S^1$, and the corresponding critical Green's function depends on both frequency and momentum through the relativistic combination $k^2-(\omega+i0)^2$. At zero spatial momentum, however, the resulting fermion self-energy has precisely the same characteristic frequency dependence:
\begin{equation}
\Sigma_\chi(\omega)\sim
\begin{cases}
\omega\log\omega, & \nu=0 ,\\
\omega^{1-2\nu}, & 0<\nu<\frac12 .
\end{cases}
\end{equation}
The first line is the marginal-Fermi-liquid scaling, while the second is the corresponding power-law non-Fermi-liquid behaviour, for which the self-energy is sublinear in frequency. In the notation of ref.~\cite{Faulkner:2010tq}, the correspondence is
\begin{equation}
\Delta=1-\nu,
\qquad
\Delta-\frac12
=\frac12-\nu=|j| .
\end{equation}

Upon continuation to Lorentzian signature, the logarithm also generates an imaginary part. At $k=0$, and up to normalization and self-energy sign conventions,
\begin{equation}
\operatorname{Re}\Sigma_\chi(\omega)
\sim \omega\log|\omega|,
\qquad
\operatorname{Im}\Sigma_\chi(\omega)
\sim-|\omega| ,
\end{equation}
again reproducing the characteristic zero-temperature marginal-Fermi-liquid self-energy. In the present case, this behaviour arises in an exactly tractable Gaussian bulk-defect model, rather than from a strongly interacting holographic sector.

We leave a full calculation of the bulk-defect correlation function $\langle \psi(x) \bar\chi(0) \rangle$ as an exercise, but the preceding arguments make clear what should happen. 
The form of the bulk-defect correlator is fixed by conformal symmetry at the endpoints of the RG flow.  In the UV, $\chi$ should behave as a free field of dimension $1/2$, and in the IR as a field of dimension $1/2 + \nu$.

A final short check of this interpretation is a verification of the $b$-theorem \cite{Jensen:2015swa}.
For conformal field theories with surface defects, the trace of the stress tensor in a curved background has a distributional contribution which is proportional to the Ricci curvature of the surface defect:
\begin{equation}
T^\mu_\mu = -\frac{b}{24 \pi} \, R \delta^{(2)} (x_\perp)  + \ldots \ .
\end{equation}
The coefficient of this term is sometimes called $b$, and it must decrease under RG flow.  
From \cite{Bianchi:2021snj}, the value of $b$ for these monodromy defects is 
\begin{equation}
b_\Delta = \begin{cases}
    -(\nu-1)^2(\nu^2-2\nu-1) \ , \; \; \; & \Delta = \frac{1}{2} + \nu  \\
    -\nu^2 (\nu^2-2) \ , \; \; \; & \Delta = \frac{3}{2}-\nu 
\end{cases}\ .
\end{equation}
The special case $\Delta = \frac{1}{2}$ corresponds to a free 2d fermion, $b_\chi = b_{\frac{1}{2}}=1$.
In the UV, we have the $\Delta = \frac{3}{2}-\nu$ monodromy theory and a free 2d fermion.  In the IR, the claim is we have the $\Delta = \frac{1}{2}+\nu$ theory.  Indeed,
\begin{equation}
    \Delta b =b_{\chi} + b_{\frac{3}{2}-\nu} - b_{\frac{1}{2}+\nu} = 2 \nu^2 (3-2\nu)  > 0
\end{equation}
in the physically interesting range $0 <\nu<\frac{1}{2}$.  Once the RG flow takes us to the $\Delta = \frac{1}{2} + \nu$ theory, there is still a relevant deformation, which corresponds to the mass-type operator $\bar \chi_{\rm IR} \chi_{\rm IR}$ with dimension $1 + 2\nu$, provided $\nu<\frac{1}{2}$.  
Deforming by this operator is expected to return us to the $\Delta = \frac{3}{2} - \nu$ theory.  Indeed, the flow is consistent with the $b$-theorem:
\begin{equation}
\Delta b' = b_{\frac{1}{2}+\nu} - b_{\frac{3}{2}-\nu} =
(1-2\nu) (1+2\nu-2\nu^2) > 0
\end{equation}
provided $\nu < 1/2$, as was observed originally in \cite{Bianchi:2021snj}. 
Note the total effect of the two RG flows $\Delta b + \Delta b' = b_\chi$ is to eliminate the free defect fermion $\chi$ that was introduced in the UV through the bilinear coupling.

\subsection{Quartic coupling}

The monodromy model with the bilinear coupling $g( \bar \chi u + \bar u \chi)$ and a quartic coupling $\frac{h}{2} (\bar \chi \chi)^2$ can be addressed using bosonization. Two-dimensional Fierz identities relate the quartic term to a current bilinear $(\bar \chi \chi)^2 = - \frac{1}{2} (\bar \chi \gamma_a \chi)^2$.
Bosonization relates the fermionic current $\bar \chi \gamma_a \chi$ to the bosonic momentum current $\frac{1}{\sqrt{\pi} }\epsilon^{ab}\partial_b \Phi$, and thus this quartic coupling has the net effect of renormalizing the kinetic term for $\Phi$, 
$\frac{1}{2}(\partial \Phi)^2 \to 
\frac{1}{2}(1+ \frac{ h}{2\pi}) (\partial \Phi)^2$. 

The renormalization of the scalar kinetic term produces a shift in the scaling dimension of $\bar \chi \chi$ \cite{Coleman:1974bu},
\begin{equation}
    \Delta_{\bar \chi \chi} = \left(1 + \frac{h}{2 \pi} \right)^{-1}
    \ .
\end{equation}
    The fermion itself is represented by the Mandelstam soliton operator \cite{Mandelstam:1975hb}, whose scaling dimension is
\begin{equation}
    \Delta_\chi = 
    \frac{1}{4} 
    \left( 1 + \frac{h}{2 \pi} + \frac{1}{1 + \frac{h}{2 \pi}} \right) \ .
\end{equation} 
Note that  $\Delta_\chi \geq \frac{1}{2}$.  Thus, the bilinear $g \bar \chi u$ that we added before will be relevant now only if $\Delta_\chi + \frac{3}{2} - \nu < 2$, or equivalently 
\begin{equation}
    4 \pi (\nu -\sqrt{\nu^2+\nu}) < h <  4 \pi (\nu + \sqrt{\nu^2+\nu})  \ .
\end{equation}
For small $\nu$, this means $|h| 
\lesssim 4 \pi \sqrt{\nu}$.  

It is tempting to conjecture that in this range for $h$,  at the endpoint of the RG flow, the field $\chi$ is absorbed and the bulk fermion has the same $\frac{1}{2}+\nu$ boundary condition as before.  It would be interesting to see if the model can be solved exactly for arbitrary couplings $g$ and $h$.

\subsection{Introducing flavour and  beta functions}
We next introduce a flavour index to both the bulk and defect fermions, $\psi^i$ and $\chi^i$, where $i=1,...,N_f$. The new action is 
\begin{equation}
    \begin{split}
        S&= \int d^4x \,\bar{\psi}^i(x) \slashed{\partial}\psi^i(x) +\delta^{(2)}(x_\perp) \Big(\bar{\chi}^i(x_\parallel)\slashed{\partial}\chi^i(x_\parallel) \\
     &\qquad\qquad+\,g\left(\bar{\psi}^i(x)\chi^i(x_\parallel)+\bar{\chi}^i(x_\parallel)\psi^i(x)\right)+\frac{1}{2}h\left(\bar{\chi}^i(x_\parallel)\chi^i(x_\parallel)\right)^2\Big)\\
    \end{split}
\end{equation}
where the sum is performed over $i$. We want to calculate the beta functions for the couplings $g$ and $h$. To do this, we use dimensional regularisation in both the defect dimension and codimension: $d_\parallel=2+\epsilon_\parallel$ and $d_\perp=2+\epsilon_\perp$. The couplings then become $g=\mu^{-\frac{\epsilon_\perp}{2}}\tilde{g}$ and $h=\mu^{-\epsilon_\parallel}\tilde{h}$, where we introduce the dimensionless couplings $\tilde{g}$ and $\tilde{h}$ and the energy scale $\mu$. Starting with the bare action (with bare quantities denoted by subscript 0) and introducing the counter-term action, we have  
\begin{equation}
\begin{split}
     S&= \int d^dx \,\bar{\psi}^i_0(x) \slashed{\partial}\psi^i_0(x) +\delta^{(d_\perp)}(x_\perp) \Big(\bar{\chi}^i_{0}(x_\parallel)\slashed{\partial}\chi^i_{0}(x_\parallel) \\
     &\qquad\qquad+\,\tilde{g}_0\left(\bar{\psi}^i_0(x)\chi^i_{0}(x_\parallel)+\bar{\chi}^i_{0}(x_\parallel)\psi^i_0(x)\right)+\frac{1}{2}\tilde{h}_0\left(\bar{\chi}^i_{0}(x_\parallel)\chi^i_{0}(x_\parallel)\right)^2\Big)\\
     &=\int d^dx \,\bar{\psi}^i(x) \slashed{\partial}\psi^i(x) +\delta^{(d_\perp)}(x_\perp) \Big(\bar{\chi}^i(x_\parallel)\slashed{\partial}\chi^i(x_\parallel) \\
     &\qquad\qquad +\,\mu^{-\frac{\epsilon_\perp}{2}}\tilde{g}\left(\bar{\psi}^i(x)\chi^i(x_\parallel)+\bar{\chi}^i(x_\parallel)\psi^i(x)\right)+\frac{1}{2}\mu^{-\epsilon_\parallel}\tilde{h}\left(\bar{\chi}^i(x_\parallel)\chi^i(x_\parallel)\right)^2\Big)\\
     &\qquad+\,\delta_{Z_\psi}\bar{\psi}^i(x) \slashed{\partial}\psi^i(x) +\delta^{(d_\perp)}(x_\perp) \Big(\delta_{Z_\chi}\bar{\chi}^i(x_\parallel)\slashed{\partial}\chi^i(x_\parallel)\\
     &\qquad\qquad+\,\mu^{-\frac{\epsilon_\perp}{2}}\delta_{\tilde{g}}\left(\bar{\psi}^i(x)\chi^i(x_\parallel)+\bar{\chi}^i(x_\parallel)\psi^i(x)\right)+\frac{1}{2}\mu^{-\epsilon_\parallel}\delta_{\tilde{h}}\left(\bar{\chi}^i(x_\parallel)\chi^i(x_\parallel)\right)^2\Big)\\
\end{split}
\end{equation}
where we have defined 
\begin{equation} \label{renormalization}
    \begin{split}
        &\psi^i_0=Z_\psi^{\frac{1}{2}}\psi^i\ , \; \; \;
        \chi^i_{0}=Z_\chi^{\frac{1}{2}}\chi^i\\
        &\delta_{Z_\psi}=Z_\psi-1\ , \; \; \;
        \delta_{Z_\chi}=Z_\chi-1\\
        &\mu^{-\frac{\epsilon_\perp}{2}}\delta_{\tilde{g}}=\tilde{g}_0Z_\psi^\frac{1}{2}Z_\chi^\frac{1}{2}-\mu^{-\frac{\epsilon_\perp}{2}}\tilde{g}\\
        &\mu^{-\epsilon_\parallel}\delta_{\tilde{h}}=\tilde{h}_0Z_\chi^2-\mu^{-\epsilon_\parallel}\tilde{h}.\\
    \end{split}
\end{equation}
\begin{figure} [H]
    \centering
    \includegraphics[width=50mm,scale=1]{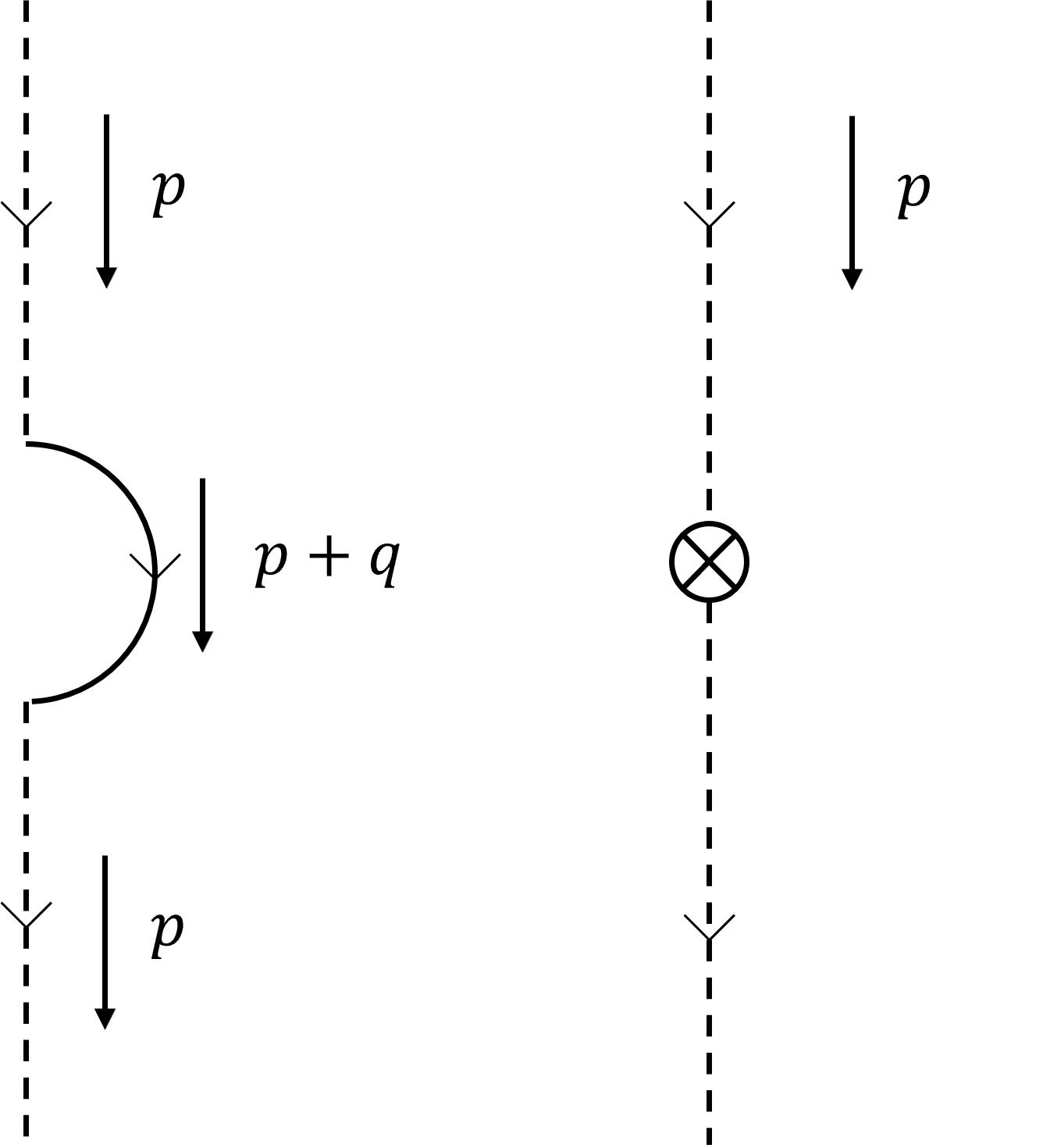}
    \caption{Left: the second order in $\tilde{g}$ divergent contribution to the defect fermion propagator. The dashed lines represent the defect fermions and the solid lines the bulk fermions. $p$ is momentum parallel to the defect and $q$ normal. Right: the counter-term diagram to cancel the divergence.}
    \label{wavefunction renormalization}
\end{figure}
We first calculate $\delta_{Z_\chi}$ using the Feynman diagrams in Fig.\ \ref{wavefunction renormalization}. (The loop correction with the $\tilde{h}$ vertex vanishes.) Using the Feynman rules in Euclidean spacetime, the first diagram is evaluated to give (after being amputated)
\begin{equation}
\begin{split}
  \Pi =   (\mu^{-\frac{\epsilon_\perp}{2}}\tilde{g})^2
  \int\frac{d^{d_\perp}q}{(2\pi)^{d_\perp}} \frac{(\slashed{q}+\slashed{p})}{q^2+p^2}
 = \tilde{g}^2\slashed{p}\,
 \frac{1}{4\pi}\left(\frac{p^2}{4\pi\mu^2}\right)^\frac{\epsilon_\perp}{2}\Gamma\left(-\frac{\epsilon_\perp}{2}\right).
\end{split}
\end{equation}
We can expand the gamma function and $\left(\frac{p^2}{4\pi\mu^2}\right)^\frac{\epsilon_\perp}{2}$ (since it is dimensionless) in terms of $\epsilon_\perp$:
\begin{equation}
\begin{split}
 \Pi   = -\tilde{g}^2\slashed{p}\,
 \frac{1}{4\pi}\left(\frac{2}{\epsilon_\perp}+\gamma+\ln\left(\frac{p^2}{4\pi\mu^2}\right)\right)+\mathcal{O}(\epsilon_\perp).
\end{split}
\end{equation}
Applying the Feynman rules to the second diagram in Fig.\ \ref{wavefunction renormalization} gives $\slashed{p}
\delta_{Z_\chi}$, and so in the MS scheme, we find
\begin{equation}
    \delta_{Z_\chi}=\frac{\tilde{g}^2}{2\pi\epsilon_\perp}.
\end{equation}
There is no vertex renormalization of $\tilde{g}$ and so $\delta_{\tilde{g}}=0$. Furthermore, there is no divergent correction to the bulk fermion propagator ($\delta_{Z_\psi}=0$). Using the relation in eq.\ \eqref{renormalization} between the renormalized $\tilde g$ and unrenormalized $\tilde g_0$ couplings,
and expanding perturbatively in $\tilde g$, we deduce 
\begin{equation}
\begin{split}
      &\mu^{-\frac{\epsilon_\perp}{2}}\tilde{g}\left(1-\frac{\tilde{g}^2}{4\pi\epsilon_\perp} + \ldots \right)=\tilde{g}_0.\\
\end{split}
\end{equation}
To find the beta function for $\tilde{g}$, defined as $\beta(\tilde{g})=\frac{d\tilde{g}}{d\ln{\mu}}$, we differentiate this expression by $\ln{\mu}$, and expand perturbatively in $\tilde g$:
\begin{equation}
    \begin{split}
&\beta(\tilde{g})=\frac{\epsilon_\perp\tilde{g}}{2}+\frac{\tilde{g}^3}{4\pi} + \ldots .
    \end{split}
\end{equation}
To find the beta function for $\tilde{h}$, we must consider the one-loop correction to the defect fermion four-point function. 
We find 
\begin{equation}
    \delta_{\tilde{h}}=\frac{\tilde{h}^2(N_f-1)}{\pi\epsilon_\parallel}.
\end{equation}
Going back to eq.\ \eqref{renormalization} and proceeding in the same manner as before, we obtain
\begin{equation}
    \begin{split}
   & \mu^{-\epsilon_\parallel} \tilde h \left(1 + \frac{\tilde h(N_f-1)}{\pi \epsilon_\parallel} - \frac{\tilde g^2 }{\pi \epsilon_\perp} + \ldots \right) = \tilde h_0  \ .
    \end{split}
\end{equation}
Differentiating by $\ln{\mu}$, using the expression for $\beta(\tilde g)$ obtained above and expanding perturbatively in the couplings 
gives
\begin{equation}
\begin{split}    &\beta({\tilde{h}})=\epsilon_\parallel\tilde{h}-\frac{\tilde{h}^2(N_f-1)}{\pi}+\frac{\tilde{h}\tilde{g}^2}{\pi} + \ldots .\\
\end{split}
\end{equation}

\subsubsection*{Fixed point analysis}
By setting the beta functions for $\tilde{g}$ and $\tilde{h}$ to zero, we find that there are four fixed points
\begin{equation}
    (\tilde{g}^*,\tilde{h}^*)=(0,0),\left(0,\frac{\pi \epsilon_\parallel}{N_f-1}\right),(\sqrt{-2\pi\epsilon_\perp},0),\left(\sqrt{-2\pi\epsilon_\perp},\frac{\pi (\epsilon_\parallel-2\epsilon_\perp)}{N_f-1}\right),
\end{equation}
where we take $\epsilon_\perp<0$ and $\epsilon_\parallel>0$.
The stability of these fixed points can be found by studying the eigenvalues of the matrix 
\begin{equation}
    \left. \begin{pmatrix}
        \partial_{\tilde{{g}}}\beta(\tilde{g})&\partial_{\tilde{{h}}}\beta(\tilde{g})\\
        \partial_{\tilde{{g}}}\beta(\tilde{h})&\partial_{\tilde{{h}}}\beta(\tilde{h})\\
    \end{pmatrix}
    \right|_{(\tilde{g}^*,\tilde{h}^*)}\quad.\end{equation}
If both eigenvalues are negative, we have a UV fixed point, if both are positive, an IR fixed point, and if the signs are different, an unstable saddle point.
\renewcommand{\arraystretch}{1.5}
\begin{center}
    \begin{tabular}{|c|c|c|}
    \hline
    Fixed point $(\tilde{g}^*,\tilde{h}^*)$ & Eigenvalues&Nature of fixed point \\
    \hline
       (0,0)& $\epsilon_\parallel$,\,$\frac{\epsilon_\perp}{2}$ &Saddle point \\     $\left(0,\frac{\pi \epsilon_\parallel}{N_f-1}\right)$&-$\epsilon_\parallel$,\,$\frac{\epsilon_\perp}{2}$&UV fixed point \\ 
     $(\sqrt{-2\pi\epsilon_\perp},0)$&$\epsilon_\parallel-2\epsilon_\perp$,\,$-\epsilon_\perp$&IR fixed point\\
     $\left(\sqrt{-2\pi\epsilon_\perp},\frac{\pi (\epsilon_\parallel-2\epsilon_\perp)}{N_f-1}\right)$&$-\epsilon_\parallel+2\epsilon_\perp$,\,$-\epsilon_\perp$& Saddle point\\
     \hline
    \end{tabular}
\end{center}

The stability eigenvalues can also be interpreted as the deviations from marginality, $\Delta_{\mathcal O}-d_\parallel$, of the corresponding eigen-deformations.  
For example, at the Gaussian fixed point, $(\bar \chi \chi)^2$ has scaling dimension $2 + 2 \epsilon_\parallel$, which is $\epsilon_\parallel$ away from the marginal value $2 + \epsilon_\parallel$. Similarly $\bar \psi \chi$ has scaling dimension $\frac{\epsilon_\perp}{2}$ away from  $2 + \epsilon_\parallel$. 
At the IR fixed point there is a simple consistency check on the eigenvalue $-\epsilon_\perp$, which corresponds to varying $\tilde g$. 
At $\tilde g_* \neq 0$, the $\chi$ equation of motion identifies the perturbing operator $(\bar \psi \chi + \bar \chi \psi)$ with the kinetic operator
$\bar \chi \slashed{\partial} \chi$, up to a total derivative.  From the wave-function renormalization we find
\begin{equation}
    \gamma_\chi = \frac{1}{2} \frac{\partial \log Z_\chi}{\partial \log \mu} = \frac{\tilde g^2}{4 \pi}  \  \; \; \; \mbox{where}
    \; \; \; \left. \frac{\tilde g^2}{4 \pi}  \right|_{\tilde g_*^2 = -2 \pi \epsilon_\perp} = - \frac{\epsilon_\perp}{2} \ .
\end{equation}
Thus
\[
\Delta_\chi=\frac{d_\parallel-1}{2}-\frac{\epsilon_\perp}{2},
\]
so that the kinetic operator has scaling dimension
\[
\Delta_{\bar\chi\slashed{\partial}\chi}
=2\Delta_\chi+1
=d_\parallel-\epsilon_\perp.
\]
Its RG eigenvalue is therefore
\[
\Delta_{\bar\chi\slashed{\partial}\chi}-d_\parallel=-\epsilon_\perp,
\]
in agreement with the corresponding eigenvalue of the stability matrix. 
At leading order in perturbation theory, there is
 little difference whether we introduce $\epsilon_\perp$ or the monodromy parameter $\nu$.  
 With the identification $\nu = -\frac{\epsilon_\perp}{2}$, 
 the two lead to identical shifts in anomalous dimension and a similar RG flow between the Gaussian fixed point and the IR fixed point.

Our interest in what follows is in the saddle-point where both $\tilde h$ and $\tilde g$ are nonzero.  Interestingly, this fixed point persists even setting $\epsilon_\parallel = 0$.  
\subsection*{Candidate three-point function}

At the mixed fixed point $g^* \neq 0$ and $h^* \neq 0$ (henceforth dropping the tildes), we can consider the three-point functions
$\langle \psi(x_1) O_{2n}(x_2) O_{2n+1}(x_3) \rangle$ and $\langle \psi(x_1) O_{2n+1}(x_2) O_{2n+2}(x_3) \rangle$, where
\begin{equation}
O_{2n} = (\bar \chi \chi)^n \ , \; \; \;
O_{2n+1} = (\bar \chi \chi)^n \bar \chi \ . 
\end{equation}
The simplest nontrivial example would be $\langle \psi(x_1) \bar \chi(x_2) {:} \bar \chi \chi (x_3) {:}  \rangle$.  
Note the case $n=1$ is exceptional, because $O_3$ is a descendant by the equation of motion.  
From \cite{Ghosh:2015opa}, we know the anomalous dimensions
at leading order in $h$ at the standard $g=0$ Gross-Neveu fixed point.   These anomalous dimensions are related to the anomalous dimensions at the mixed fixed point at this leading order, since at $h=0$, the theory with $g \neq 0$ is Gaussian and produces no extra anomalous dimensions of normal ordered composites.  Corrections to $O(h)$ vertex anomalous dimensions involving $g$ begin at $O(hg^2)$.  In particular, the differences
\begin{equation}
\gamma_{2n}^{(h)} \equiv \Delta_{2n} - 2n \Delta_\chi  \; \; \;
\mbox{and} \; \; \;
\gamma_{2n+1}^{(h)} \equiv \Delta_{2n+1} - (2n+1) \Delta_\chi \ ,
\end{equation}
are the same at leading order in $h$ at both fixed points where\footnote{%
Note the $N_f$ of ref.\ \cite{Ghosh:2015opa} is twice ours.
}
\begin{eqnarray}
\gamma_{2n}^{(h)}  &=& -\frac{n(2N_f-n)}{2\pi} h \ , \\
\gamma_{2n+1}^{(h)} &=& -\frac{n(2N_f-n-1)}{2\pi} h \ .
\end{eqnarray}
(Note that the result for $\gamma_4^{(h)}$ is consistent with the eigenvalues of the 
beta function matrix we found above.)
We find then that
\begin{equation}
    \Delta_{2n+1} - \Delta_{2n} - \Delta_\chi =
    \gamma_{2n+1}^{(h)} - \gamma_{2n}^{(h)} =
    \frac{n h}{2\pi} +O(h^2,hg^2)\neq 0 \ 
\end{equation}
(or $h(1+n-2N_f)/(2 \pi)$ for the example involving the $O_{2n+1}$  and $O_{2n+2}$ operators),
which violates the double twist condition at leading order in $h$.
 Thus, this three-point function should be regular in the interior through the mechanism discussed in section \ref{sec:bulkdefectdefect}, with contributions from both members of the shadow pair, rather than one of the double-twist forms obtained when only a single defect-OPE channel is allowed. 
 At the mixed fixed point, $\chi$ has the dimension of the $d$ member of the shadow pair, while the equation $G_u^{-1}u=g\chi$ fixes the three-point coefficient involving $u$ in terms of that involving $\chi$, where $u$ is the bulk field $\psi$ restricted to the defect.  
The shadow relation fixes the relative coefficient of the two solutions precisely so that the unphysical singularity at $\myupsilon=1$ cancels.
Focusing on the first family $(2n, 2n+1)$ and setting $\epsilon_\parallel=0$, 
the amount by which the double twist condition is violated,
\begin{equation}
\frac{nh}{2 \pi} = -\frac{n\epsilon_\perp}{N_f-1} = \frac{2n\nu}{N_f-1} \ ,
\end{equation}
relies on having a small parameter (either $\epsilon_\perp$ or $\nu$) that makes the bilinear $\bar \psi \chi$ slightly relevant.  The double-twist condition returns in the limit $\nu \to 0$ or $\epsilon_\perp \to 0$. 

We have, then, an explicit interacting example in which a bulk-defect-defect three-point function evades the generalized-free double-twist condition, through the coexistence of a shadow pair in the defect OPE of the bulk fermion.

\section{Discussion}
\label{sec:discussion}

The general theme of this work is the strong constraints the Dirac equation places on correlation functions in defect conformal field theory involving a free massless spinor in the bulk.  
The Dirac shortening condition restricts the spectrum of operators in the defect OPE of the bulk spinor, and imposes a double-twist condition on bulk-defect-defect three-point functions, whenever only one member of the shadow pair is allowed by the OPE.
While we demonstrated these facts in the 2d/4d case, we expect them to hold more generally.

As mentioned in the introduction, the theory of differential equations and their boundary conditions provides an elegant way to conceptualize these constraints on three (and higher) -point functions. 
We start by assuming $\langle \psi(x_1) {\mathcal O}_2(x_2) {\mathcal O}_3(x_3) \rangle$ satisfies the Dirac equation when acting on the bulk spinor $\psi(x_1)$.  We further assume the spinor operator ${\mathcal W}$ is present in the OPE of $\psi$; this assumption allows us to impose
boundary data on $\langle \psi(x_1) {\mathcal O}_2(x_2) {\mathcal O}_3(x_3) \rangle$, namely the value of
$\langle {\mathcal W}(x) {\mathcal O}_2(x_2) {\mathcal O}_3(x_3) \rangle$.
Further, we
impose a regularity condition in the interior on $\langle \psi(x_1) {\mathcal O}_2(x_2) {\mathcal O}_3(x_3) \rangle$: the absence of a singularity on the locus of points where the cross-ratio $\myupsilon=1$.  Then the theory of differential equations gives us a unique solution which should let us read off the conjugate boundary condition, i.e.\ the correlation function involving the shadow dual operator
$\langle {\mathcal S}(x) {\mathcal O}_2(x_2) {\mathcal O}_3(x_3) \rangle$. 
If the boundary condition is that ${\mathcal S}$ is absent, then the shadow transform will, in general, imply that $\langle {\mathcal W}(x) {\mathcal O}_2(x_2) {\mathcal O}_3(x_3) \rangle$ must vanish also.  The only loophole is the double twist condition, where the shadow transform itself annihilates its three-point function argument.  

 In the case of the monodromy defects, the only ${\mathcal S}$ that were allowed by unitarity sat at the bottom of the tower.
 Then ${\mathcal S}$ and ${\mathcal W}$ could be identified with the  $d^{(l)}$
 and $u^{(l)}$ operators respectively, with $l = -\nu$ and $l = 1-\nu$.  In this case, the shadow transform provides a relation between the defect three-point functions $\langle d^{(l)}(x) {\mathcal O}_2(x_2) {\mathcal O}_3(x_3) \rangle$ and $\langle u^{(l)}(x) {\mathcal O}_2(x_2) {\mathcal O}_3(x_3) \rangle$.  Indeed, it will provide a relation between all higher point functions involving $d^{(l)}$ and $u^{(l)}$ as well, something which could be investigated in future work.  The applications to four-point functions and crossing symmetry look particularly intriguing, as was also emphasized already in the scalar case by ref.\ \cite{Behan:2020nsf}.  
 
 In addition to these formal results, we explored several examples which exemplify these features.  We solved the  monodromy case, with a free fermion $\chi$ living on the defect, along with a bilinear coupling $g (\bar \psi \chi + \bar \chi \psi)$.  In particular, we showed that the model flows to an IR fixed point where $\chi$ is absorbed by the bulk $\psi$ and shifts the boundary condition, replacing the $u^{(l)}$ at the bottom of the tower with its shadow dual $d^{(l)}$.  We then explored a model with a quartic coupling $h (\bar \chi \chi)^2$ on the defect, and using bosonization argued that there is a range of $h$ for which the bilinear coupling remains relevant.  We conjectured the IR fixed point is the same -- the boundary conditions of $\psi$ are shifted and $\chi$ is absorbed -- but it would be nice and perhaps feasible to solve this model exactly.  Finally, we increased the number of defect fermions $\chi$ from 1 to $N_f$.
At the mixed fixed point, the anomalous dimensions violate the double-twist condition, while the equation of motion relates the two shadow channels with precisely the structure required to cancel the spurious bulk singularity. 
Indeed, the takeaway message from this work is that when one of the shadow dual pair of operators is below the unitarity bound, a double twist relation is imposed on three-point functions involving the remaining operator, while when both operators are allowed, their correlation functions are related by the shadow transform.

\section*{Acknowledgments}

We would like to thank ChatGPT and Claude for discussion.
C.H.\ thanks the Oxford Maths and Physics Departments for hospitality during his sabbatical.  
The authors would like to thank the INI for Mathematical Sciences, Cambridge, for support and hospitality during the programme ``BIDS in QFT'', where work on this paper commenced. This work was supported in part by 
STFC grant ST/X000753/1  and EPSRC grant EP/Z000580/1.

\appendix

\section{Embedding space}
\label{sec:embeddingspace}
Since the conformal group in $d$-dimensional Euclidean space is $SO(d+1,1)$, we can embed coordinates into two dimensions higher, and so work with just linear transformations. The embedded coordinates are defined to satisfy \cite{Weinberg_2010,Costa_2011}
\begin{equation}
    \begin{split}
        \eta_{MN} P^M P^N&=-(P^0)^2+\delta_{\mu\nu}P^\mu P^\nu+(P^{d+1})^2\\
        &=\delta_{\mu\nu}P^\mu P^\nu-P^+P^-\\
        &=0
    \end{split}
\end{equation}
with $\mu,\nu=1,...,d$ and 
\begin{equation}
    P^\pm=P^0\pm P^{d+1}.
\end{equation}
The embedded coordinates are related to the original coordinates by
\begin{equation}
    (P^+,P^-,P^\mu)=(1,x^2,x^\mu) \ ,
\end{equation}
and so satisfy
\begin{equation}
    -2P_1\cdot P_2 =(x_{12})^2.
\end{equation}
Since we are working in Euclidean space, the gamma matrices are defined by 
\begin{equation}
    \{\gamma_\mu,\gamma_\nu\}=2\delta_{\mu\nu}.
\end{equation}
In the embedding space, the gamma matrices must satisfy 
\begin{equation}
     \{\Gamma_M,\Gamma_N\}=2\eta_{MN} \ ,
\end{equation}
and we can choose the basis
\begin{equation}
    \Gamma_\mu=\begin{pmatrix}
        \gamma_\mu&0\\0&-\gamma_\mu 
    \end{pmatrix},
    \ \Gamma_0=\begin{pmatrix}
        0&\mathbb{I}\\-\mathbb{I} &0
    \end{pmatrix},
    \ \Gamma_{d+1}=\begin{pmatrix}
        0&\mathbb{I}\\\mathbb{I}&0
    \end{pmatrix}.
\end{equation}
Spinors in the embedding space must satisfy the following relations, with the second relation reducing the number of degrees of freedom \cite{Weinberg_2010,Iliesiu_2016,Isono_2017}:
\begin{equation}
\begin{split}
    \psi(\lambda P)&=\lambda^{-\Delta-\frac{1}{2}}\psi(P) \ , \\
    P^M \Gamma_M \,\psi(P)&=0,\\
\end{split}
\end{equation}
To remove the spinor indices, we can contract $\psi$ with a polarisation spinor $\bar{s}$ in real space, and $\bar{S}$ in the embedding space. In order to satisfy the transversality condition $P^M \Gamma_M\, S=0$, we require that 
\begin{equation} \label{s spinor relation}
    \Bar{S}=\bar{s}\begin{pmatrix}
        \slashed{x}&1
    \end{pmatrix}, \ S=\begin{pmatrix}
        1\\-\slashed{x}
    \end{pmatrix} s.
\end{equation}
The transversality condition also means that  
\begin{equation} \label{condition}
    \bar{S}_1 \slashed{P}_1 S_2 =\bar{S}_1 \slashed{P}_2 S_2=0.
\end{equation}
Using eq.\ \eqref{s spinor relation}, we find the replacement rules from embedding to real space: 
\begin{equation}
\begin{split}
    \bar{S}_1S_2 &\rightarrow \bar{s}_1\slashed{x}_{12}s_2 \ , \\
   \bar{S}_1 \slashed{P}_3 S_2 &\rightarrow \bar{s}_1\slashed{x}_{13}\slashed{x}_{32}s_2.\\
\end{split}
\end{equation}
\subsection{Defect spinors}
We split coordinates $P^M$ into two sets, those parallel to the defect and those normal with $M=(A,I)$, $A=0,...,p+1$ and $I=1,...,q$ (so the extra two dimensions are parallel to the defect). Explicitly we have 
\begin{equation}
    P^A=(P^+,P^-,P^a)=(1,x^ax_a+x^ix_i,x^a),
\end{equation}
with 
\begin{equation}
    P^\pm=P^0\pm P^{p+1}, 
\end{equation}
and 
\begin{equation}
    P^I=x^i,
\end{equation}
where $a=1,...,p$ and $i=1,...,q$.
There are two different inner products to consider: one that is invariant under $SO(p+1,1)$ and another which is invariant under $SO(q)$ \cite{Bill__2016}:
\begin{equation}
\begin{split}
     &P_1 \bullet P_2=-2\eta_{AB}P^A_1P^B_2 \ , \\
     &P_1 \circ P_2=-2 \delta_{IJ}P_1^I P_2^J \ . \\
\end{split}
\end{equation}
In real space, these become 
\begin{equation}
\begin{split}
     &P_1 \bullet P_2\rightarrow(x_{12})^a(x_{12})_a+(x_1)^i(x_1)_i+(x_2)^i(x_2)_i \ , \\
     &P_1 \circ P_2\rightarrow -2 (x_1)^i(x_2)_i \ .
\end{split}
\end{equation}
Since $P^2=P\bullet P+P\circ P=0$, it follows that $P\bullet P=-P\circ P$. In section \ref{sec:conformalconstraints},  we use the shorthand $P\circ P$ to denote the positive quantity $-\frac{1}{2}P\circ P$, which reduces to $x^ix_i$ in real space.
\par In the $q=2$ case, there are two normal directions to the defect, which we label by $n_1$ and $n_2$. We denote the gamma matrices in these directions in real space as $\gamma^{n_1}$and $\gamma^{n_2}$, and in embedding space as $\Gamma^{n_1}$and $\Gamma^{n_2}$. 
For spinors that live on the defect, we want to reduce the degrees of freedom from $SO(d+1,1)$ to $SO(p+1,1)$.
This is done by introducing projectors in real and embedding space, that take bulk objects ($\rho$) and reduce them to defect ones ($\rho_\pm$): 
\begin{align}
   & \mathbb{P_{\pm}}=\frac{1}{2}\left(\mathbb{I}\pm i \gamma^{n_1}\gamma^{n_2}\right), & \Pi_\pm=\frac{1}{2}\left(\mathbb{I}\pm i \Gamma^{n_1}\Gamma^{n_2}\right),
\end{align}
with   
\begin{align}
    & \rho_\pm(x)=\mathbb{P_{\pm}}\rho(x),&\rho_\pm(P)=\Pi_{\pm}\rho(P).
\end{align}
The $\pm$ corresponds to defect spinors that either spin clockwise or anticlockwise about the defect. It follows that 
\begin{align}
   & i \gamma^{n_1}\gamma^{n_2} \rho_\pm(x)=\pm\rho_\pm(x),& i \Gamma^{n_1}\Gamma^{n_2} \rho_\pm(P)=\pm\rho_\pm(P).
\end{align}
To define the Dirac adjoint, we use a semi-Lorentzian version of conjugation which combines complex conjugation with time reversal
\begin{equation}
    \bar{\rho}_\pm(x)=\rho_\pm^\ddagger(x) \gamma^1.
\end{equation}
In the embedding space, there is a new timelike direction, so the Dirac adjoint is given by 
\begin{equation}
    \bar{\rho}_\pm(P)=\rho_\pm^\ddagger(P) \Gamma^1 \Gamma^0.
\end{equation}
Therefore, we have
\begin{align}
     &  \bar{\rho}_\pm(x)i \gamma^{n_1}\gamma^{n_2}=\pm\bar{\rho}_\pm(x),&  \bar{\rho}_\pm(P)i \Gamma^{n_1}\Gamma^{n_2}=\pm\bar{\rho}_\pm(P).
\end{align}
Hence, unlike in \cite{Herzog_2023}, there is no exchange in helicity eigenvalue between real and embedding space. This also means that the polarisation spinors satisfy (using the same notation as for the spinors)
\begin{align}
   & i \gamma^{n_1}\gamma^{n_2} s_\pm(x)=\pm s_\pm(x),& i \Gamma^{n_1}\Gamma^{n_2} S_\pm(P)=\pm S_\pm(P),\\
     &  \bar{s}_\pm(x)i \gamma^{n_1}\gamma^{n_2}=\pm\bar{s}_\pm(x),&  \bar{S}_\pm(P)i \Gamma^{n_1}\Gamma^{n_2}=\pm\bar{S}_\pm(P) \ ,
\end{align}
which can be seen from requiring consistency in e.g. $\bar{s}_\pm \rho_\pm=\pm \bar{s}_\pm\,(i \gamma^{n_1}\gamma^{n_2})\rho_\pm $.
 Writing $S_\alpha$ and $s_\alpha$ where $\alpha=\pm1$, we have, for example
\begin{equation}
    \bar S_1S_{2,\alpha}=\bar S_1\Pi_\alpha S_2\rightarrow \bar{s}_1\slashed{x}_{12}\mathbb{P}_\alpha s_2=\bar{s}_1\slashed{x}_{12} s_{2,\alpha}.
\end{equation}

\subsection{Levi-Civita in Embedding Space}

We specialize now to $q=2$.
To form invariant and covariant objects in embedding space, the Levi-Civita tensor is an important object invariant under $SO(n)$ transformations: 
\begin{equation}
    \epsilon^{i_1...i_n}=n!\,V_1^{[i_1}...V_n^{i_n]}
\end{equation}
for an orthonormal basis $\{V_1,...,V_n\}$. In embedding space, we denote the basis vectors parallel to the defect as $V_1,...,V_{p+2}$ and those normal as $V_{n_1},V_{n_2}$, with $V_N^M=\delta_N^M$. For the symmetries of our system, in addition to the identity, there are three possible invariant structures to consider: $V_1^{[M_1}...V_{p+2}^{M_{p+2}]}$ for the $SO(p+1,1)$ symmetry, $V_{n_1}^{[M_1}V_{n_2}^{M_{2}]}$ for the $SO(q)$ symmetry, and $V_1^{[M_1}...V_{p+2}^{M_{p+2}]}V_{n_1}^{[M_{n_1}}V_{n_2}^{M_{n_2}]}$, the product of these. However, since our correlation functions involve bulk spinors (and projections of these to give defect spinors) contracted with the polarisation spinors, the invariant structures must form scalars under $SO(d+1,1)$. Therefore, they will be sandwiched by the polarisation spinors and contracted with the gamma matrices, e.g.:
\begin{equation}
\begin{split}
&\bar{S}_1\Gamma_{M_1}V_1^{[M_1}...\Gamma_{M_{p+2}}V_{p+2}^{M_{p+2}]} S_2\\
    &=\frac{1}{(p+2)!}\bar{S}_1 \Gamma_{M_1}...\Gamma_{M_{p+2}}\epsilon^{M_1...M_{p+2}}S_2\\
    &=\bar{S}_1 \Gamma_1...\Gamma_{p+2}S_2,\\
\end{split}
\end{equation}
where in the final line, we used the anticommutative property of the gamma matrices.

\section{Finding the OPE operators} \label{sec:OPEops}
As an example, we can find the operator ${\mathcal D}_{\shortuparrow +}^{(l)}$ in eq.\ \eqref{OPEspinup} by substituting the OPE into $\braket{\psi_+(x_1)\bar{\rho}^{(l)}_{\shortuparrow++}(\vec{x}_2)}$:
\begin{equation}
    \braket{\psi_+(x_1)\bar{\rho}^{(l)}_{\shortuparrow++}(\vec{x}_2)}={\mathcal D}_{\shortuparrow +}^{(l)}(z,\bar{z},\partial_{\vec{x}_1})\braket{\rho^{(l)}_{\shortuparrow+-}(\vec{x}_1)\bar{\rho}^{(l)}_{\shortuparrow++}(\vec{x}_2)}.
\end{equation}
The sum over $l$ disappears as the correlation function is only nonzero if both $\rho_{\shortuparrow+-}$ and $\rho_{\shortuparrow++}$ have the same value of $l$. Using the known expressions for the correlation functions gives (dropping the constants)
\begin{equation}
    \frac{1}{\bar{z}^l|x_1-\vec{x}_2|^{4-2l}}\begin{pmatrix}
            \bar{z}\\i \bar{w}_{21}
        \end{pmatrix}=\begin{pmatrix}
            \left({\mathcal D}_{\shortuparrow +}^{(l)}\right)_1(z,\bar{z},\partial_{\vec{x}_1})\\\left({\mathcal D}_{\shortuparrow +}^{(l)}\right)_2(z,\bar{z},\partial_{\vec{x}_1})
        \end{pmatrix}\frac{i\bar{w}_{21}}{|\vec{x}_1-\vec{x}_2|^{4-2l}}.
\end{equation}
Without loss of generality, we can set $\vec{x}_2=(w_2,\bar{w}_2)=0$ and drop the subscript on the $x_1$ coordinates:
\begin{equation}
    \frac{1}{\bar{z}^{l}(|w|^2+|z|^2)^{2-l}}\begin{pmatrix}
            \bar{z}\\-i\bar{w}
        \end{pmatrix}=\begin{pmatrix}
            \left({\mathcal D}_{\shortuparrow +}^{(l)}\right)_1(z,\bar{z},\partial_{\vec{x}_1})\\\left({\mathcal D}_{\shortuparrow +}^{(l)}\right)_2(z,\bar{z},\partial_{\vec{x}_1})
        \end{pmatrix}\frac{-i\bar{w}}{|w|^{4-2l}}.
\end{equation}
The left hand side can be written using a series expansion:
\begin{equation}
     \frac{1}{(|w|^2+|z|^2)^{\alpha}}=\frac{1}{|w|^{2\alpha}}\sum_{j=0}^{\infty}\frac{(\alpha)_j}{j!}\left(-\left|\frac{z}{w}\right|^2\right)^j
\end{equation}
and so we find
\begin{equation}
    \begin{split}
         &\frac{1}{\bar{z}^{l-1}}\sum_{j=0}^{\infty}\frac{(2-l)_j}{j!}\frac{(-|z|^2)^j}{w^{2-l+j}{\bar{w}^{2-l+j}}}=\left({\mathcal D}_{\shortuparrow +}^{(l)}\right)_1(z,\bar{z},\partial_{w},\partial_{\bar{w}})\left(\frac{-i}{w^{2-l}\bar{w}^{1-l}}\right)\\
         &\frac{1}{\bar{z}^{l}}\sum_{j=0}^{\infty}\frac{(2-l)_j}{j!}\frac{(-|z|^2)^j}{w^{2-l+j}{\bar{w}^{1-l+j}}}=\left({\mathcal D}_{\shortuparrow +}^{(l)}\right)_2(z,\bar{z},\partial_{w},\partial_{\bar{w}})\left(\frac{1}{w^{2-l}\bar{w}^{1-l}}\right).\\
    \end{split}
\end{equation}
From this, the operators can be deduced to be 
\begin{equation}
    \begin{split}
        &\left({\mathcal D}_{\shortuparrow +}^{(l)}\right)_1(z,\bar{z},\partial_{w},\partial_{\bar{w}})=\frac{-i}{\bar{z}^{l-1}}\sum_{j=0}^{\infty}\frac{(-|z|^2)^j}{(1-l)_{j+1}\,j!}\partial_w^j\partial_{\bar{w}}^{j+1}\\
        &\left({\mathcal D}_{\shortuparrow +}^{(l)}\right)_2(z,\bar{z},\partial_{w},\partial_{\bar{w}})=\frac{1}{\bar{z}^{l}}\sum_{j=0}^{\infty}\frac{(-|z|^2)^j}{(1-l)_{j}\,j!}\partial_w^j\partial_{\bar{w}}^{j}.\\
    \end{split}
\end{equation}

\section{Finding the conformal blocks}\label{sec:conformalblocks}
Taking part of the $_{12}$ component of eq.\ \eqref{conformal block matrix} as an example:  
\begin{equation}
   \begin{split}
   \left( {\mathcal D}^{(l)}_{\shortuparrow +}\right)_1&(x_1) \left(\overline{\mathcal D}^{(l)}_{\shortuparrow -} \right)_2(x_2) 
\langle d^{(l)}_{\shortuparrow +}(\vec x_1) \bar d^{(l)}_{\shortuparrow -} (\vec x_2) \rangle\\
        &=\frac{1}{\bar{z}_1^{l-1}z_2^{l}}\sum_{m,n}\frac{(-|z_1|^2)^m(-|z_2|^2)^n}{(1-l)_{m+1}m!(1-l)_{n}n!}\partial^m_{w_1}\partial^{m+1}_{\bar{w}_1}\partial^n_{w_2}\partial^{n}_{\bar{w}_2}\frac{1}{(w_{21})^{2-l}(\bar{w}_{21})^{1-l}}\\&=\frac{1}{\bar{z}_1^{l-1}z_2^{l}}\sum_{m,n}\frac{(-|z_1|^2)^m(-|z_2|^2)^n}{(1-l)_{m+1}m!(1-l)_{n}n!}\frac{(1-l)_{m+n+1}(2-l)_{m+n}}{(w_{21})^{2-l+m+n}(\bar{w}_{21})^{2-l+m+n}}
       \\
       &=\frac{1}{\bar{z}_1^{l-1}z_2^{l}}\sum_{m,n}\frac{1-l+n}{1-l}\frac{\left((2-l)_{m+n}\right)^2}{(2-l)_{m}m!(2-l)_{n}n!}\frac{(-|z_1|^2)^m(-|z_2|^2)^n}{|w_{12}|^{2(2-l+m+n)}}
       \\
       &=\frac{1}{\bar{z}_1^{l-1}z_2^{l}}\frac{1-l+\frac{1}{2} (z_2 \partial_{z_2} + \bar{z}_2 \partial_{\bar{z}_2})}{1-l}\sum_{m,n}\frac{\left((2-l)_{m+n}\right)^2}{(2-l)_{m}m!(2-l)_{n}n!}\frac{(-|z_1|^2)^m(-|z_2|^2)^n}{|w_{12}|^{2(2-l+m+n)}} \ ,
   \end{split}
\end{equation}
where we used properties of the Pochhammer symbol in the penultimate line. We can convert the double sum into a single one using the following identity ((B.8) of \cite{Dolan_2001}):
\begin{equation}
    \sum_{m,n=0}^{\infty}
\frac{1}{m!\,n!}
\frac{(\lambda)_{m+n}(\kappa)_{m+n}}{(\kappa)_m(\kappa)_n}
\frac{A^m B^n}{D^{2(\lambda+m+n)}}
=
\frac{1}{C^{\lambda}}
\sum_{m=0}^{\infty}
\frac{(\lambda)_{2m}}{m!(\kappa)_m}
\left(\frac{AB}{C^2}\right)^m \ ,
\end{equation}
where $D^2=A+B+C$. If we take $A=-|z_1|^2$, $B=-|z_2|^2$ and $D^2=|w_{12}|^2$, we find that 
\begin{equation}
    C=|w_{12}|^2+|z_1|^2+|z_2|^2=2|z_1||z_2|\chi \ ,
\end{equation}
where $\chi$ is the cross-ratio defined in eq.\ \eqref{chi}. This leads to
\begin{equation}
    \begin{split}
         \left( {\mathcal D}^{(l)}_{\shortuparrow +}\right)_1&(x_1) \left(\overline{\mathcal D}^{(l)}_{\shortuparrow -} \right)_2(x_2) 
\langle d^{(l)}_{\shortuparrow +}(\vec x_1) \bar d^{(l)}_{\shortuparrow -} (\vec x_2) \rangle \\&=\frac{1}{\bar{z}_1^{l-1}z_2^{l}}\frac{1-l+\frac{1}{2} (z_2 \partial_{z_2} + \bar{z}_2 \partial_{\bar{z}_2})}{1-l}\left(\frac{1}{(2|z_1||z_2|\chi)^{2-l}}\sum_{m=0}^\infty\frac{(2-l)_{2m}}{m!(2-l)_{m}}\left(\frac{1}{4\chi^2}\right)^m\right)\\
        &=\frac{1}{\bar{z}_1^{l-1}z_2^{l}}\frac{1-l+\frac{1}{2} (z_2 \partial_{z_2} + \bar{z}_2 \partial_{\bar{z}_2})}{1-l}\left(\frac{1}{(2|z_1||z_2|\chi)^{2-l}}\sum_{m=0}^\infty\frac{\left(\frac{2-l}{2}\right)_{m}\left(\frac{3-l}{2}\right)_{m}}{m!(2-l)_{m}}\left(\frac{1}{\chi^2}\right)^m\right)\\
        &=\frac{1}{\bar{z}_1^{l-1}z_2^{l}}\frac{1-l+\frac{1}{2} (z_2 \partial_{z_2} + \bar{z}_2 \partial_{\bar{z}_2})}{1-l}\left(\frac{1}{(2|z_1||z_2|\chi)^{2-l}}\,{}_2F_1\left(\frac{2-l}{2},\frac{3-l}{2},2-l,\frac{1}{\chi^2}\right)\right)
    \end{split}
\end{equation}
using the definition of the hypergeometric function in the final line.

\section{Bulk-defect-defect OPE results}
\label{sec:bddexpansions}

We present expressions for several bulk-defect-defect correlation functions in the limit $|x_3| \to \infty$.  
The results are:\footnote{%
These were simplified using the identity ${}_2F_1(a,b;c;z)-{}_2F_1(a-1,b;c;z)-\frac{b}{c}z\,{}_2F_1(a,b+1;c+1;z)=0$}
\begin{equation}
\begin{split}
&\left<\psi_+\bar{\rho}^{(l_2)}_{++}\bar \phi^{(l_3)}\right>_2\\
\\
&\quad=\begin{cases}
\begin{aligned}
a^{(l)}_{\shortdownarrow +} \left({\mathcal{U}}^{(l)}_{\downarrow +}\right)_2 \left<{u}_{\downarrow +}^{(l)}(x_1)\bar{\rho}^{(l_2)}_{++}(x_2)\bar \phi^{(l_3)}(\infty)\right>
\end{aligned}
& \text{if }l=l_2+l_3>0,
\\[1.2em]
\begin{aligned}
b^{(l)}_{\shortuparrow +} \,\left(\mathcal{D}^{(l)}_{\shortuparrow +} \right)_2\left<d_{\shortuparrow +}^{(l)}(x_1)\bar{\rho}^{l_2)}_{++}(x_2)\bar \phi^{(l_3)}(\infty)\right>
\end{aligned}
& \text{if }l=l_2+l_3<1.
\end{cases}\\
\\
&\quad=\begin{cases}
\begin{aligned}
i a^{(l)}_{\shortdownarrow +} c_{u \shortdownarrow +}\frac{\left(1+2l+2\Delta_{23}\right)}{4l}\bar w_{12}   |x_{12}|^{-\frac{5}{2} - \Delta_{23}} H(l,v)\left(\frac{z}{|z|} \right)^{l}
\end{aligned}
& \text{if }l>0,
\\[1.5em]
\begin{aligned}
-i b^{(l)}_{\shortuparrow +} c_{d \shortuparrow +}\bar w_{12}   |x_{12}|^{-\frac{5}{2} - \Delta_{23}} H(-l,v)\left(\frac{z}{|z|} \right)^{l}
\end{aligned}
& \text{if }l<1.
\end{cases}
\end{split}
\end{equation}

\begin{equation}
\begin{split}
&\left<\psi_-\bar{\rho}^{(l_2)}_{++}\bar \phi^{(l_3)}\right>_1\\
\\
&\quad=\begin{cases}
\begin{aligned}
a^{(l)}_{\shortdownarrow -} \left( {\mathcal{U}}^{(l)}_{\shortdownarrow -} \right)_1\left< u_{\shortdownarrow -}^{(l)}(x_1)\bar{\rho}^{(l_2)}_{++}(x_2)\bar \phi^{(l_3)}(\infty)\right>
\end{aligned}
& \text{if }l=l_2+l_3>0,
\\[1.2em]
\begin{aligned}
b^{(l)}_{\shortuparrow -} \, \left(\mathcal{D}^{(l)}_{\shortuparrow -} \right)_1 \left<d^{(l)}_{\shortuparrow -} (x_1)\bar{\rho}^{(l_2)}_{++}(x_2)\bar \phi^{(l_3)}(\infty)\right>
\end{aligned}
& \text{if }l=l_2+l_3<1.
\end{cases}\\
\\
&\quad=\begin{cases}
\begin{aligned}
i a^{(l)}_{\shortdownarrow -}  c_{u \shortdownarrow -}\,\bar{z}\,\bar{w}_{12}   |x_{12}|^{-\frac{7}{2} - \Delta_{23}} (1-v)^{\frac{1}{2}} (-v)^{-\frac{1}{2}} H(l-1,v)\left(\frac{z}{|z|} \right)^{l}
\end{aligned}
& \text{if }l>0,
\\[1.5em]
\begin{aligned}
i b^{(l)}_{\shortuparrow -} c_{d \shortuparrow -}\frac{\left(3 + 2 \Delta_{23} -2l\right)}{4(1-l)}\,\bar{z}\,\bar{w}_{12}   |x_{12}|^{-\frac{7}{2} - \Delta_{23}} (1-v)^{\frac{1}{2}} (-v)^{-\frac{1}{2}} H(1-l,v)\left(\frac{z}{|z|} \right)^{l}
\end{aligned}
& \text{if }l<1.
\end{cases}
\end{split}
\end{equation}

\par For the $\alpha=-1$ case we have 
\begin{equation}
\begin{split}
&\left<\psi_+\bar{\rho}^{(l_2)}_{-+}\bar \phi^{(l_3)}\right>_2\\
\\
&\quad=\begin{cases}
\begin{aligned}
a^{(l)}_{\shortuparrow +} \left({\mathcal{U}}^{(l)}_{\shortuparrow +}\right)_2 \left<{u}_{\shortuparrow +}^{(l)}(x_1)\bar{\rho}^{(l_2)}_{-+}(x_2)\bar \phi^{(l_3)}(\infty)\right>
\end{aligned}
& \text{if }l=l_2+l_3>-1,
\\[1.2em]
\begin{aligned}
b^{(l)}_{\shortdownarrow +} \,\left(\mathcal{D}^{(l)}_{\shortdownarrow +} \right)_2\left<d_{\shortdownarrow +}^{(l)}(x_1)\bar{\rho}^{l_2)}_{-+}(x_2)\bar \phi^{(l_3)}(\infty)\right>
\end{aligned}
& \text{if }l=l_2+l_3<0.
\end{cases}\\
\\
&\quad=\begin{cases}
\begin{aligned}
-i a^{(l)}_{\shortuparrow +} c_{u \shortuparrow +}\frac{\left(3 + 2 \Delta_{23} +2l\right)}{4(1+l)}\,z \,\bar{w}_{12}   |x_{12}|^{-\frac{7}{2} - \Delta_{23}} (1-v)^{\frac{1}{2}} (-v)^{-\frac{1}{2}} H(1+l,v)\left(\frac{z}{|z|} \right)^{l}
\end{aligned}
& \text{if }l>-1,
\\[1.5em]
\begin{aligned}
-i b^{(l)}_{\shortdownarrow +} c_{d \shortdownarrow +}z\,\bar{w}_{12}   |x_{12}|^{-\frac{7}{2} - \Delta_{23}} (1-v)^{\frac{1}{2}} (-v)^{-\frac{1}{2}} H(-l-1,v)\left(\frac{z}{|z|} \right)^{l}
\end{aligned}
& \text{if }l<0.
\end{cases}
\end{split}
\end{equation}

\begin{equation}
\begin{split}
&\left<\psi_-\bar{\rho}^{(l_2)}_{-+}\bar \phi^{(l_3)}\right>_1\\
\\
&\quad=\begin{cases}
\begin{aligned}
a^{(l)}_{\shortuparrow -} \left( {\mathcal{U}}^{(l)}_{\shortuparrow -} \right)_1\left< u_{\shortuparrow -}^{(l)}(x_1)\bar{\rho}^{(l_2)}_{-+}(x_2)\bar \phi^{(l_3)}(\infty)\right>
\end{aligned}
& \text{if }l=l_2+l_3>-1,
\\[1.2em]
\begin{aligned}
b^{(l)}_{\shortdownarrow -} \, \left(\mathcal{D}^{(l)}_{\shortdownarrow -} \right)_1 \left<d^{(l)}_{\shortdownarrow -} (x_1)\bar{\rho}^{(l_2)}_{-+}(x_2)\bar \phi^{(l_3)}(\infty)\right>
\end{aligned}
& \text{if }l=l_2+l_3<0.
\end{cases}\\
\\
&\quad=\begin{cases}
\begin{aligned}
i a^{(l)}_{\shortuparrow -}  c_{u \shortuparrow -}\,\bar w_{12}   |x_{12}|^{-\frac{5}{2} - \Delta_{23}} H(l,v)\left(\frac{z}{|z|} \right)^{l}
\end{aligned}
& \text{if }l>-1,
\\[1.5em]
\begin{aligned}
i b^{(l)}_{\shortdownarrow -} c_{d \shortdownarrow -}\,\frac{\left(1-2l+2\Delta_{23}\right)}{4l}\bar w_{12}   |x_{12}|^{-\frac{5}{2} - \Delta_{23}} H(-l,v)\left(\frac{z}{|z|} \right)^{l}
\end{aligned}
& \text{if }l<0.
\end{cases}
\end{split}
\end{equation}
As stated earlier, if $\alpha$ swaps sign, the sign of $l$ in the solutions to the Dirac equation also swaps:
\begin{eqnarray} \label{bulk-defect-defect f1 alpha minus}
f_{1}(v) &=& C_{1-} H(l, v) + C_{2-} H(-l,v) \ , \\
f_{4}(v) &=& \frac{4(v-1) f_{1}'(v)
+ (-2l(1-\frac{1}{v}) + (2 \Delta_{23}-3  )) f_{1}(v) }{ 2\Delta_{23}-2l+1} \ ,
\end{eqnarray}
\begin{eqnarray}
f_{2}(v) &=& (1-v)^{\frac{1}{2}}  (-v)^{-\frac{1}{2}}\left[  C_{3-} H(1+l, v)+  C_{4-} H(-l-1,v) \right] \ , \\
f_{3}(v) &=&\frac{4v(v-1) f_{2}'(v)
+ ((2 \Delta_{23}+2l-1) v+2(-l-2)) f_{2}(v) }{(2\Delta_{23}+2l+3)(v-1)} \ .
\end{eqnarray}

\subsection{Expansions of bulk-defect-defect functions near $\zeta =1$}
\label{sec:singularity}

\par 
Here are expansions of bulk-defect-defect three-point functions near the singular locus $\zeta = 1$.
To remove the singular behaviour from $f_2$ and $f_3$ in the $\alpha = +$ case, we consider
\begin{equation}
\begin{split}
&\left<\psi_-\bar{\rho}^{(l_2)}_{++}\bar \phi^{(l_3)}\right>_1
\sim\\
\\
&\quad\begin{cases}
\begin{aligned}
i C_{4+}\ \frac{2^{-\frac{1}{2}-\Delta_{2}-\Delta_{3}}\, \Gamma\!\left(l\right)}
{\Gamma\!\left(\frac{1}{4}(3+2l+2\Delta_{23})\right)\,
 \Gamma\!\left(\frac{1}{4}(1+2l-2\Delta_{23})\right)\,\epsilon}
\end{aligned}
& \text{if }l=l_2+l_3>0,
\\[1.2em]
\begin{aligned}
i C_{3+}\ \frac{2^{-\frac{1}{2}-\Delta_{2}-\Delta_{3}}\, \Gamma\!\left(2-l\right)}
{\Gamma\!\left(\frac{1}{4}(7-2l+2\Delta_{23})\right)\,
 \Gamma\!\left(\frac{1}{4}(5-2l-2\Delta_{23})\right)\,\epsilon}
\end{aligned}
& \text{if }l=l_2+l_3<1,
\end{cases}
\end{split}
\end{equation}

\begin{equation}
\left<\psi_-\bar{\rho}^{(l_2)}_{++}\bar \phi^{(l_3)}\right>_2\sim -i\left<\psi_-\bar{\rho}^{(l_2)}_{++}\bar \phi^{(l_3)}\right>_1.
\end{equation}

Rewriting these in terms of the OPE defect operator dimensions, we have 
\begin{equation}
    \begin{alignedat}{2}
        \Delta_3&=-1+\Delta_{u \shortdownarrow -}+\Delta_2+2n\ , \qquad
        &\Delta_2&=-2+\Delta_{u \shortdownarrow -}+\Delta_3+2n \ , \\
        \Delta_3&=-2+\Delta_{d \shortuparrow -}+\Delta_2+2n\ , \qquad
        &\Delta_2&=-1+\Delta_{d \shortuparrow -}+\Delta_3+2n \ ,\\
    \end{alignedat}
\end{equation}
for $n=1,2,3...$. For the third of these, the $n=1$ case, $\Delta_3=\Delta_{d \shortuparrow -}+\Delta_2$,  cannot be seen using the series expansion, as it corresponds to the special case in eq.\ \eqref{n=0 f3} where the solution is regular, even though the general solution for $f_3$ diverges. In terms of the conformal weights, we have 
\begin{equation}
    \begin{alignedat}{2}
        h_3&=-1+h_{u\shortdownarrow -}+h_2+n\ , \qquad
        &h_2&=-1+h_{u\shortdownarrow -}+h_3+n \ ,\\
        h_3&=-1+h_{d \shortuparrow -}+h_2+n\ , \qquad
        &h_2&=h_{d \shortuparrow -} +h_3+n \ .\\
    \end{alignedat}
\end{equation}

\par Similarly, for the $\alpha=-1$ case we have
\begin{equation}
\begin{split}
&\left<\psi_+\bar{\rho}^{(l_2)}_{-+}\bar \phi^{(l_3)}\right>_1
\sim\\
\\
&\quad\begin{cases}
\begin{aligned}
C_{3-}\ \frac{2^{-\frac{1}{2}-\Delta_{2}-\Delta_{3}}\, \Gamma\!\left(2+l\right)}
{\Gamma\!\left(\frac{1}{4}(7+2l+2\Delta_{23})\right)\,
 \Gamma\!\left(\frac{1}{4}(5+2l-2\Delta_{23})\right)\,\epsilon}
\end{aligned}
& \text{if }l=l_2+l_3>-1,
\\[1.2em]
\begin{aligned}
C_{4-}\ \frac{2^{-\frac{1}{2}-\Delta_{2}-\Delta_{3}}\, \Gamma\!\left(-l\right)}
{\Gamma\!\left(\frac{1}{4}(3-2l+2\Delta_{23})\right)\,
 \Gamma\!\left(\frac{1}{4}(1-2l-2\Delta_{23})\right)\,\epsilon}
\end{aligned}
& \text{if }l=l_2+l_3<0,
\end{cases}
\end{split}
\end{equation}
\begin{equation}
\left<\psi_+\bar{\rho}^{(l_2)}_{-+}\bar \phi^{(l_3)}\right>_2\sim -i\left<\psi_+\bar{\rho}^{(l_2)}_{-+}\bar \phi^{(l_3)}\right>_1.
\end{equation}
\begin{equation}
    \begin{alignedat}{2}
        \Delta_3&=-2+\Delta_{u \shortuparrow +}+\Delta_2+2n\ , \qquad
        &\Delta_2&=-1+\Delta_{u \shortuparrow +}+\Delta_3+2n \ , \\
        \Delta_3&=-1+\Delta_{d \shortdownarrow +}+\Delta_2+2n\ , \qquad
        &\Delta_2&=-2+\Delta_{d \shortdownarrow +}+\Delta_3+2n \ ,\\
    \end{alignedat}
\end{equation}
\begin{equation}
    \begin{alignedat}{2}
        h_3&=-1+h_{u\shortuparrow +}+h_2+n\ , \qquad
        &h_2&=h_{u\shortuparrow +}+h_3+n \ ,\\
        h_3&=-1+h_{d \shortdownarrow +}+h_2+n\ , \qquad
        &h_2&=-1+h_{d \shortdownarrow +} +h_3+n \ .\\
    \end{alignedat}
\end{equation}

\begin{equation}
\begin{split}
&\left<\psi_-\bar{\rho}^{(l_2)}_{-+}\bar \phi^{(l_3)}\right>_1
\sim\\
\\
&\quad\begin{cases}
\begin{aligned}
i C_{1-}\ \frac{2^{-\frac{1}{2}-\Delta_{2}-\Delta_{3}}\, \Gamma\!\left(1+l\right)}
{\Gamma\!\left(\frac{1}{4}(5+2l+2\Delta_{23})\right)\,
 \Gamma\!\left(\frac{1}{4}(3+2l-2\Delta_{23})\right)\,\epsilon}
\end{aligned}
& \text{if }l=l_2+l_3>-1,
\\[1.2em]
\begin{aligned}
i C_{2-}\ \frac{2^{-\frac{1}{2}-\Delta_{2}-\Delta_{3}}\, \Gamma\!\left(1-l\right)}
{\Gamma\!\left(\frac{1}{4}(5-2l+2\Delta_{23})\right)\,
 \Gamma\!\left(\frac{1}{4}(3-2l-2\Delta_{23})\right)\,\epsilon}
\end{aligned}
& \text{if }l=l_2+l_3<0,
\end{cases}
\end{split}
\end{equation}
\begin{equation}
\left<\psi_-\bar{\rho}^{(l_2)}_{-+}\bar \phi^{(l_3)}\right>_2\sim -i\left<\psi_-\bar{\rho}^{(l_2)}_{-+}\bar \phi^{(l_3)}\right>_1.
\end{equation}
\begin{equation}
    \begin{alignedat}{2}
        \Delta_3&=-1+\Delta_{u \shortuparrow -}+\Delta_2+2n\ , \qquad
        &\Delta_2&=-2+\Delta_{u \shortuparrow -}+\Delta_3+2n \ , \\
        \Delta_3&=-2+\Delta_{d \shortdownarrow -}+\Delta_2+2n\ , \qquad
        &\Delta_2&=-1+\Delta_{d \shortdownarrow -}+\Delta_3+2n \ ,\\
    \end{alignedat}
\end{equation}
\begin{equation}
    \begin{alignedat}{2}
        h_3&=-1+h_{u\shortuparrow -}+h_2+n\ , \qquad
        &h_2&=-1+h_{u\shortuparrow -}+h_3+n \ ,\\
        h_3&=-1+h_{d \shortdownarrow -}+h_2+n\ , \qquad
        &h_2&=h_{d \shortdownarrow -} +h_3+n \ .\\
    \end{alignedat}
\end{equation}

\bibliographystyle{JHEP}
\bibliography{bibliography}

@article{Neretin:2024,
    author = "Neretin, Yury A.",
    title = "{On the Dotsenko{\textendash}Fateev complex twin of the Selberg integral and its extensions}",
    eprint = "2212.09112",
    archivePrefix = "arXiv",
    primaryClass = "math.CA",
    doi = "10.1007/s11139-023-00804-3",
    journal = "Ramanujan J.",
    volume = "64",
    number = "1",
    pages = "37--55",
    year = "2024"
}

@article{PhysRevLett.63.1996,
    author = "Varma, C. M. and Littlewood, P. B. and Schmitt-Rink, S. and Abrahams, E. and Ruckenstein, A. E.",
    title = "{Phenomenology of the normal state of Cu-O high-temperature superconductors}",
    doi = "10.1103/PhysRevLett.63.1996",
    journal = "Phys. Rev. Lett.",
    volume = "63",
    pages = "1996--1999",
    year = "1989"
}

@article{Moy:2026iey,
    author = "Moy, Benjamin and Fradkin, Eduardo",
    title = "{Chern insulator boundary criticality}",
    eprint = "2608.18199",
    archivePrefix = "arXiv",
    primaryClass = "cond-mat.str-el",
    month = "8",
    year = "2026"
}

@article{Faulkner:2010tq,
    author = "Faulkner, Thomas and Polchinski, Joseph",
    title = "{Semi-Holographic Fermi Liquids}",
    eprint = "1001.5049",
    archivePrefix = "arXiv",
    primaryClass = "hep-th",
    reportNumber = "NSF-KITP-10-010",
    doi = "10.1007/JHEP06(2011)012",
    journal = "JHEP",
    volume = "06",
    pages = "012",
    year = "2011"
}

@article{Mandelstam:1975hb,
    author = "Mandelstam, S.",
    editor = "Stone, M.",
    title = "{Soliton Operators for the Quantized Sine-Gordon Equation}",
    reportNumber = "Print-75-0198 (UC,BERKELEY)",
    doi = "10.1103/PhysRevD.11.3026",
    journal = "Phys. Rev. D",
    volume = "11",
    pages = "3026",
    year = "1975"
}

@article{Coleman:1974bu,
    author = "Coleman, Sidney R.",
    editor = "Stone, M.",
    title = "{The Quantum Sine-Gordon Equation as the Massive Thirring Model}",
    reportNumber = "Print-74-1722 (HARVARD)",
    doi = "10.1103/PhysRevD.11.2088",
    journal = "Phys. Rev. D",
    volume = "11",
    pages = "2088",
    year = "1975"
}

@article{Ghosh:2015opa,
    author = "Ghosh, Sudip and Gupta, Rajesh Kumar and Jaswin, Kasi and Nizami, Amin A.",
    title = "{$\epsilon$-Expansion in the Gross-Neveu model from conformal field theory}",
    eprint = "1510.04887",
    archivePrefix = "arXiv",
    primaryClass = "hep-th",
    doi = "10.1007/JHEP03(2016)174",
    journal = "JHEP",
    volume = "03",
    pages = "174",
    year = "2016"
}

@article{Herzog:2017xha,
    author = "Herzog, Christopher P. and Huang, Kuo-Wei",
    title = "{Boundary Conformal Field Theory and a Boundary Central Charge}",
    eprint = "1707.06224",
    archivePrefix = "arXiv",
    primaryClass = "hep-th",
    doi = "10.1007/JHEP10(2017)189",
    journal = "JHEP",
    volume = "10",
    pages = "189",
    year = "2017"
}

@article{Fraser-Taliente:2024lea,
    author = "Fraser-Taliente, Ludovic and Herzog, Christopher P. and Shrestha, Abhay",
    title = "{A nonlocal Schwinger model}",
    eprint = "2412.02514",
    archivePrefix = "arXiv",
    primaryClass = "hep-th",
    doi = "10.1007/JHEP06(2025)252",
    journal = "JHEP",
    volume = "06",
    pages = "252",
    year = "2025"
}

@article{Jensen:2015swa,
    author = "Jensen, Kristan and O'Bannon, Andy",
    title = "{Constraint on Defect and Boundary Renormalization Group Flows}",
    eprint = "1509.02160",
    archivePrefix = "arXiv",
    primaryClass = "hep-th",
    reportNumber = "OUTP-15-19P, YITP-SB-15-33",
    doi = "10.1103/PhysRevLett.116.091601",
    journal = "Phys. Rev. Lett.",
    volume = "116",
    number = "9",
    pages = "091601",
    year = "2016"
}

@article{Bianchi:2021snj,
    author = "Bianchi, Lorenzo and Chalabi, Adam and Proch{\'a}zka, Vladim{\'\i}r and Robinson, Brandon and Sisti, Jacopo",
    title = "{Monodromy defects in free field theories}",
    eprint = "2104.01220",
    archivePrefix = "arXiv",
    primaryClass = "hep-th",
    reportNumber = "UUITP- 16/21",
    doi = "10.1007/JHEP08(2021)013",
    journal = "JHEP",
    volume = "08",
    pages = "013",
    year = "2021"
}

@article{Chai:2021wac,
    author = "Chai, Noam and Chakraborty, Soumangsu and Goykhman, Mikhail and Sinha, Ritam",
    title = "{Long-range fermions and critical dualities}",
    eprint = "2110.00020",
    archivePrefix = "arXiv",
    primaryClass = "hep-th",
    doi = "10.1007/JHEP01(2022)172",
    journal = "JHEP",
    volume = "01",
    pages = "172",
    year = "2022"
}

@article{DiPietro:2019hqe,
    author = "Di Pietro, Lorenzo and Gaiotto, Davide and Lauria, Edoardo and Wu, Jingxiang",
    title = "{3d Abelian Gauge Theories at the Boundary}",
    eprint = "1902.09567",
    archivePrefix = "arXiv",
    primaryClass = "hep-th",
    doi = "10.1007/JHEP05(2019)091",
    journal = "JHEP",
    volume = "05",
    pages = "091",
    year = "2019"
}

@article{Lauria:2020emq,
    author = "Lauria, Edoardo and Liendo, Pedro and van Rees, Balt C. and Zhao, Xiang",
    title = "{Line and surface defects for the free scalar field}",
    eprint = "2005.02413",
    archivePrefix = "arXiv",
    primaryClass = "hep-th",
    reportNumber = "CPHT-RR003.012021, DESY-20-082",
    doi = "10.1007/JHEP01(2021)060",
    journal = "JHEP",
    volume = "01",
    pages = "060",
    year = "2021"
}

@article{Behan:2020nsf,
    author = "Behan, Connor and Di Pietro, Lorenzo and Lauria, Edoardo and Van Rees, Balt C.",
    title = "{Bootstrapping boundary-localized interactions}",
    eprint = "2009.03336",
    archivePrefix = "arXiv",
    primaryClass = "hep-th",
    reportNumber = "CPHT-RR112.122020",
    doi = "10.1007/JHEP12(2020)182",
    journal = "JHEP",
    volume = "12",
    pages = "182",
    year = "2020"
}

@article{Behan:2021tcn,
    author = "Behan, Connor and Di Pietro, Lorenzo and Lauria, Edoardo and van Rees, Balt C.",
    title = "{Bootstrapping boundary-localized interactions II. Minimal models at the boundary}",
    eprint = "2111.04747",
    archivePrefix = "arXiv",
    primaryClass = "hep-th",
    doi = "10.1007/JHEP03(2022)146",
    journal = "JHEP",
    volume = "03",
    pages = "146",
    year = "2022"
}

@article{Herzog_2023,
    author = "Herzog, Christopher P. and Schaub, Vladimir",
    title = "{Fermions in boundary conformal field theory: crossing symmetry and E-expansion}",
    eprint = "2209.05511",
    archivePrefix = "arXiv",
    primaryClass = "hep-th",
    doi = "10.1007/JHEP02(2023)129",
    journal = "JHEP",
    volume = "02",
    pages = "129",
    year = "2023"
}

@article{Bill__2016,
    author = "Bill{\`o}, Marco and Gon{\c{c}}alves, Vasco and Lauria, Edoardo and Meineri, Marco",
    title = "{Defects in conformal field theory}",
    eprint = "1601.02883",
    archivePrefix = "arXiv",
    primaryClass = "hep-th",
    doi = "10.1007/JHEP04(2016)091",
    journal = "JHEP",
    volume = "04",
    pages = "091",
    year = "2016"
}

@article{Herzog_2022,
    author = "Herzog, Christopher P. and Shrestha, Abhay",
    title = "{Conformal surface defects in Maxwell theory are trivial}",
    eprint = "2202.09180",
    archivePrefix = "arXiv",
    primaryClass = "hep-th",
    doi = "10.1007/JHEP08(2022)282",
    journal = "JHEP",
    volume = "08",
    pages = "282",
    year = "2022"
}

@book{itzykson1980quantum,
  title={Quantum Field Theory},
  author={Itzykson, C. and Zuber, J.B.},
  isbn={9780070320710},
  lccn={lc78040977},
  series={International series in pure and applied physics},
  year={1980},
  publisher={McGraw-Hill International Book Company}
}

@article{Dolan_2001,
    author = "Dolan, F. A. and Osborn, H.",
    title = "{Conformal four point functions and the operator product expansion}",
    eprint = "hep-th/0011040",
    archivePrefix = "arXiv",
    reportNumber = "DAMTP-2000-125",
    doi = "10.1016/S0550-3213(01)00013-X",
    journal = "Nucl. Phys. B",
    volume = "599",
    pages = "459--496",
    year = "2001"
}

@article{Weinberg_2010,
    author = "Weinberg, Steven",
    title = "{Six-dimensional Methods for Four-dimensional Conformal Field Theories}",
    eprint = "1006.3480",
    archivePrefix = "arXiv",
    primaryClass = "hep-th",
    reportNumber = "UTTG-04-10",
    doi = "10.1103/PhysRevD.82.045031",
    journal = "Phys. Rev. D",
    volume = "82",
    pages = "045031",
    year = "2010"
}

@article{Costa_2011,
    author = "Costa, Miguel S. and Penedones, Joao and Poland, David and Rychkov, Slava",
    title = "{Spinning Conformal Correlators}",
    eprint = "1107.3554",
    archivePrefix = "arXiv",
    primaryClass = "hep-th",
    reportNumber = "LPTENS-11-22, NSF-KITP-11-128",
    doi = "10.1007/JHEP11(2011)071",
    journal = "JHEP",
    volume = "11",
    pages = "071",
    year = "2011"
}

@article{Iliesiu_2016,
    author = "Iliesiu, Luca and Kos, Filip and Poland, David and Pufu, Silviu S. and Simmons-Duffin, David and Yacoby, Ran",
    title = "{Bootstrapping 3D Fermions}",
    eprint = "1508.00012",
    archivePrefix = "arXiv",
    primaryClass = "hep-th",
    reportNumber = "PUPT-2480",
    doi = "10.1007/JHEP03(2016)120",
    journal = "JHEP",
    volume = "03",
    pages = "120",
    year = "2016"
}

@article{Isono_2017,
    author = "Isono, Hiroshi",
    title = "{On conformal correlators and blocks with spinors in general dimensions}",
    eprint = "1706.02835",
    archivePrefix = "arXiv",
    primaryClass = "hep-th",
    doi = "10.1103/PhysRevD.96.065011",
    journal = "Phys. Rev. D",
    volume = "96",
    number = "6",
    pages = "065011",
    year = "2017"
}

@article{Barkeshli:2025cjs,
    author = "Barkeshli, Maissam and Fechisin, Christopher and Komargodski, Zohar and Zhong, Siwei",
    title = "{Disclinations, Dislocations, and Emanant Flux at Dirac Criticality}",
    eprint = "2501.13866",
    archivePrefix = "arXiv",
    primaryClass = "cond-mat.str-el",
    doi = "10.1103/kfd3-qtk7",
    journal = "Phys. Rev. X",
    volume = "16",
    number = "1",
    pages = "011017",
    year = "2026"
}

\end{document}